\documentclass[utf8,latin1,showpacs,3p,preprint,floats,superscriptaddress,10pt,sort&compress]{elsarticle}
\usepackage{dcolumn}
\usepackage{epsfig}

\usepackage[english]{babel}
\usepackage{amsmath,amssymb,euscript,wasysym}
\usepackage{mathtools}
\usepackage{placeins}
\usepackage{comment}

\usepackage{xspace} 
\newcommand{\rfrac}[2]{{{}^{#1}\!/_{#2}}} 

\usepackage{ifthen}
\newcommand{\hideandshow}[1]{%
 \ifthenelse{\isundefined{\showme}}{}{#1}}
\newcommand{\showandhide}[1]{%
 \ifthenelse{\isdefined{\showme}}{}{#1}}

\usepackage{color}
\usepackage{ulem}
\renewcommand{\emph}[1]{\textit{#1}}
\definecolor{darkblue}{rgb}{0,0,0.5}
\definecolor{darkgreen}{rgb}{0,0.5,0}
\definecolor{darkred}{rgb}{.7,0,0}
\definecolor{purple}{rgb}{0.8,0,1}
\definecolor{orange}{rgb}{1,0.5,0}
\definecolor{grey}{rgb}{.6,.6,.6}
\definecolor{lightpink}{rgb}{1,0.7,0.75}
\definecolor{pink}{rgb}{1,0.4,0.58}
\definecolor{deeppink}{rgb}{1,0.08,0.58}

\newcommand{{\HHM}}{{3HHM}\xspace} 

\newcommand{\AHM}{{\rm AHM}}

\newcommand{\bath}{{\rm bath+hyb}}
\newcommand{\interact}{{\rm int}}
\newcommand{\charge}{{\rm ch}}
\newcommand{\spin}{{\rm sp}}
\newcommand{\orb}{{\rm orb}}

\newcommand{\flavor}{{\rm fl}}
\newcommand{\pdag}{{\phantom{\dagger}}}

\newcommand{\Tkspin}{\ensuremath{T_{\rm K}^\spin}\xspace}
\newcommand{\Tkorb}{\ensuremath{T_{\rm K}^\orb}\xspace}

\newcommand{\Sec}[1]{Sec.~\ref{#1}}

\newcommand{\Eq}[1]{Eq.~\eqref{#1}}
\newcommand{\Eqs}[1]{Eqs.~\eqref{#1}}

\newcommand{\Fig}[1]{Fig.~\ref{#1}}
\newcommand{\Figs}[1]{Figs.~\ref{#1}}
\newcommand{\FIG}[1]{Figure~\ref{#1}}

\def\imag{\mathrm{Im}\,} 
\def\real{\mathrm{Re}\,} 

\def\refA{\hyperref[sec:spinfreeze]{(D1)}\xspace} 
\def\refB{\hyperref[sec:Janus_intro]{(D2)}\xspace} 
\def\refC{\hyperref[sec:prox2MIT]{(D3)}\xspace} 
\def\refD{\hyperref[sec:sos]{(D4)}\xspace} 

\newcommand{\gone}{g1}  
\newcommand{\gtwo}{g2}  
\newcommand{\eone}{e1}  
\newcommand{\etwo}{e2}  

\RequirePackage[
   hyperindex,colorlinks,bookmarksnumbered,
  plainpages=true,pdfstartview=FitH]{hyperref}
 \hypersetup{linkcolor=blue,urlcolor=blue,citecolor=blue}
\usepackage{hyperref}

\begin{document}

\title{
Hundness versus Mottness in a three-band Hubbard-Hund model:\\
On the origin of strong correlations in Hund metals
\tnoteref{t1}
}
\tnotetext[t1]{
K.M.S., A.W., and J.v.D. acknowledge support
from the excellence initiative NIM; A.W. was also supported by
WE4819/1-1 and WE4819/2-1 until 12/2017
and by US DOE under contract number
DE-SC0012704 since. G.K. was supported by National
Science Foundation grant DMR-1733071. 
Author contributions:  K.M.S. and G.K. proposed this project;
K.M.S. performed the DMFT+NRG calculations;  A.W. developed the 
NRG code and assisted K.M.S. in the initial stages of the
DMFT+NRG computation.
K.M.S. drafted the manuscript with the help of G.K., A.W and J.v.D.}


%
\author[label1]{K. M. Stadler\corref{cor1}} 
\ead{Katharina.M.Stadler@physik.uni-muenchen.de}
 \cortext[cor1]{Corresponding author.}
\address[label1]{Physics
  Department, Arnold Sommerfeld Center for Theoretical Physics and
  Center for NanoScience, Ludwig-Maximilians-Universit\"at M\"unchen,
  80333 M\"unchen, Germany} 
  \author[label2]{G. Kotliar} 
  \ead{kotliar@physics.rutgers.edu}
  \address[label2]{Department of Physics and Astronomy, 
  Rutgers University, Piscataway, NJ 08854, USA} 
 \author[label1,label3]{A. Weichselbaum} 
\ead{weichselbaum@bnl.gov}
\address[label3]{Condensed Matter Physics and Materials Science Department,
  Brookhaven National Laboratory, Upton, New York 11973, USA}

\author[label1]{J. von Delft} 
\ead{vondelft@lmu.de}

  %
\begin{abstract}

  Hund metals are multi-orbital systems with moderate
  Coulomb interaction, $U$, among charges and sizeable Hund's rule
  coupling, $J(<U)$, that aligns the spins in different orbitals. They
  show strong correlation effects, like very low Fermi-liquid
  coherence scales and intriguing incoherent transport regimes,
  resulting in bad-metallic behavior. But to what extent are these
  strong correlations governed by Mottness, i.e. the blocking of
  charge fluctuations close to a Mott insulator transition (MIT)
  induced by $U$, or by Hundness, a new route towards strong
  correlations induced by $J$?  To answer this question, we study the
  full phase diagram of a degenerate three-band Hubbard-Hund model on
  a Bethe lattice at zero temperature using single-site dynamical
  mean-field theory and the numerical renormalization group as
  efficient real-frequency multi-band impurity solver. Hund metal
  behavior occurs in this minimal model for a filling close to
  $n_d=2$, moderate $U$ and sizeable $J$, the ``Hund-metal regime".
  In particular, strong correlations manifest themselves there by an unusually low
  quasiparticle weight.  Generalizing previous results on this model, we show that
 ``spin-orbital separation'' (SOS) is a generic Hund's-coupling-induced feature in the whole
  metallic regime of the phase diagram for $1<n_d<3$ and sizeable $J$.
  There orbital screening always occurs at much higher energies than
  spin screening below which Fermi-liquid behavior sets in.  The low
  quasiparticle weight can then be directly explained in terms of the
  Hund's-coupling-reduced Fermi-liquid scale.  We carefully analyze
  the effect of $J$ (Hundness), and the effect of the MIT at $n_d=2$
  and $n_d=3$ (Mottness) on the energy scales and the nature of SOS.
  \textit{In the Hund-metal regime, far from any MIT, Hundness -- the
    localization of large spins -- is shown to be the key player to
    induce strong correlations}. There, physical properties are
  governed by a broad incoherent energy regime of SOS where intriguing
  Hund metal physics occurs: large, almost unscreened spins are
  \textit{coupled} to screened orbital degrees of freedom.  With
  increasing proximity to an MIT correlations are further enhanced and
  the Fermi-liquid scale is further reduced. However, in the
  Hund-metal regime, this effect of Mottness is minor. In contrast,
  very close to the MIT at $n_d=2$, the incoherent spin-orbital
  separation regime is strongly downscaled and becomes 
 negligibly small, whereas Mottness -- the localization of charges -- becomes dominant
  in inducing strong correlations.  Close to the MIT at $n_d=3$, the
  SOS regime widens up because the orbital degrees of freedom get
  blocked by the formation of an $S{=}\rfrac{3}{2}$
   impurity spin, but its nature changes: the orbital and spin dynamics 
   get decoupled.  Our results confirm Hundness as a distinct mechanism towards strong
  correlations in the normal state of Hund metals, leading to various
  interesting implications for the nature of electronic transport.
\end{abstract}


\maketitle


\section{Introduction and Motivation} 
\label{sec:Motiv}

\subsection{Bad-metal superconductors}
Iron-based high-temperature superconductors   \citep{Hosono2006,
Hosono2008} (HTSCs) are ``bad metals".  On the one hand,  in their
superconducting state (with critical temperatures up to 56K
 \citep{Ren2008,Ren2008a,Wu2009}), they are perfect conductors with
dissipationless supercurrents; on the other hand,  in their normal
state they conduct surprisingly badly. But which fundamental
physical mechanism causes this bad-metallic behavior? 
Interestingly, this bad-metallic behavior is not found in
conventional BCS-like superconductors, but it is reminiscent of the
unconventional normal state of (doped) cuprate HTSCs. These  are
known to be strongly correlated and  the conventional
superconducting mechanism based on electron-phonon coupling is most
likely not strong enough to generate their high critical
temperatures.

There is firm evidence that strong correlation effects play a key
role in iron-based HTSCs, as well. In their paramagnetic phase,
these materials exhibit anomalous and bad transport properties that
are characterized by very low Fermi-liquid (FL) coherence scales
 \citep{Haule2009,Yin2012,Yi2013,Hardy2013}.
Above the FL
scale puzzling non-Fermi-liquid (NFL) behavior
 \citep{Haule2009,Yin2012,Yi2013,Hardy2013,Haule2008,Liebsch2010,Ishida2010,Aichhorn2010,Werner2012,Schafgans2012,Fink2013,Yi2015} 
occurs  in a large intermediate (paramagnetic)  energy window,
typically at or slightly below room temperature, 
together with poorly screened,
large fluctuating local moments,
as observed in observed in X-ray emission spectroscopy measurements \citep{Gretarsson2011, Werner2016b,Lafuerza2017}.
At higher temperatures, the
resistivity reaches unusually large values  that exceed the
Mott-Ioffe-Regel limit \citep{Haule2009,Hardy2013}.
In accordance,
various experiments
revealed particularly large mass enhancements \citep{Yi2013,Hardy2013,Fink2013,Yi2015,Qazilbash2009,Terashima2010,Terashima2010a,Tamai2010,Yamasaki2010,Borisenko2010,Terashima2013,Yoshida2014}.

\subsection{Hundness versus Mottness  in multi-orbital bad metals}

Since  the ``standard model'' of a Fermi liquid in
condensed matter theory breaks down
in the presence of strong correlations, both the superconducting and
the bad-metal normal state are still poorly understood in the
iron-based HTSCs. In particular, one widely but controversially
debated fundamental question pertains to the origin of strong
correlations: is it  ``Hundness" or ``Mottness"?

Cuprate HTSCs are widely considered as doped charge-transfer Mott
insulators \citep{Orenstein2000,Lee2006}. Strong correlations arise here due to Mottness:  the
proximity to a Mott-insulator transition (MIT), i.e. a transition 
at a critical interaction strength $U_c$
from an (increasingly correlated) metal to an insulator, which is
driven by a large Coulomb repulsion, $U$. In theoretical
descriptions,  the original multi-band electronic structure of
cuprates is usually reduced to a low-energy effective (two-dimensional) one-band
Hubbard model, such that $U$  acts only between electrons in one
orbital per lattice site and the MIT occurs at half-filling for
undoped cuprates.

In contrast, doped \textit{and} undoped iron-based HTSCs are (bad)
metallic materials with an effective \textit{multi}-band description that
allows for an additional type of interaction: Hund's rule coupling,
$J$ (Hundness),  which favors the alignment of spins in different
orbitals on the same (iron) atom and {consequently correlates the
electron hopping in terms of a non-trivial interplay of orbital and
spin degrees of freedom
 \citep{Haule2009,Yin2012,Aron2015}. 
In iron-based HTSCs and other multi-band
materials,  the strong correlation effects may thus be caused by either
Hundness, or Mottness, or a combination of both.  

Therefore, the following question has been raised
 \citep{Yin2012,Fanfarillo2015}:
what is the role of  ``Hundness versus  
Mottness'' as origin of strong correlations
in multi-orbital bad metals? Here we address this question 
from a fundamental model-based
point of view: we investigate the zero-temperature properties of 
a toy model, the degenerate three-band
Hubbard-Hund model  Hamiltonian (\HHM)  \citep{Yin2012,Aron2015,Stadler2015}, 
using single-site dynamical mean-field theory (DMFT) and
a highly-efficient multi-band numerical renormalization group (NRG)
impurity solver \citep{Stadler2015,Weichselbaum2007, Weichselbaum2012a, Weichselbaum2012b}
to tackle the correlated many-body problem. 
A central theme of our work is spin-orbital separation (SOS).
It was first revealed and argued to be related to anomalous power 
law behavior for the Matsubara self-energy in Ref.~\citep{Yin2012},
further analyzed using perturbative scaling arguments in Ref.~\citep{Aron2015}, 
and conclusively established by a detailed DMFT+NRG analysis in Ref.~\citep{Stadler2015}.
(For a complementary study, where we focussed on finite-temperature 
properties not addressed in this paper, see Ref.~\citep{Deng2018}.)

We next summarize the state of research on multi-band models 
motivating and providing the basis for this article.


\section{Scope and Aim} 
\label{sec:Intro}

\subsection{Hund metals}

For a long time strong electronic correlations in materials have
exclusively been associated with the proximity to a MIT evoked by
$U$, i.e. to the suppression of charge fluctuations. 
The MIT was extensively studied in one-band systems
 \citep{Mott1968,Rozenberg1995,Kotliar1999,Kotliar2002}, 
including the cuprate HTSCs. But a MIT also occurs at
any integer filling of multi-orbital materials. Examples are various
3d (and 4d) transition metal oxides 
with the prototypal Mott material $V_2O_3$ \citep{Deng2018,McWhan1969,
McWhan1973, McWhan1973a, Hansmann2013}.

Soon after the discovery of the iron pnictides \citep{Hosono2006,
Hosono2008}, it was realized that the special
multi-orbital character of these HTSCs (and many other strongly
correlated materials) allows for a new mechanism towards heavy
effective masses: Hundness \citep{Haule2009}.  This new class of materials was dubbed
``Hund metals"  \citep{Yin2011,Georges2013} and includes multi-orbital
materials like iron pnictides and chalcogenides \citep{Haule2009,Yin2012,Werner2012,
Schafgans2012,Yin2011,Georges2013,Lanata2013,
Bascones2015a,deMedici2017},
as well as various transition metal oxides of the 3d and 4d
series, such as ruthenates  \citep{Yin2012,Georges2013,Werner2008,Mravlje2011,deMedici2011,Stricker2014,Mravlje2016}.
Hund metals are characterized by rather broad bands leading to
sizeable Hund's coupling strengths compared to only moderate Coulomb
interactions, which are strongly screened in these materials due to
the large spatial extension of the correlated orbitals
 \citep{Mravlje2011,Kroll2008}. 

Interestingly, bad-metal behavior can be  found in essentially all these Hund metals.
Although the importance of Hund's
coupling in realistic materials is increasingly being appreciated
there is still an ongoing debate whether Hundness or Mottness is the
key player in renormalizing the electron masses of Hund metals. This
debate  is strongly driven by the fact that, indeed, striking
analogies in the (doping-temperature) phase diagrams of cuprate and
iron-based HTSCs hint towards a common framework. For instance, in
both cases superconductivity emerges in the vicinity of an
incoherent metallic regime with NFL properties and unconventional
spin dynamics.
So, ultimately,  understanding the normal state of Hund metals might
lead to deeper insights into  the superconducting mechanism in
HTSCs. 

\subsection{Hund models}

A very basic approach to address the issue of ``Hundness versus
Mottness" in Hund metals is to study the paramagnetic phase diagram
of Hubbard-Kanamori-like model Hamiltonians (for a review, see
Ref.~\citep{Georges2013}). These
take into account two spin and \textit{multiple} ($N_c$) orbital
degrees of freedom, a Coulomb interaction, $U$, and, most
importantly, a finite ferromagnetic Hund's coupling, $J$. Hund-metal
physics is then captured by these models for a filling, $n_d$, close 
to one charge away from half-filling: $n_d\approx N_c\pm1$. This is
motivated by the particle-hole asymmetry of real Hund materials. For
instance, the average occupancy of the five Fe $3d$ orbitals is $d6$
for the  undoped stoichiometric parent compounds of almost all
iron-based HTSC families \citep{deMedici2017}. Small to moderate
crystal field splittings, as well as hole or electron doping lead
to variations in the occupancy, such that the electron densities
can range between $5.5$ and $6.3$ electrons per iron
atom \citep{Yin2011,deMedici2017}.
Assuming a fully filled $e_g$ duplet, this leaves $n_d\sim2$
electrons for three active $t_{2g}$ orbitals.
Similarly, ruthenates have an average filling of  approximately
four electrons  in 
$t_{\rm 2g}$-orbitals. 

Here we study the minimal model  \citep{Yin2012,Aron2015} with relevance
for Hund metals, the \HHM,  presented in Sec.~\ref{sec:Models}. It
involves three degenerate orbitals.  We thus
fully neglect any material-specific details like crystal-field
splitting or realistic band structures, although undoubtedly present
in real materials. Our aim in this study is to focus attention on the
most generic aspects of Hundness and Mottness in the maximally simple
context of full orbital degeneracy, in order to reveal which many-body
effects can be understood on this simple model level and which ones
require full information of the electronic structure.
Since Hund's rule coupling is only effective for a site 
occupation that is larger than one electron (and smaller than one hole),
we simulate fillings $1<n_d<3$ with particular emphasis on $n_d=2$.
(By the particle-hole symmetry of the model with respect to half-filling,
this also describes the fillings $3<n_d<5$.)

\subsection{ Phase diagram and bad-metal regime}

Our work is motivated by the results of various single-site dynamical
mean-field theory (DMFT)  \citep{Georges2013,Werner2008,deMedici2011,
  deMedici2011a} and slave-boson  \citep{Fanfarillo2015,deMedici2017,
  deMedici2016} studies of degenerate three-band
Hubbard-Kanamori-type models that reproduced basic Hund metal physics: in
the $n_d$-$U$ phase diagram at finite $J$, they found strongly
correlated, bad-metallic behavior in an extended region around a
filling of $n_d=2$, which we dub ``Hund-metal regime'' 
(hatched area in Fig.~\ref{fig:sketch_phasediagram} at moderate $U\ll U_c^{(2)}$).
Naturally, bad-metallic behavior (light regions in
Fig.~\ref{fig:sketch_phasediagram}) occurs close to the MIT at
$n_d=2$, but interestingly, it also ranges down to rather small
Coulomb interaction strengths $U\ll U_c^{(2)}$, provided that
  Hund's coupling $J$ is sizable (in a sense
defined at the end of Sec.~\ref{sec:Hloc2:mult}).  Most Hund
metals can be placed there, around one charge away from half-filling
and at moderate $U$.  Further, the bad-metallic regime (light area)
also reaches out to the insulating state at half-filling, $n_d=3$,
where the MIT develops already at a very low critical interaction
strength, $U_c^{(3)}$.

\begin{figure}
\centering
\includegraphics[width=0.5\linewidth, trim=8mm 0mm 0mm 0mm, clip=true]{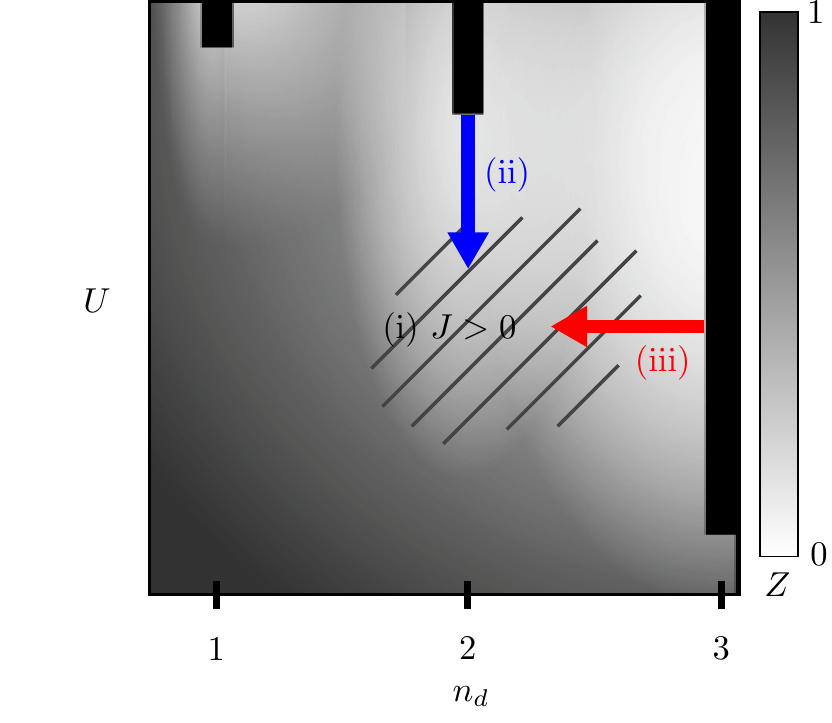}
\caption{
  Schematic sketch of
  the $n_d$-$U$ phase diagram for the 
  \HHM at finite
  $J$. We only show half of the phase diagram, as it is symmetric
  with respect to half-filling.
  The shading reflects the quasiparticle (QP) weight $Z$. 
  Darker regions (large $Z$) indicate
  good metallic, lighter regions (small $Z$) bad-metallic behavior.
  The black bars mark Mott insulating phases.  At all integer
  fillings, a MIT  occurs above a (different) critical interaction
  strength, $U_c$. Interestingly, an extended light region exists
  also at moderate $U\ll U_c^{(2)}$ around $n_d=2$ (and reaches to
  $n_d=3$). In this ``Hund-metal regime''
  (hatched area), 
  where most Hund metals can be placed \citep{deMedici2011}, strong
  electronic correlations might either be induced by Hundness, (i) the presence
  of sizeable $J$, or Mottness, (ii) the   
  influence of the MIT at $n_d=2$ (blue
  arrow), and (iii) the influence of the MIT at $n_d=3$ (red arrow), or a
  combination of these scenarios.
}
\label{fig:sketch_phasediagram}
\end{figure}

Bad-metal behavior 
manifests itself by
a small quasiparticle (QP) weight $Z$. Based on a coherent FL
QP picture, $Z$ quantifies the weight of the coherent
quasiparticle peak (QPP) of the local spectral function (correlated
density of states). Within DMFT and slave-boson methods, the inverse
QP weight is equivalent to the electronic mass
enhancement, $Z^{-1}=m^*/m$, and thus serves as  measure for strong
electronic correlations. For Hund models with $N_c>3$ bands,
equivalent regions of low $Z$ (hatched area) were revealed around all integer
fillings $1<n_d<N_c$ \citep{Fanfarillo2015,Georges2013,deMedici2016}, 
but they are most prominent at $n_d=N_c\pm1$ (see
supplement of Ref.~\citep{Fanfarillo2015}).

We note that in the presence of crystal-field splitting $Z$ and the
filling can acquire an orbital dependence: various simulations
 \citep{Yin2012,Liebsch2010,Ishida2010,Fanfarillo2015,Yin2011,Lanata2013,deMedici2009,Yu2012,
Yu2013,Misawa2012,Bascones2012,Huang2012,deMedici2014,Werner2016,Kugler2018} and measurements
 \citep{Yi2013,Hardy2013,Yi2015,Terashima2013,Yoshida2014}
suggest the occurrence of  orbital differentiation  and
even orbital selective Mott phases (OSMP), depending on the type and
strength of the splitting \citep{Huang2012,Kugler2018}.  In the phase
diagram, both effects seem to intensify with increasing $J$,
increasing $U$, and decreasing distance to half-filling.  
A thorough understanding of the physics of degenerate multi-orbital
models is a prerequisite for exploring these effects
of orbital selectivity.
However, pronounced orbital differentiation is considered
to be relevant only for some Hund metals, e.g. for ruthenates. In
contrast, for iron-pnictides, it might be less important. In the
class of the iron-based HTSCs, only iron chalcogenides are expected
to be at the verge of an orbital-selective MIT \citep{Yin2011,
Yu2013}.  Interestingly, many materials with large orbital
differentiation  are either no superconductors or exhibit only low
transition temperatures: indeed, large  orbital differentiation is
argued to be harmful for
superconductivity \citep{Yin2011}.

\subsection{The Hund-metal problem}

In principle, three scenarios seem possible to induce strong
correlation effects and to lead to the bad-metallic behavior (low
$Z$) in the Hund-metal regime (hatched area in 
Fig.~\ref{fig:sketch_phasediagram}) of  a \HHM  
also sufficiently far way from any Mott insulating state
(black bars in \Fig{fig:sketch_phasediagram}):
\begin{itemize} 

\item[(i)] Hundness: sizeable $J$ is the key player to induce
strong correlations with considerable electronic mass enhancements.

\item[(ii)]  Mottness at $n_d=2$:  the interaction-induced  MIT at
one charge away from half-filling,  $n_d=2$, triggers the strong
correlations (blue arrow in Fig.~\ref{fig:sketch_phasediagram}). 

\item[(iii)] Mottness at $n_d=3$: the strong correlations are emanated
by  the half-filled Mott insulator (red arrow in
Fig.~\ref{fig:sketch_phasediagram}). 

\end{itemize}

Scenario (i) suggests a new route towards strong correlations:
Hundness. Sizeable Hund's
rule coupling, $J$, leads to the formation of high-spin states and to
the suppression of $Z$.  It goes back to Ref.~\citep{Haule2009}
and is supported in various
publications \citep{Yin2012,Schafgans2012,Stadler2015,Deng2018,
Yin2011,Georges2013,Mravlje2011,deMedici2011,Mravlje2016,Yin2011a}. 

Scenario (ii) is not much discussed in the literature, as $U_c^{(N_c-1)}$ is
large  while $U$ has moderate values for
Hund metals. 

Scenario (iii) is motivated by the cuprate picture of doped
half-filled Mott-insulators and advocated by several
authors \citep{Ishida2010,Fanfarillo2015,deMedici2017,deMedici2016}.
 In this scenario the existence of  finite
$J$ would have a subordinate role in correlating the electrons 
by lowering $U_{c1}^{(3)}$.

Although all the model calculations cited above confirmed that
strong correlation effects dominate the Hund-metal regime of the
phase diagram their  origin and nature have been under debate  even
for this toy model until today, either based on different physical
interpretations or just on inconsistent terminology. In particular,
scenarios (i)-(iii) have been discussed   
in the context of \refA 
the existence of a spin-freezing phase
 \citep{Werner2008}, \refB 
the ``Janus-faced" influence of Hund's rule
coupling  \citep{Georges2013,deMedici2011,deMedici2011a}, and
\refC various proximity
effects of the half-filled MIT  \citep{Fanfarillo2015,deMedici2017,deMedici2016}, 
such as Hund's-coupling-induced Fermi-liquid
instabilities \citep{deMedici2016}. In this work we will 
elucidate the role of another very fundamental effect:
\refD spin-orbital separation (SOS) \citep{Yin2012,Aron2015,Stadler2015}. 
We will show that the phenomena
\refA, \refB, and \refC 
are directly connected to  \refD.
Based on this insight, we will study scenarios (i)-(iii)
by revisiting \refA-\refC 
from the perspective of SOS
in Sec.~\ref{sec:Results}.

\subsubsection*{(D1) Spin-freezing phase}\label{sec:spinfreeze} 

The so-called spin-freezing phase characterizes the Hund-metal
regime in terms of a spin-spin correlation function with an
unusually slow (imaginary-time) decay, which does not approach zero
but a constant at finite temperature. In this picture, scattering
off Hund's-coupling-induced large composite and very long-lived (or
even frozen, static) magnetic moments leads to the incoherent
transport behavior. 

The spin-freezing scenario was introduced in 2008 in a first
(finite-temperature) DMFT study \citep{Werner2008} of the
$n_d$-$U$ phase diagram of a degenerate three-band Hubbard-Kanamori
model using a  Quantum Monte Carlo (QMC)
impurity solver. Later it was extended to (realistic) five-band
calculations for iron-pnictides  \citep{Liebsch2010,Ishida2010,
Werner2012, Werner2016b} (demonstrating the importance of Hund's
rule coupling and electronic correlations for the formation of local
moments in the paramagnetic phase \citep{Werner2016b}) and to models
with crystal-field-splitting  \citep{Werner2016} and spin-orbit
coupling  \citep{Werner2017}. In 2015 it led to the proposal of a
fluctuating-moment-induced spin-triplet superconducting mechanism
for strontium ruthenates and uranium compounds \citep{Werner2015}.

The transition into the Hund-metal regime was first interpreted as a
quantum phase transition from a paramagnetic metallic FL phase (at
small $n_d$ and/or small $U$) to an incoherent metallic NFL phase with
frozen local moments (at larger $n_d$ and/or larger
$U$) \citep{Werner2008}. Since 2011, the existence of a FL ground state
(with fully screened local moments) has been anticipated in the
Hund-metal regime and has led to the picture of a spin-freezing
crossover at finite temperatures -- although the complete decay of the
imaginary-time spin-spin correlation function to zero has not
been explicitly demonstrated until recently \citep{Kowalski2018},
because, in general,  QMC solvers do not have access to low
enough temperatures \citep{Werner2012,deMedici2011}. However, a new QMC
  technique using (super) state-sampling \citep{Kowalski2018} was able
  to show the FL ground state in the spin freezing-phase for fillings
  up to $n_d=2.63$.

Spin-freezing has been assumed to originate, in principle, from (i)
Hundness.  However, similar to $Z$, the spin-freezing phenomenon is
considered to be strongly doping dependent and is very pronounced in
the vicinity of the half-filled Mott insulator  \citep{Werner2008,
Werner2015}.  Interestingly, the crossover towards spin-freezing  near
$n_d=N_c-1$ is characterized by a steep drop of $Z$ as a function of
$n_d$. A detailed quantitative analysis if and how the spin-freezing
phenomenon is connected to $Z$, induced by (i) Hundness and/or
influenced by Mottness of kind (ii) or (iii) has not yet been
performed. One reason for this is that the mass enhancement could only be 
computed in an approximate manner because the QMC solver did not
reach the FL regime  \citep{Werner2015}. Further, data was only
available on the imaginary Matsubara frequency axis. 

\subsubsection*{(D2) Janus-faced influence of Hund's rule coupling} 
\label{sec:Janus_intro}

The ``Janus-faced" influence of Hund's rule coupling was a  major
result of a first more detailed DMFT+
QMC study of the phase diagram
of the degenerate three-band Hubbard-Kanamori model
 \citep{Georges2013,deMedici2011,deMedici2011a} (including a realistic
classification of various 3d and 4d transition metal oxides via
their mass enhancements).  A detailed exploration in terms of the
QP weight, $Z$, revealed that Hund's coupling induces
apparent conflicting tendencies at $n_d=2$.
On the one hand, increasing $J$
promotes metallicity by shifting the critical interaction strength,
$U_c^{(2)}$, of the MIT at $n_d=2$ to higher values. On the other hand, at moderate
$U$, increasing $J$ reduces $Z$, supporting scenario (i) that
Hund's-coupling-induced strong correlations lead to  bad-metallic
behavior far from a Mott phase.
Together, 
this Janus-faced behavior results in an interesting MIT
for sizeable $J$ upon increasing $U$
that is qualitatively different from the MIT of one-band and
multi-band Hubbard models without Hund's coupling: starting from a
weakly correlated metal at small $U$, the system first evolves into
a strongly correlated metal which is stable  for an extended range
of $U$ values and characterized by very small $Z$, before it
eventually reaches the Mott insulating phase at large $U_c^{(2)}$.  

The degenerate three-band study of Refs.~\citep{deMedici2011,deMedici2011a}
was followed  by similar analyses for up to five bands, both
with \citep{Fanfarillo2015,Georges2013, deMedici2016}
and without \citep{deMedici2016,Yu2012,Bascones2012,Bascones2010} 
orbital degeneracy, revealing qualitatively similar
behavior as in the three-band case.  For degenerate models,
Janus-faced behavior emerges for any integer filling away from
single and half-filling. 

But even for the degenerate three-band model the origin of the
Janus-faced behavior has not been fully revealed.  
Obviously both the QPP itself and the opening of the
insulating Mott gap  are affected at the same time by changing $J$.
Previous studies \citep{Georges2013,deMedici2017,deMedici2011,deMedici2011a}
quantified these changes  by performing a Hubbard-I-type analysis for the
gap dependence and by calculating $Z$ to characterize the
QPP. However, without access to (reliable)
real-frequency spectral data, the Hubbard-I predictions could never
be explicitly verified and the physical origin of the low $Z$ could
only be speculated about. A connection to the low coherence scale in
Hund metals was assumed but never proven, and the nature of the
incoherent regime remained unclear. Although considered, a clear
connection between  spin-freezing and the Janus-faced behavior has
not yet been demonstrated.  Moreover, we note that the value of $Z$
can have an error of up to $10\%$ in these DMFT+QMC simulations (see
supplement of Ref.~\citep{deMedici2011}), also strongly
affecting the values of $U_c$.

We therefore conclude that both scenarios (i) and (ii) should be
revisited. In particular, the Janus-faced behavior has to be
disentangled by identifying a measure for Mottness (ii)
which does not change with $J$, in order to study the pure effect of
Hundness (i), and to analyze the difference in nature between
strongly correlated Hund metals at moderate $U$ and strongly-correlated 
systems close to the MIT.  Scenario (iii) will be
considered in the context of \refC. 

\subsubsection*{(D3) Proximity to the half-filled MIT}
\label{sec:prox2MIT}

At half-filling, $n_d=3$, $U_c^{(3)}$ is strongly reduced. The region of
low $Z$ in Fig.~\ref{fig:sketch_phasediagram}  directly starts at
the border of the MIT at $n_d=3$ 
and extends, even at moderate $U$,
from there to $n_d=2$ with $Z$ slightly increasing when passing  from
$n_d=3$ to $n_d=2$. Such a filling-dependence is  observed  in
simulations and experiments of iron-based superconductors: their
correlations are enhanced with hole-doping (i.e approaching
half-filling) \citep{Werner2016b,Terashima2013,deMedici2014,Sudayama2011,Hardy2016}. 
Furthermore, also the spin-freezing
phenomenon \citep{Werner2008} is strongly doping dependent: the spin
freezing phase occurs in the vicinity of the half-filled MIT.

Motivated by this behavior it has been argued in
Refs.~\citep{Fanfarillo2015,deMedici2017,deMedici2016} that the
suppression of $Z$ around $n_d=2$ at moderate $U$ is connected to
the MIT at half-filling, $n_d=3$. In particular, the effect of
suppressing intra-orbital double occupancy by $J$ has been regarded
as a direct link to the MIT at $n_d=3$ \citep{Fanfarillo2015}.
However, it has been noted that in contrast to  the  one-band
Hubbard model, the reduction of $Z$ in Hund metals does not imply
the general suppression of charge fluctuations (far from the MIT,  
as shown in Ref.~\citep{Deng2018})
and $Z$ is thus not a good measure for the latter: the origin of low
$Z$ and its filling dependence is subtle. Again, DMFT+NRG real
frequency data can help to further investigate this issue by
complementing the slave-boson approaches of
Refs.~\citep{Fanfarillo2015,deMedici2017} and quantitatively
revealing the connection between spin-freezing and $Z$. 

We note that for non-degenerate models, low $Z$ is argued to be
induced by the ``proximity to a half-filled MIT", as well, but here,
the half-filled MIT denotes an orbital selective Mott transition:
when an orbital is individually half-filled it can become
insulating, independently of the other orbitals \citep{deMedici2014}.
This orbital decoupling effect is enhanced by Hund's coupling, but
will not be discussed further in this work.

In a slave-boson study \citep{deMedici2016} of degenerate and
non-degenerate multi-band Hund models, a zone of negative
compressibility,
$\kappa_{\rm el}=\frac{\partial n_d}{\partial \mu}<0$, is observed
at zero temperature for nonzero $J$ in the $n_d$-$U$ phase
diagram, above $U\ge U_c$, reaching (depending on $N_c$) from
half-filling towards $n_d=N_c+1$. The transition from
$\kappa_{\rm el}>0$ to $\kappa_{\rm el}<0$ is realized through a
divergence of the compressibility, which occurs in the phase diagram
together with a strong reduction in $Z$. In the absence of symmetry
breaking in the model, this divergence is interpreted as a genuine
thermodynamic Hund's-coupling-induced instability towards a phase
separation. The enhancement of $\kappa_{\rm el}$ has even been argued
to be directly connected to the enhanced critical $T_c$ of
HTCS \citep{deMedici2011a,deMedici2016,deMedici2009,Huang2012}.
This strong statement of a negative compressibility is solely the
  result of slave-boson approaches (rotationally invariant form of the
  Kotliar-Ruckenstein slave-bosons for the full Hubbard-Kanamori model
  involving two bands, and slave-spin mean-field approximation for the
  Hubbard-Kanamori model without spin flip and pair hopping term
  involving up to five bands).  It has so far not been validated by
  another (zero-temperature) method.

In order to investigate if the suppression of $Z$ in the Hund-metal regime is mediated by 
the MIT at half-filling  and to  check if a negative compressibility is a generic 
Hund's-coupling-induced effect (i.e. independent of details of the model and the method),
  we will  also study scenario (iii), the effect of the MIT at
$n_d=3$ on $Z$ and $\kappa_{\rm el}$.

\subsubsection*{(D4) Spin-orbital separation (SOS)}
\label{sec:sos}

Besides the phenomena \refA, \refB, and \refC, 
also a Hund's-coupling-induced
coherence-incoherence crossover  with increasing temperature has
been discussed as a new and generic normal state property of Hund
metals in the literature  \citep{Haule2009, Yin2012}. 
Further an incoherent
frequency regime with anomalous power-law exponents in the 
Matsubara self-energy was
revealed for $1.5\lesssim n_d\lesssim2.5$, which is most pronounced
at $n_d=2$ \citep{Yin2012,Werner2008}. The incoherent  temperature and
frequency regime was proposed to be induced by two degrees of
freedom that behave in different ways: the orbital degrees of
freedom are quenched and fluctuate very rapidly
while the spin degrees of freedom are unquenched
and fluctuate albeit 
slowly (accordingly the local spin susceptibility has Curie-Weiss
form and a large static value) \citep{Yin2012, Deng2018}. An analytic
RG analysis in the Kondo regime \citep{Aron2015} provided a simple
understanding of the origin of the incoherent regime and established
how the Kondo scales depend on the representations of the spin and
orbital operators.

However, still, several issues needed to be clarified: in particular, the
DMFT+QMC calculations could not reach sufficiently low temperatures
to fully reveal the FL phase. To settle this issue, zero- (and
finite-) temperature, real-frequency DMFT+NRG calculations were
performed in 2015 in Ref.~\citep{Stadler2015} for the
\HHM of Eq.(\ref{eq:HU-Hloc}) at $n_d=2$. These
calculations clearly confirmed that, at zero temperature, finite Hund's
coupling leads to SOS 
[see Fig.~\ref{fig:sos_sketch}(a)] -- a two-stage screening process, in
which orbital screening occurs at much higher energies than spin
screening --  thus strongly reducing the coherence scale below which
a FL ground state is formed. Importantly, at intermediate
energies above the coherence scale, a broad incoherent regime opens
up involving  screened, delocalized orbitals which are non-trivially
coupled to almost unscreened, large, localized spins. The incoherent
frequency regime is strongly particle-hole asymmetric and displays
approximate power-law behavior in the self-energy for positive 
real frequencies only,
leading to apparent fractional power laws  on the imaginary
Matsubara axis.  SOS also occurs in pure
impurity calculations without DMFT self-consistency. With
increasing temperature, SOS in frequency space
translates to a coherence-incoherence crossover for
temperature-dependent quantities. Only recently, this two-stage
crossover was  confirmed in realistic DFT+DMFT+QMC simulations of
the temperature dependence of the thermopower, entropy \citep{Mravlje2016} 
and the local spin and
orbital susceptibilities \citep{Deng2018} for ${\rm Sr}_2{\rm RuO}_4$.
SOS is thus considered to be relevant not only
for degenerated toy models but also for realistic Hund materials
featuring tetragonal crystal-field splitting of the $t_{2g}$
orbitals.

However, in Ref.~\citep{Stadler2015} SOS
was studied  only at $n_d=2$ for a small set of parameters $U$ and
$J$, which (as will be shown in Fig.~\ref{fig:P1_nd2_1fig}) lie at
the border of the coexistence region of the phase diagram, thus
close to the MIT. Therefore many open questions remained: Is
SOS a generic phenomenon of Hund metals? Where
does it occur in the phase diagram and how is it influenced by $J$
and the proximity to the MIT at $n_d=2$ and $n_d=3$? How is it
connected to the phenomena of \refA-\refC  
and how to the low $Z$ in the
Hund-metal regime? And most importantly, what is the origin of
SOS, scenario (i), (ii), or (iii), or a combination of these?

\subsection{Aim of this paper}

The aim of this work is to identify the origin of strong
correlations in the Hund-metal regime of the \HHM,
based on real-frequency data, and to develop
from this a global, unified and consistent scenario for strong
correlation effects in Hund metals. For this we study scenarios
(i)-(iii), i.e. ``Hundness versus Mottness", by scanning the full phase
diagram of the \HHM at zero temperature, using
DMFT+NRG.  In DMFT the lattice model (the \HHM) is mapped self-consistently
onto a quantum impurity model [the Anderson-Hund model (AHM) 
of Eq.~(\ref{eq:AHM})], which we solve with NRG, a powerful
\textit{real-frequency} multi-band impurity solver. NRG is well
suited for the investigation of Hund and Mott physics as it both
reveals the spectral properties  of Hund metals down to its very low
coherence scales and still captures the main features of the Hubbard
side bands.  We thus provide, for the first time, detailed and
\textit{unbiased} real-frequency spectral data in a large parameter
space of the phase diagram instead of only measuring the strength of
strong correlations by analyzing the behavior of $Z$, as done in
previous studies \citep{Fanfarillo2015,Georges2013,deMedici2011,deMedici2011a}.
This allows us to reveal the origin of those correlations and the
physical nature of the incoherent regime in Hund metals.

The paper is structured as follows. In Sec.~\ref{sec:Models} we give
a detailed description of our model and discuss its local multiplet 
level structure at $n_d=2$ (in particular its dependency on $J$) 
and at $n_d=3$.
The DMFT+NRG method is introduced in Appendix~\ref{appendix}.     
In Sec.~\ref{sec:Results} we present our main insights: we will show
that the low $Z$ in the Hund-metal regime results directly from the
suppression of the coherence scale due to SOS. SOS
therefore forms the basis of our main study
and scenarios (i)-(iii) will be investigated from that perspective.
In particular, we follow a three-fold approach in
Sec.~\ref{sec:Results}. We revisit \refA
the spin-freezing phase in Sec.~\ref{sec:spinfreezing}, \refB 
the Janus-faced influence of Hund's
rule coupling in Sec.~\ref{sec:Janus}, and the influence of \refC  
the  MIT at half-filling in Sec.~\ref{sec:half-filled-MIT},
and explain these aspects step by step within the SOS  framework.

\subsection{Model and Methods}
\label{sec:Models}

For our \HHM we use the Hamiltonian of 
Refs.~ \citep{Yin2012,Aron2015,Stadler2015,Deng2018} \vspace{0mm} in the form
\begin{subequations}
\label{eq:HU-Hloc}
\begin{eqnarray}
  \hat{H}_{\rm HHM} & = & \sum_{i} \left(  - \mu \hat{n}_i
+ \hat{H}_\interact [\hat d^\dag_{i\nu}] \right)  \label{eq:HU}
+\sum_{\langle ij\rangle \nu} t\,
  \hat{d}^{\dagger}_{i\nu}\hat{d}^{\phantom{\dagger}}_{j\nu}  , 
\\
\hat{H}_\interact[\hat d^\dag_{i\nu} ] \label{eq:Hloc1}
& = & U \sum_{\langle m} \hat n^\dag_{im\uparrow}\hat{n}^{\phantom{\dagger}}_{im\downarrow}  
  +  (U-J) \sum_{ m\neq m'} \hat n^\dag_{im\uparrow}\hat{n}^{\phantom{\dagger}}_{im'\downarrow} 
   +  (U-2J) \sum_{ m < m',\sigma} \hat n^\dag_{im\sigma}\hat{n}^{\phantom{\dagger}}_{im'\sigma}  \notag \\
   & - &  J \sum_{m \neq m'} \hat d^\dag_{im\uparrow}\hat{d}^{\phantom{\dagger}}_{im\downarrow}\hat d^\dag_{im'\downarrow}\hat{d}^{\phantom{\dagger}}_{im'\uparrow}\\
\label{eq:HUloc2}
& = & \tfrac{1}{2} \underbrace{\left( U-\tfrac{3}{2}J \right)
}_{\equiv \tilde{U}}
\hat{n}_i (\hat{n}_i -1)-{J}\hat{\mathbf S}_i^2
+ \tfrac{3}{4} {J}  \hat{n}_i .  \qquad 
\end{eqnarray}
\end{subequations}
This is a minimal version of the generalized Kanamori Hamiltonian of Ref.~\citep{Georges2013},
with $\text{U(1)}_\charge\times
\text{SU(2)}_\spin \times\text{SU(3)}_\orb$ symmetry for its charge
(\charge), spin (\spin) and orbital (\orb) degrees of freedom.
$\hat d^\dagger_{i \nu}$ creates an electron on site $i$ of
flavor (fl) $\nu = (m\sigma)$, which is composed of a
spin ($\sigma \! = \uparrow,\downarrow$) and an orbital ($m=1,2,3$) index.
$\hat n_{i\nu} \equiv 
\hat{d}^{\dagger}_{i\nu}\hat{d}^\pdag_{i\nu}$
counts the electrons of flavor $\nu$ on site $i$.
$\hat{n}_i \equiv \sum_{\nu}\hat n_{i\nu}$ is
the total number operator for site $i$
with $n_d \equiv \langle \hat{n}_i \rangle$,
and $\hat{\mathbf S}_i$ its
total spin, with components $\hat S_i^\alpha =
\sum_{m\sigma\sigma'}\hat{d}^{\dagger}_{i m\sigma}
\tfrac{1}{2}\sigma^\alpha_{\sigma\sigma'}\hat{d}_{i m\sigma'}$, where
$\sigma^{\alpha}$ are Pauli matrices. 
We study a Bethe lattice with degenerate bands, each of bandwidth
$W=4t$, i.e we assume negligible crystal field splitting and a
uniform hopping amplitude $t$ 
restricted to
nearest-neighbor hopping between the same kind of orbital and spin
degrees of freedom.  Both the chemical potential $\mu$ and the
hopping amplitude $t$ are then equal for all flavors, leading to a
locally $\text{SU(6)}_{\flavor}$ symmetric kinetic term in
Eq.~(\ref{eq:HU}).  $t=1$ serves as energy unit.

The onsite interaction term, $\hat H_\interact$,  incorporates Hund's rule 
and Mott physics in its most basic form and reduces the
symmetry to $\text{SU(2)}_\spin \times\text{SU(3)}_\orb$ for $J>0$. It was
first introduced by Dworin and Narath in a generalization of the
Anderson impurity model  to study magnetic impurities
 \citep{Dworin1970}. The first three terms of Eq.~(\ref{eq:Hloc1}) are
density-density interactions.  $U$ is the intraorbital Coulomb
interaction between electrons with opposite spins in the same
orbital, $U-J<U$ the interorbital Coulomb interaction between
electrons with opposite spins in different orbitals, and $U-2J$
the Coulomb interaction between electrons with parallel spins in
different orbitals, where the interorbital Coulomb interaction is
further reduced by the  ferromagnetic coupling $J$  due to Hund's
first rule that  favors the alignment of spins. The last term of
Eq.~(\ref{eq:Hloc1}) is a spin exchange term. 

The generalized Kanamori Hamiltonian of Ref.~\citep{Georges2013}
involves some additional terms not present in Eq.~\eqref{eq:HU-Hloc},
which reduce the $\text{SU(3)}_\orb$ symmetry in the orbital 
sector to $\text{SO(3)}_\orb$. However, these additional terms 
do not affect the low-energy physics, since they are 
irrelevant in a renormalization group sense \citep{Horvat2016}.
  
Eq.~(\ref{eq:HUloc2}) is a more compact notation of Eq.~(\ref{eq:Hloc1}) 
and summarizes the two main aspects of our model. The first term 
is known to trigger Mott physics, whereby $U$ penalizes double
occupancy of orbitals.
The second term  
directly reflects Hund's first rule: it favors a large spin per site
for $J>0$.  Note that the third term only shifts the chemical
potential, $\mu$. 

We choose $\mu$ such that we obtain a total filling per lattice
site, $n_d=\langle \hat{n}_i \rangle$, of $1\leq n_d\leq3$.
For $n_d>1$, Hund's first rule reduces the atomic ground state
degeneracy and thus strongly influences the physics of the system.
The orbital and spin degrees of freedom  of electrons  can show very
distinct behavior and conspire in a highly non-trivial way, leading
to striking new phenomena like spin-orbital
separation \citep{Stadler2015}.
In contrast, at half-filling,  $n_d=3$, a fundamentally different
ground state emerges: a large spin state is formed and orbital
degrees of freedom are fully blocked \citep{Georges2013}.

We treat the \HHM of Eq.~(\ref{eq:HU-Hloc}) with 
single-site DMFT and use  full-density-matrix (fdm)NRG  \citep{Weichselbaum2007} 
as real-frequency impurity solver. For methodological details and further definitions 
of physical quantities used in the main paper, see Appendix~\ref{appendix}.

\subsubsection{Multiplet structure at filling $n_d=2$}
\label{sec:Hloc2:mult}
The physical behavior of the system depends in a crucial
manner on the multiplet structure of the local Hamiltonian, 
and can change in dramatic ways when parameters are tuned such
that level crossings occur \citep{Aichhorn2009}. This section is therefore
devoted to a detailed discussion of this multiplet structure.

The local Hamiltonian of a single site $i$ is given by
$ \hat{H}_\mathrm{loc}^{(i)} \equiv
  \hat{H}_\interact[\hat d^\dag_{i\nu} ] - \mu \hat{n}_i
$.
With focus on the specific filling $n_d=2$,
this Hamiltonian can be written as
\begin{eqnarray}
   \hat{H}_\mathrm{loc}^{(i)} &=&
   \tfrac{\tilde{U}}{2} (\hat{n}_i - 2)^2 - J \hat{\mathbf{S}}_i^2
 - \underbrace{(\mu-\mu_2)}_{\equiv \delta\mu_2} \hat{n}_i
 - 2\tilde{U}
\text{ .}\label{eq:Hloc2}
\end{eqnarray}
with $\mu_2 \equiv \frac{3}{2}(U-J)$. Here the Coulomb
interaction in the first term on the r.h.s. has been written
such that for $\mu=\mu_2$, i.e. $\delta\mu=0$ and small
$J$, this Hamiltonian clearly favors the desired filling of
$n_d=2$.
By writing the local states space in terms of symmetry
multiplets, the above Hamiltonian reduces to one-dimensional
multiplet blocks and hence already becomes diagonal. The
symmetry labels of SU(3) follow the Dynkin convention where
the irreducible representation $q=(q_1,q_2)\equiv (q_1 q_2)$
corresponds to a Young diagram with $q_1+q_2$ ($q_2$) boxes
in its first (second) row.

For the case $\mu=\mu_2$, the multiplet structure of
the local Hamiltonian in \Eq{eq:Hloc2} is sketched in
\Fig{fig:Hloc:mult}.  There the two low-energy multiplets at
$n_d=2$ are labeled by $\gone$ and $\gtwo$, also referred to
as the $g$-levels. The actual ground state multiplet $\gone$
is in triplet configuration across two out of the three
orbitals. The singlet configuration $\gtwo$, split off by an
energy $2J$, also includes the pair singlets within a single
orbital.  This therefore results in a total of $d_{\gtwo}=6$
symmetric states described by the single irreducible
multiplet $q=(20)$.  By removing an electron, this leads to
the hole-like level, denoted by $h$. It  
contains just one electron, $n_d=1$, 
which can be in any spin and orbital,
hence $S=\rfrac{1}{2}$ and the defining representation
$q=(10)$.  Conversely, by adding a particle to the
$g$-multiplets, one obtains half-filling $n_d=3$. This
allows states with one particle per orbital, resulting in
one $S{=}\rfrac{3}{2}$ multiplet, labeled $\eone$ with
$(S,q)=(\frac{3}{2},00)$, and two $S{=}\rfrac{1}{2}$
multiplets.  By symmetry, the latter ones need to be grouped
with the six $S=\rfrac{1}{2}$ multiplets with a double and a
singly occupied orbital into the single SU(3) multiplet
$q=(11)$ with $8$ states total, forming the single multiplet
$\etwo$.

\begin{figure}
\centering
\includegraphics[width=0.5\linewidth]{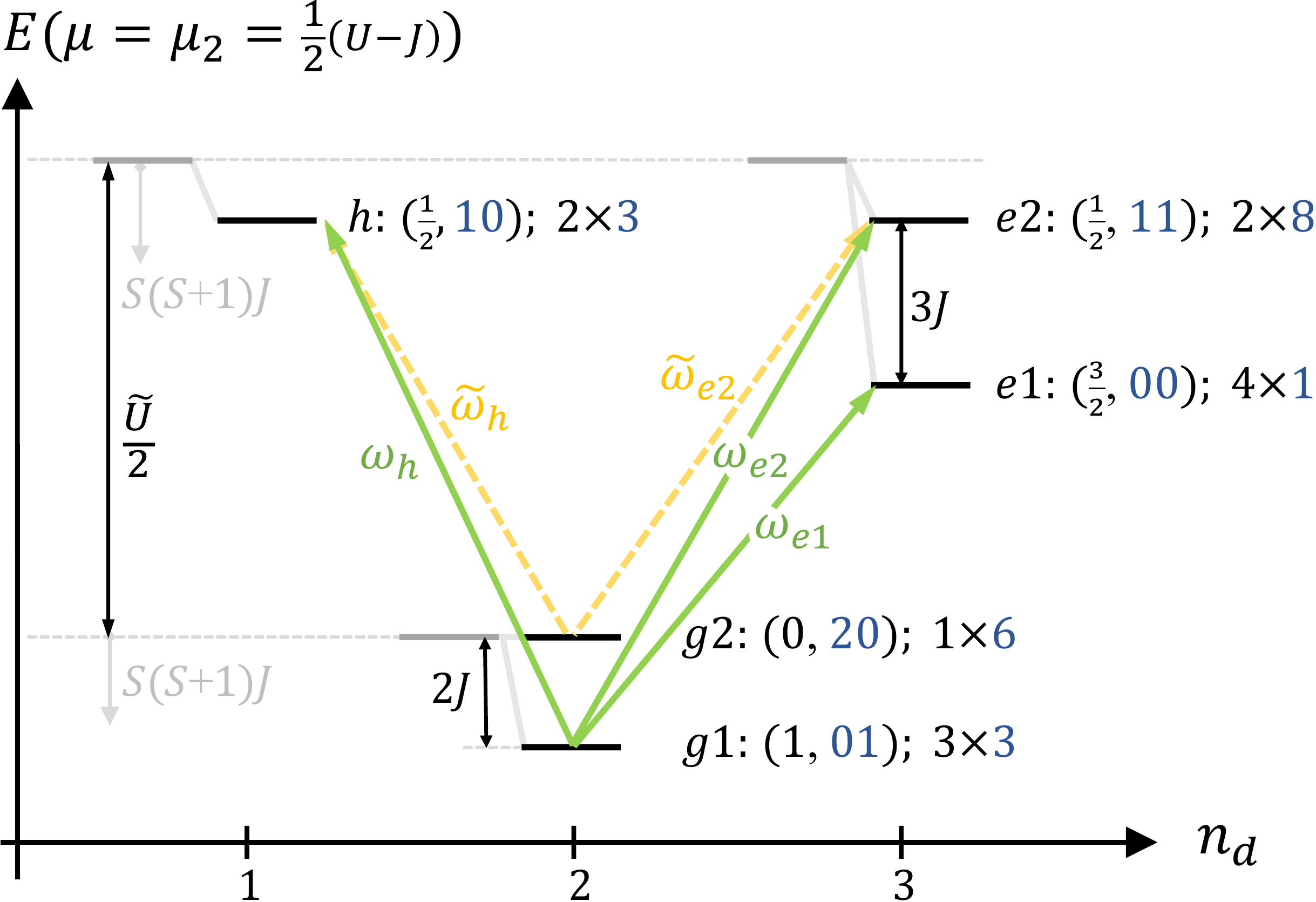}
\caption{
  Local multiplet structure of a single {\HHM} site at
   filling $n_d=2$ using $\mu=\mu_2$,  with $\tilde{U}\equiv
   U-\frac{3}{2}J$ as specified in \Eq{eq:Hloc2}. The energies
   for $J=0$ are indicated by the thick 
   grey levels, which are split when turning on $J$ as indicated.
   The individual multiplets are given labels
   $g$ for ``ground state'', $h$ for hole-like, and $e$ for
   electron (particle) like, which are specific to the current
   filling, here $n_d=2$. 
   Each multiplet is followed by 
   its symmetry labels $(S,q)$ and
   the combined multiplet dimension 
   of spin times SU(3), 
   with the SU(2) spin $S$ and SU(3) representation
   $q\equiv(q_1,q_2)\equiv(q_1 q_2)$. 
   The grey downward arrows indicate a lowering of the energy 
   levels by the Hund's term $-J\hat{\mathbf S}_{(i)}^2$.
   The shown multiplet structure is complete for $n_d=1,2,3$. Together with
   the vacuum state at $n_d=0$ and the symmetry relative to
   half-filling, this yields a total number of states (respective
   to $n_d=0,\,6$ and $1,\,5$ and $2,\,4$ and $3$) of
   $2\cdot(1\times1)+2\cdot(2\times3)+2\cdot(1\times6+3\times3)
   +1\cdot(2\times8+4\times1) = 64 = 4^3$, 
   i.e. the complete state space of three spinful fermionic levels.
   Note that 1-particle excitations from $\gtwo$ (yellow lines) cannot
   reach the $S=\tfrac{3}{2}$ multiplet $\eone$. 
}
\label{fig:Hloc:mult}
\end{figure}

In what follows, we now slightly alter the chemical
potential towards finite $\delta\mu_2$ in \Eq{eq:Hloc2},
using the specific choice $\delta\mu_2 = - \frac{3J}{2}$.
This raises the $e$-levels in \Fig{fig:Hloc:mult} and lowers
the $h$-level by equal amounts relative to the $g$-levels at
$n_d=2$, to the extent that level $h$ and $\eone$ become
aligned, i.e. degenerate.  This simplified setting is the
reason for our choice of $\delta\mu_2$.

The resulting excitation energies from the ground state
multiplet $\gone$ can be simply determined from
\Fig{fig:Hloc:mult} while also accounting for the plain
shift due to $\delta\mu_2$ in \Eq{eq:Hloc2},
\begin{subequations}
\label{eq:Hloc2:exc}
\begin{align}
  \ocircle && \omega_{\eone}^{(2)} &\equiv +(E_{\eone} - E_{\gone})  = 
\tfrac{U}{2} - J\,, &&
\notag \\
  +   && \omega_{\etwo}^{(2)} &\equiv +(E_{\etwo} - E_{\gone}) 
 = \tfrac{U}{2} + 2J\,,  &&
\label{eq:Hloc2:exc:g1} \\
  \triangle  && \omega_h^{(2)} &\equiv
  -(E_{h} - E_{\gone}) = - \omega_{\eone}^{(2)}\,,&& \notag
\end{align}
where we added the superscript $(2)$ to these transition
frequencies for later reference to emphasize the current setting
of having $n_d=2$ (this filling is implicit for the $g$-, $e$-, and
$h$-multiplet labels in the present discussion, for readability).
The signs in \Eqs{eq:Hloc2:exc:g1} are taken in consistency with
the definition of the spectral function $A(\omega)$,
and is thus opposite for particle- and hole-like excitations.
The symbols to the left 
will be used in \Sec{sec:Janus} and \Sec{sec:half-filled-MIT}
to mark the positions of the multiplet excitation energies 
in the spectral function $A(\omega)$.

Similarly, also the transition energies w.r.t.
level $\gtwo$ are simply derived from \Fig{fig:Hloc:mult},
\begin{align}
\label{eq:Hloc2:exc:g2}
  +  && \tilde{\omega}_{\etwo}^{(2)} &\equiv
  +(E_{\etwo} - E_{\gtwo}) = \omega_{\etwo}^{(2)} - 2J = \tfrac{U}{2}\,, &&
\\
  \triangle  && \tilde{\omega}_h^{(2)} &\equiv
  -(E_{h} - E_{\gtwo}) = \omega_{h}^{(2)} + 2J = - (\tfrac{U}{2} - 3J ) \,, &&
\notag
\end{align}
\end{subequations}
where we note that the transition $\tilde{\omega}_{\eone}^{(2)} =
-\tilde{\omega}_{h}^{(2)}$ is forbidden for 1-particle spin-half
excitation processes.

The above picture of well-separated ground-state multiplets
breaks down entirely, once $\omega_{\eone}^{(2)}$ in
\Eqs{eq:Hloc2:exc:g1} becomes negative, i.e. levels $h$ and
$\eone$ cross $\gone$ as the new ground state.  Hence we
will mostly constrain our discussion to the regime
$J/U<0.5$. This regime, nevertheless, already reaches up to
extraordinarily large Hund's coupling from a materials point
of view where one typically encounters $J/U\lesssim 0.2$
 \citep{Georges2013}.

For $J\ll U$, the $g$-levels are typically considered
well-separated from the $e$- and $h$-levels. However, this
picture already breaks down earlier,
namely once the degenerate $e1$- and $h$-levels pass across $\gtwo$.
According to the excitation energies in \Eqs{eq:Hloc2:exc:g2},
this occurs at $\tilde{\omega}_h^{(2)}
 = 0$ which defines the
crossover energy scale  $J^\ast\equiv\frac{U}{6}$.  The
regime $J\gtrsim J^\ast$ quantifies what we mean by {\it
sizeable} Hund's coupling in the \HHM at $n_d=2$.
There for $J\gtrsim J^\ast$, we expect a qualitative
change in the emerging physics of the \HHM.
   
\subsubsection{Multiplet structure at filling $n_d=3$}
\label{sec:Hloc3:mult}

We now focus on the filling $n_d=3$ with the Hamiltonian
\begin{eqnarray}
\hat{H}_\mathrm{loc}^{(i)}
&=& \tfrac{\tilde{U}}{2} 
(\hat{n}_i - 3)^2 - J \hat{\mathbf{S}}_i^2
- \underbrace{(\mu-\mu_3)}_{\equiv \delta\mu_3}
\hat{n}_i  - \tfrac{9}{2}\tilde{U}
\text{ ,}\label{eq:Hloc3}
\end{eqnarray}
and $\mu_3 \equiv \frac{5}{2}U-3J$. By construction, $\mu=\mu_3$, 
i.e. $\delta\mu_3=0$ directly leads 
to a particle-hole symmetric excitation spectrum,
and therefore to exact 
half-filling at $n_d=3$.
The multiplets in \Fig{fig:Hloc:mult}
are shifted relative to each other for different $n_d$
such that $n_d=3$ 
becomes the new ground state
symmetry sector with the lowest energy excitations
in $n_d=2$ and $4$ split off symmetrically by
$\tilde{U}/2$ at $J=0$. Hence the $g$- and $e$-multiplets
in the previous discussion for $n_d=2$ as in \Fig{fig:Hloc:mult}
acquire the new respective labels $h$ and $g$ here at $n_d=3$.

In the following we only focus on 
the case of sizeable $J$,
and there, for simplicity, only on the lowest
levels $h$, $g$, and $e$ at $n_d=2$, $3$, $4$, respectively.
The level $g$ has 
maximal 
spin $S= \rfrac{3}{2}$ linked with 
an orbital singlet configuration $q=(00)$
[level $\eone$ in \Fig{fig:Hloc:mult}].
The lowest hole level $h$ at $n_d=2$ has $(S,q)=(1,01)$
[i.e. level $\gone$ in \Fig{fig:Hloc:mult}]. The lowest particle
level $e$ at $n_d=4$ is given by $(S,q)=(1,10)$,
i.e. the particle-hole transformed level $h$.  

The excitation energies from the ground state multiplet
$g$ at $\mu=\mu_3$  
can be simply determined from \Eq{eq:Hloc3},
analogous to \Eq{eq:Hloc2:exc:g1},
\begin{align}
*& &\omega_{e}^{(3)} &\equiv +(E_{e} - E_{g}) 
 = \tfrac{U}{2} + J \,, &&
\label{eq:Hloc3:exc:g1} \\
\diamond&&\omega_h^{(3)} &\equiv -(E_{h} - E_{g}) = 
- \omega_{e}^{(3)}\,,&& \notag
\end{align}
where the reference point of a filling of $n_d=3$ is implied,
yet also explicitly indicated with the superscript in the
transition frequencies.
We will refer to them in  \Sec{sec:half-filled-MIT}.


\subsection{Overview of Results}
\label{sec:Results}

In the following three sections we present our real-frequency-based DMFT+NRG results for 
the {\HHM}. 
In Sec.~\ref{sec:spinfreezing} we reveal
the connection between SOS and spin-freezing.
We argue that while both terminologies describe in principle the
same Hund physics, the latter term has the drawback that it was
proposed based on QMC results that did not account for a Fermi-liquid
ground state.  In Sec.~\ref{sec:Janus} we study the $U$-$J$-phase
diagram at $n_d=2$ and systematically disentangle the Janus-faced
effects of (i) Hundness and (ii) Mottness. Thereby we quantitatively
explain the existence of the low QP weight, $Z$, by
SOS, which is revealed to occur in the whole
metallic regime, but at different scales. We  explain the difference
between Hund- and Mott-correlated systems. In particular, we show
that sizeable $J$ leads to low $Z$ also far away from the MIT at
$n_d=2$ and opens up a \textit{large} incoherent frequency regime
where intriguing Hund-correlated physics occurs: large, almost
unscreened spins are coupled to screened orbital degrees of freedom.
In  Sec.~\ref{sec:half-filled-MIT}, we study the doping-dependence
of $Z$ and the compressibility, $\kappa_{el}$. We demonstrate that,
in principle, SOS also occurs and determines the 
low $Z$ behavior at intermediate fillings,  $1<n_d<3$. 
We give evidence that SOS is
generically based on a  two-stage screening process involving the
formation and the full screening of effective $\rfrac{3}{2}$ spins. The
details of this process, however, vary with filling.  $\kappa_{\rm
el}$ is shown to be positive at finite $J$ for all fillings and values of $U$, that we have studied.
Thus we assume that no Hund's-coupling-induced instabilities emerge in the system. 

Overall, we scan the parameter space of the phase diagram in two
orthogonal directions (indicated by the arrows in
Fig.~\ref{fig:sketch_phasediagram}): we either vary $n_d$ (along
the horizontal direction of the red arrow)
for different parameter sets of
$U$ and $J$ as in Sec.~\ref{sec:spinfreezing} and
Sec.~\ref{sec:half-filled-MIT}, or we vary $U$ (along the
vertical direction
of the blue arrow) and $J$ for fixed $n_d=2$ as in
Sec.~\ref{sec:Janus}.

To summarize, we will develop  a global picture of spin-orbital
separation  that strongly supports (i) Hundness as a new mechanism
towards strong correlations in the normal state of Hund metals.

\section{Spin-freezing and spin-orbital separation - two terminologies for the same Hund physics}
\label{sec:spinfreezing}

To set the scene, we first revisit SOS
 \citep{Stadler2015} and explain its connection to the spin-freezing
theory introduced in 2008 in a finite-temperature DMFT+QMC
study \citep{Werner2008} of the $n_d$-$U$ phase diagram of a
degenerate three-band Hund model.

\subsection{Spin-orbital separation at $n_d=2$ revisited}
\label{sec:sosrevisited}

We calculate the dynamical real-frequency spin and orbital
susceptibilities
\begin{subequations}
\label{eq:chi}
\begin{eqnarray}
\chi_\spin & =& \tfrac{1}{3}\sum_\alpha \langle \hat S^\alpha \mbox{$\parallel$}
\hat S^\alpha \rangle_\omega,\\
\chi_\orb & = &\tfrac{1}{8}\sum_a \langle \hat T^a\mbox{$\parallel$} \hat T^a \rangle_\omega,  \qquad 
\end{eqnarray}
\end{subequations}
respectively, where $\hat T^a =
\sum_{mm'\sigma}\hat{d}^{\dagger}_{m\sigma} \tfrac{1}{2}
\tau^a_{mm'}\hat{d}_{m'\sigma}$ are the impurity orbital
operators with the SU(3) Gell-Mann matrices, $\tau^a$, normalized as
${\rm Tr}[ \tau^a\tau^b ] =2\delta_{ab}$. 


\begin{figure}
\centering
\includegraphics[width=0.65\linewidth, trim=0mm 0mm 0mm 0mm, clip=true]{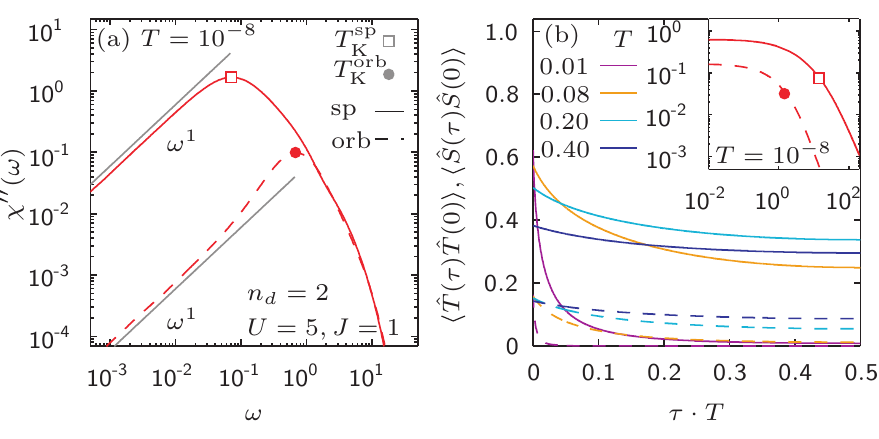}
\caption{
(a) The imaginary part of  the dynamical real-frequency  orbital
   $\chi''_{\orb}$ (dashed) and spin $\chi''_{\spin}$ (solid)
   susceptibility  for $U=5$, $J=1$, $n_d=2$ and $T=0$.   The
   orbital Kondo  scale $\Tkorb$ (filled circle) and  the spin Kondo
   scale $\Tkspin$ (open square) are defined as the peak positions
   of $\chi''_\orb$ and $\chi''_\spin$, respectively, and show
   SOS, i.e. $\Tkorb \gg \Tkspin$. Grey guide-to-the-eye
   lines indicate Fermi-liquid power laws. 
(b) The imaginary-time impurity orbital-orbital $\langle \hat{
   T}(\tau)\hat{T}(0)\rangle$ (dashed) and spin-spin
   $\langle \hat{S}(\tau)\hat{S}(0)\rangle$ (solid)
   correlator plotted as a function of the rescaled imaginary time $\tau\cdot
   T$ for the same parameters as in (a), but at different
   temperatures. The solid yellow and blue curves show
   spin-freezing:  $\langle \hat{ S}(\tau)\hat{
   S}(0)\rangle$ approach large constant values at  times
   $\tau=1/(2T)$.
    The inset  shows the zero-temperature results of $\langle \hat{
   T}(\tau)\hat{ T}(0)\rangle$ (dashed) and 
   $\langle \hat{S}(\tau)\hat{ S}(0)\rangle$ (solid)
   calculated from (a) the real-frequency
   susceptibilities. Both approach zero in the FL regime at very
   large imaginary times. The filled circle and
   the open square mark
   $1/\Tkorb$ and $1/\Tkspin$, respectively. 
}
\label{fig:AoP_spin_freezing_nd2}
\end{figure}

\Fig{fig:AoP_spin_freezing_nd2}(a) depicts the zero-temperature
results of the imaginary parts, $\chi'' (\omega) \equiv
-\frac{1}{\pi}\imag \chi(\omega)$,
of the dynamical impurity
orbital (dashed curve) and spin (solid curve) susceptibilities for
$U=5$, $J=1$ and a filling of $n_d=2$. The filled circle and the
open square mark the orbital and spin Kondo scales, $\Tkorb$ and
$\Tkspin$, which are defined as the peak positions of  $\chi''_\orb$
and $\chi''_\spin$, respectively.
Clearly, these two energy scales are very distinct:  in
Fig.~\ref{fig:AoP_spin_freezing_nd2}(a) we revisit the central
result of our DMFT+NRG study of the 3CAHM -- \textit{spin-orbital
separation} [see Fig.~(3c) in Ref.~\citep{Stadler2015} and also
Fig.~\ref{fig:sos_sketch}]. \textit{Orbital screening sets in at
much higher energies than spin screening, $\Tkorb\gg\Tkspin$,
opening a non-trivial intermediate NFL regime exhibiting ``Hund
metal physics": slowly fluctuating (\textit{not} frozen),
Hund's-coupling-induced large spins are \textit{coupled} to screened
orbital degrees of freedom.} The existence of  large, composite
spins which are only poorly screened, 
manifests itself in an enhancement of $\chi''_\spin$ with decreasing
frequencies. Interestingly, the fluctuations of these spins
influence the physics of the screened orbitals, leading to an
intriguing interplay of spin and orbital degrees of freedom: below
$\Tkorb$, $\chi''_\orb$ decreases as the frequency is lowered,
indicating the screening of the orbital degrees of freedom. However,
for $\omega>\Tkspin$, $\chi''_\orb$  does not follow FL scaling, as
the orbital degrees of freedom still ``feel" the slowly
fluctuating, large local moments.  Below the very small,
Hund's-coupling-reduced coherence scale, $\Tkspin\approx0.072$,
both the spin and orbital degrees of freedom get fully screened and
FL behavior is restored [$\chi''_{\orb}(\omega)\propto\omega$ and
$\chi''_{\spin}(\omega)\propto\omega$, see
Fig.~\ref{fig:AoP_spin_freezing_nd2}(a), grey lines].

From the real-frequency orbital and spin susceptibility we also calculate the
imaginary-time impurity orbital-orbital and spin-spin correlators, 
\begin{eqnarray}
\langle\hat{ T}(\tau)\hat{ T}(0)\rangle &\equiv&
  \tfrac{1}{8}\langle\hat{\mathbf T}(\tau) \cdot \hat{\mathbf T}(0)\rangle 
= \int  {{\rm d}\omega}\, n_B(\omega)  \chi''_{\orb}(\omega)\,{\rm e}^{\omega\tau},
\notag \\
\langle\hat{ S}(\tau)\hat{ S}(0)\rangle &\equiv&
   \tfrac{1}{3}\langle\hat{\mathbf S}(\tau)  \cdot \hat{\mathbf S}(0)\rangle 
 = \int {{\rm d}\omega}\, n_B(\omega)  \chi''_{\spin}(\omega)\,{\rm e}^{\omega\tau}
,\ \quad 
\label{eq:chitau}
\end{eqnarray}
respectively, with the Bose-Einstein distribution
$n_B(\omega)=1/(\rm{e}^{\beta\omega}-1)$. In the
inset of Fig.~\ref{fig:AoP_spin_freezing_nd2}(b) we plot both
correlators for zero temperature and the same parameters as in
Fig.~\ref{fig:AoP_spin_freezing_nd2}(a). In accordance with the
real-frequency susceptibilities, the orbital-orbital correlator
(dashed curve) is much smaller than the spin-spin correlator (solid
curve). The latter approaches zero rather slowly, thus, the FL
regime is only reached at very long imaginary times, $\tau>100$. 

\subsection{Spin-freezing at $n_d=2$}

In order to understand the connection of SOS
and the spin-freezing phenomenon that was based on
\textit{finite}-temperature DMFT+QMC \citep{Werner2008} data, we have
performed similar calculations at higher temperatures  [see
Fig.~\ref{fig:AoP_spin_freezing_nd2}(b)]. For temperatures well
below  the FL coherence scale, $T<\Tkspin$,
$\langle \hat{S}(\tau)\hat{S}(0)\rangle$
decays to zero on the scale
$\tau= 1/(2T)$ (solid purple curve). For $\Tkorb\geq T \geq
\Tkspin$, in contrast,
$\langle \hat{S}(\tau)\hat{S}(0)\rangle$ 
approaches a large constant value at  times
$\tau\approx1/(2T)$ (solid yellow and blue curves).  This
finite-temperature finding -- a spin-spin correlation function which
does not decay to zero at long times -- was called ``spin freezing''
in Ref.~\citep{Werner2008} and interpreted as the existence of
\textit{frozen} local moments leading to an incoherent metallic state.

As exemplified in Fig.~\ref{fig:AoP_spin_freezing_nd2}  (a,b) and
further demonstrated in this work, spin freezing was a
phenomenological  interpretation of the spin-spin correlator  based
on a QMC solver that didn't reach low enough temperatures (or
equivalently long enough times) to reveal the FL ground state for
many parameters in the phase space. However, \textit{the spins are
\textit{not} frozen, they fluctuate slowly above $\Tkspin$ and get
fully screened in the FL regime below $\Tkspin$.}

Moreover, a detailed analysis of
$\langle \hat{ T}(\tau)\hat{T}(0)\rangle$ 
at $n_d=2$ in
Fig.~\ref{fig:AoP_spin_freezing_nd2}(b) shows that  the
orbital-orbital correlators (dashed yellow and dashed light and dark
blue curves) do not fully decay to zero in the incoherent temperature regime
$T>\Tkspin$, but remain finite, as well (as opposed to the statement
in Ref.~\citep{Werner2008}). This finding supports the
interpretation obtained from the real-frequency orbital
susceptibility and further revises the spin-freezing picture:
\textit{the orbital degrees of freedom are screened below
$T<\Tkorb$, but they are \textit{not} fully decoupled from the spin
dynamics.}

\begin{figure}
\centering 
\includegraphics[width=0.65\linewidth]{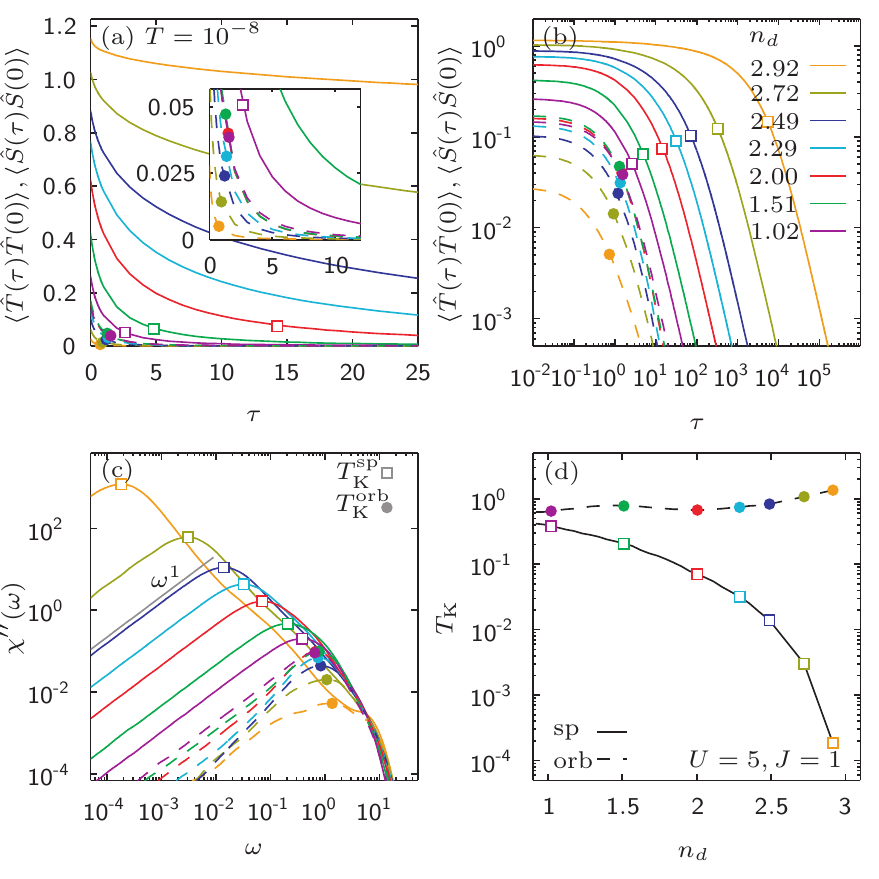}
\caption{
    (a,b) The imaginary-time impurity orbital-orbital $\langle
    \hat{T}(\tau)\hat{T}(0)\rangle$ (dashed) and
    spin-spin  $\langle \hat{S}(\tau)\hat{
    S}(0)\rangle$ (solid) correlators calculated from (c) the
    real-frequency susceptibilities for $U=5$, $J=1$, and $T=0$ at
    various fillings $n_d$. The filled circles and the open squares
    mark $1/\Tkorb$ and $1/\Tkspin$, respectively. The inset in (a)
    shows a zoom  to better resolve the orbital-orbital correlators.
    (a) For short imaginary times, the curves for $\langle
    \hat{S}(\tau)\hat{ S}(0)\rangle$ seem to remain
    constant, a phenomenon which was interpreted as spin-freezing in
    Ref.~\citep{Werner2008}. (b) In contrast, for large
    imaginary times, they clearly show FL behavior. (c) The
    imaginary parts of  the dynamical real-frequency orbital
    $\chi''_{\orb}$ (dashed) and spin  $\chi''_{\spin}$ (solid)
    susceptibilities. The orbital Kondo  scales $\Tkorb$  and  the
    spin Kondo scales $\Tkspin$ are marked as filled circles and
    open squares, respectively. (d)  The orbital Kondo  scales
    $\Tkorb$ (dashed line with filled circles) and  spin Kondo
    scales $\Tkspin$ (solid line with open squares) plotted versus
    the filling $n_d$. SOS is revealed for all $1<n_d<3$.
}
\label{fig:AoP_spin_freezing_ndvar}
\end{figure}

\subsection{Spin-freezing for varying $n_d$}

Originally, without access to the FL ground state, it was argued
that the Hund-metal regime of the phase diagram in
Fig.~\ref{fig:sketch_phasediagram} is a spin-freezing NFL phase  and
that  a quantum phase transition connects a paramagnetic FL phase
(at small $n_d$ and small $U$) and a paramagnetic NFL phase
featuring frozen local moments (at larger $n_d$ and larger $U$)
 \citep{Werner2008}.

In Fig.~\ref{fig:AoP_spin_freezing_ndvar} we revisit this transition
with our NRG solver at $T=0$. We calculate the imaginary-time
orbital-orbital and spin-spin correlators for intermediate $U=5$,
$J=1$ and vary $n_d$ from $1.02$ to $2.92$. Indeed, at short times,
$\tau\lesssim25$, our DMFT+NRG results in
Fig.~\ref{fig:AoP_spin_freezing_ndvar}(a) seem to confirm this
FL-to-NFL transition. For $n_d<2 $,
$\langle \hat{S}(\tau)\hat{S}(0)\rangle$ 
decays to zero (solid purple and
green curves) while at larger $n_d$ it grows and remains finite
(solid red to yellow curves), seemingly indicating frozen local
moments. 

However, in contrast to QMC solvers, we have direct access to
exponentially long times (low temperatures) and can explicitly
reveal the existence of a FL ground state for any given filling. In
Fig.~\ref{fig:AoP_spin_freezing_ndvar}(b) we confirm that for
sufficiently long times, $\tau \gg 1/\Tkspin$,
$\langle \hat{S}(\tau)\hat{S}(0)\rangle$ 
approaches zero for all fillings, $1<n_d<3$ (solid curves).
Equivalently, all real-frequency spin susceptibilities exhibit FL
behavior below $\Tkspin$ [$\chi''_{\spin}(\omega)\propto\omega$, see
Fig.~\ref{fig:AoP_spin_freezing_ndvar}(c), grey line].  Clearly, the
NFL regime is not governed by the proximity to a quantum critical
point. 

The general existence of a FL ground state for all fillings
 was later conjectured  \citep{Werner2012,deMedici2011, Werner2015} 
 and only recently demonstrated \citep{Kowalski2018} based on 
 DMFT+QMC Hund-model studies,  and spin
freezing was reinterpreted as the existence of long-lived magnetic
moments. Instead of a quantum phase transition, a ``spin-freezing
crossover" from a FL to a NFL state at finite temperatures was
suggested  \citep{Werner2015} (which is called
``coherence-incoherence crossover" by
others \citep{Haule2009,Yin2012,Stadler2015}).  The present work 
demonstrates directly and completely that
\textit{the
time-dependence of  orbital-orbital and spin-spin correlation
functions  reveal FL behavior in the
long-time limit for all fillings $1<n_d<3$.}

\subsection{Spin-orbital separation for varying $n_d$}

Interestingly, we observe in Fig.~\ref{fig:AoP_spin_freezing_ndvar}
that \textit{SOS, i.e. $\Tkorb\gg\Tkspin$, occurs at
all fillings $1<n_d<3$} (in Ref.~\citep{Stadler2015}, it was
only explicitly revealed at $n_d=2$). $\Tkspin$  is found to be
strongly doping dependent [see
Fig.~\ref{fig:AoP_spin_freezing_ndvar}(c,d), open squares]. It
decreases very fast with increasing filling
$n_d\rightarrow3$, such that  the decay of
$\langle \hat{S}(\tau)\hat{S}(0)\rangle$ 
with imaginary time becomes
very weak and is therefore almost invisible on  short time
scales [Fig.~\ref{fig:AoP_spin_freezing_ndvar}(a),  e.g. solid,
yellow curve]. In contrast, $\Tkorb$ is almost independent of the
filling [see Fig.~\ref{fig:AoP_spin_freezing_ndvar}(c,d), filled
circles]. It even increases 
slightly from $n_d=2$ to $n_d=3$. In
summary, this leads to an intermediate regime of SOS
that expands with larger $n_d\rightarrow3$, mainly
towards smaller energies [Fig.~\ref{fig:AoP_spin_freezing_ndvar}(d)]. 

Based on these insights we conclude that SOS
is a generic feature in the whole Hund-metal regime, evolving with
$n_d$ in the following way.  With increasing $n_d$, larger local
moments form  in the intermediate SOS regime and lead to the
increase of the maximum of $\chi''_{\spin}$ (or equivalently
$\langle \hat{S}(\tau)\hat{S}(0)\rangle$) 
[see solid curves
Fig.~\ref{fig:AoP_spin_freezing_ndvar}(a-c)]. At the same time,
$\Tkspin$ is lowered, because, heuristically, it is more difficult
to screen these larger spins. In contrast, the height
of $\chi''_{\orb}$ (or equivalently
$\langle \hat{T}(\tau)\hat{T}(0)\rangle$) 
decreases
with increasing $n_d\rightarrow3$ [see dashed curves in
Fig.~\ref{fig:AoP_spin_freezing_ndvar}(a-c) and inset of (a)]. This
reflects the reduction of the phase space for orbital fluctuations
due to the formation of large spins composed of electrons in
different orbitals. Consequently, the interplay of spin and orbital
degrees of freedom is diminished for $n_d$ close to 3.

This first crude analysis of our results with varying $n_d$ will be
refined in Sec.~\ref{sec:half-filled-MIT}. There, we will  show in
more detail how it is connected to the SOS
scenario introduced above for $n_d=2$.

\subsection{The connection between spin-freezing and spin-orbital separation} 

In sum, we argue that \textit{the two terminologies,
``spin-freezing'' and ``spin-orbital separation", ultimately
describe  the same physics of the Hund-metal regime.} The large
spins that appear as ``frozen" at short imaginary times (which are
accessible for QMC) were revealed by our real-frequency finite and
zero-temperature DMFT+NRG approach  as long-lived, slowly
fluctuating, large local moments in the incoherent regime, that get
fully screened at long imaginary times to form a FL ground state.
In this picture, the intermediate energy regime of Hund metals  with
its  incoherent transport properties is governed by scattering off
(almost) free, large and long-lived magnetic moments that are
non-trivially coupled to (almost) screened orbital degrees of
freedom. A local spin susceptibility showing Pauli behavior at low and
(quasi) Curie-Weiss behaviour at intermediate temperatures  in
Refs.~\citep{Deng2018, Stadler2019} supports this viewpoint.
 
 We note that various DMFT+QMC findings on spin-freezing, such as
 spin-freezing in (realistic) five-band calculations for
 iron-pnictides  \citep{Liebsch2010,Ishida2010,Werner2012,
 Werner2016b}, spin-freezing in models with crystal-field-splitting
  \citep{Werner2016}, and spin-orbit coupling  \citep{Werner2017}
 eventually demonstrate the importance of SOS.
 In 2015, a fluctuating-moment-induced s-wave spin-triplet
 superconducting mechanism was proposed for Hund metals, where
 equal-spin electrons  are paired in different local orbitals. It
 was shown to be connected to the emergence of local magnetic
 moments in the NFL regime \citep{Werner2015}. In 2016, it was even
 conjectured that the relevant model for cuprates, the
 single-orbital Hubbard model on the square lattice, can be mapped
 onto an effective multi-orbital problem with strong ferromagnetic
 Hund's coupling, suggesting that spin-freezing (or equivalently SOS)
 is the universal mechanism which controls
 the properties of unconventional superconductors \citep{Werner2016a}.

The insights gained above are relevant for a wide range of fillings
$n_d$ and interaction strengths $U$ and $J$, as will be further
demonstrated  in Sec.~\ref{sec:Janus} and
Sec.~\ref{sec:half-filled-MIT}. In these sections we will also
clearly show that, indeed, SOS causes the
numerically observed bad-metallic behavior in the \HHM. 
SOS therefore constitutes the framework for our
main study of Hund metals.

\section{Janus-faced influence of Hund's rule coupling: Hundness  versus Mottness at $n_d=2$}
\label{sec:Janus}

\begin{figure}[tbh!]
\centering
\includegraphics[width=0.5\linewidth]{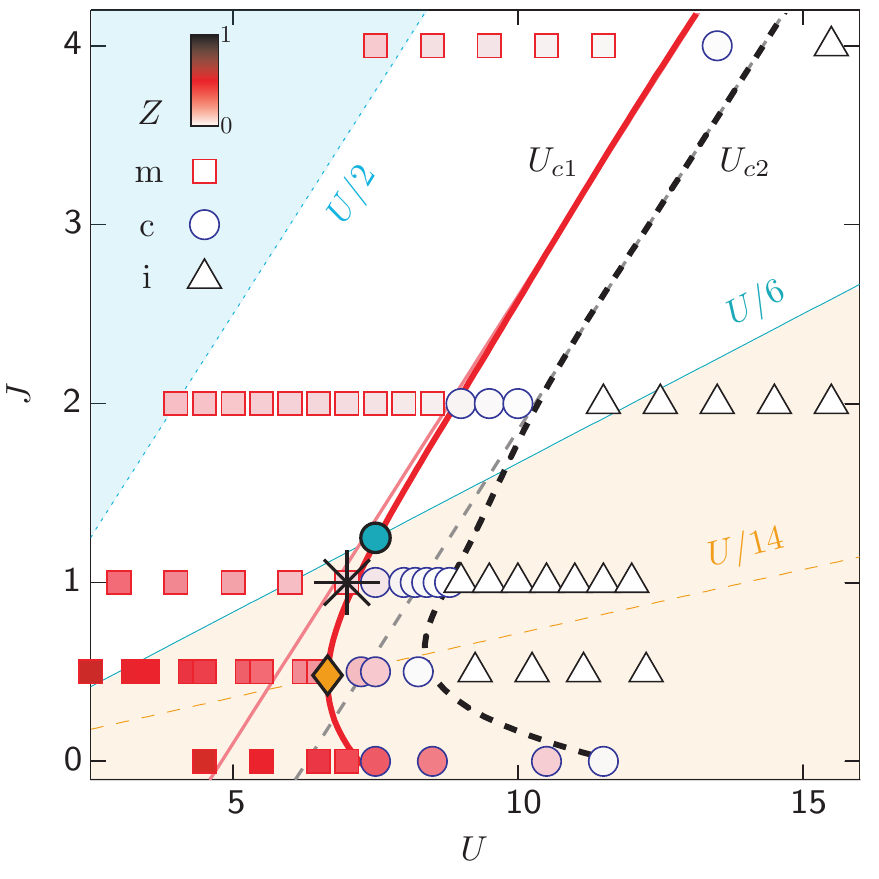}
\caption{
   The zero-temperature  phase diagram of the \HHM  at $n_d=2$
   reveals three phases in the $J$-$U$-plane: a metallic phase (squares), 
   a coexistence region (circles), and an insulating
   phase (triangles), separated by two non-monotonic phase
   transition lines, $U_{c1}$ (solid red curve) and $U_{c2}$
   (dashed black curve), obtained when initiating the DMFT
   self-consistency with an insulating and metallic seed,
   respectively. The color intensity of the symbols in the metallic 
   and the coexistence region indicates the value of $Z\in[0,1]$: 
   the lower $Z$ the more faded is the red color.
   Based on the discussion of the multiplet structure
   in \Fig{fig:Hloc:mult}, we added guides at $J=U/2$
   [$\omega_{\eone}=\omega_{h}=0$] and $J=U/6$
   [$\tilde{\omega}_{h}=0$]
   and shaded the areas separated by these.
   The crossing point of $U_{c1}$ with the $U/6$
   (cyan circle) occurs at $(U,J)\approx(7.5,1.25)$.
   We also added a guide $U/14$ (see text), whose
   crossing point with $U_{c1}$ (orange diamond) occurs
   very close to the minimum of $U_{c1}$
   at $(U,J)\approx(6.66,0.48)$.
   The black star marks the parameters for which SOS
   has first been revealed in Ref.~\citep{Stadler2015}.
   [Note that  Ref.~\citep{Stadler2015} used a slightly different 
   definition of the Coulomb energy which, while keeping the 
   definition of $J$ the same, corresponds to $U=7$ here.]
}
\label{fig:4a-AW}
\end{figure}
In this section we derive SOS as a consistent
explanation for the extended bad-metallic behavior (low $Z$) in the
phase diagram  at $n_d=2$ that reaches  from a high critical $U_c^{(2)}$
down to an unusually low $U$, i.e. we explain the Janus-faced
behavior. By introducing clear measures for (i) Hundness and (ii)
Mottness we are able to show that sizeable $J$, thus (i),  leads to
low $Z$ also far away from the MIT at $n_d =2$ and opens up a large
incoherent SOS regime with intriguing
Hund-correlated physics. 

In this section, all results are calculated at $T=0$.
Further, we note that we will neglect the superscript $(2)$ 
in $U_c^{(2)}$ because we will mainly refer to the filling,
$n_d=2$, in the following. The few exceptions  
where we refer to other fillings will be clear from the context.

\subsection{$U$-$J$ phase diagram}
\label{sec:UJphasediagram}

As an overview, \Fig{fig:4a-AW} presents the full
$U$-$J$ phase diagram for $n_d=2$ at $T=0$.   We find a metallic
(squares), coexistence (circles) and insulating (triangles) region, 
which are separated by two distinct Mott
transition lines, $U_{c1}$ (solid red line) and $U_{c2}$ (black
dashed line), respectively. We note that, so far, only $U_{c2}$ has
been studied in the context of three-band Hund models in the
literature, because it can be simply derived from the
QP weight $Z$.
The black star in \Fig{fig:4a-AW} marks the parameters of
the main result in Fig.~3 of Ref.~\citep{Stadler2015}, for which
SOS was revealed. It lies at the border of the
coexistence region close to $U_{c1}$, raising the question how
stable this feature is at lower $U$.

In Landau's Fermi-liquid theory, the quasiparticle weight
\[ 
   Z=\left(1-\partial_{\omega}\real\Sigma(\omega)|_{\omega=0}\right)^{-1}
=\tfrac{m}{m^*}
\]
is obtained from the frequency-dependent self-energy $\Sigma(\omega)$,
which is directly accessible in NRG, and measures the inverse mass
enhancement within single-site DMFT. Landau's Fermi-liquid theory is
based  on a one-to-one correspondence between  long-lived, coherent
but renormalized Landau QPs and the low-energy
excitations of a free Fermi gas. $Z\in[0,1]$ reflects the weight of
the Lorentzian-shaped coherent QPP of the
momentum-dependent local spectral function in a first order
expansion, while the additional incoherent part has weight $1-Z$. 
In \Fig{fig:4a-AW} the value of $Z$ is indicated by  the color intensity 
of the red squares and blue triangles in the conducting regime $U<U_{c2}$.

Similar to the case of the one-band Hubbard model,
the MIT shows hysteresis at low temperatures in the multi-band case. Starting with an
``insulating seed" (iS) [i.e. a real-frequency local
spectral function  $A(\omega)$, 
with an insulating Mott gap, $\Delta$, around the Fermi level],
the MIT transition  occurs at
a lower critical interaction strength, $U_{c1}$, at which $\Delta$ closes with decreasing $U$.
Starting with  a
``metallic seed" (mS)  [i.e. a metallic input spectral function with
finite weight at $\omega=0$] leads, in contrast, to a larger critical value,
$U_{c2}$, above which the QP 
resonance is lost (accordingly $Z=0$)
and a stable gap is formed with increasing $U$.
Therefore 
$Z$ can be used to quantitatively track the MIT
at $U_{c2}$ when initiating the DMFT loop with a mS.
 The coexistence region between $U_{c1}$ and $U_{c2}$ is 
 characterized by two solutions, a metallic solution for mS  and
an insulating solution for iS. This is typical for DMFT. As
mean-field approach with an iterative solution  scheme  it can have
more than one stable fixed point, depending on the initialization.
\Fig{fig:4a-AW} demonstrates that the coexistence
region is broad at $J=0$, reaching from moderate to large values of
$U$; for finite but small $J$, it strongly narrows, shifting to
lower $U$ values; and  at $J>1$, it eventually approaches a fixed
width while shifting linearly with $J$ to ever larger $U$
values \citep{Florens2002,Inaba2006}.
It is known that for  $J=0$ both
$U_{c1}$ and $U_{c2}$ grow as a function of $N_c$ at \textit{all}
fillings of multi-orbital models \citep{Florens2002}. In contrast, for given $N_c$, the
effect of a finite $J$ on $U_{c1}$ and $U_{c2}$ is strongly filling
dependent \citep{Georges2013,deMedici2011}. At half-filling $U_{c1}$
and $U_{c2}$ is strongly reduced, as finite $J$ increases
correlations by forming large $S={N_c}/{2}$ spin states that block
the orbitals. For one electron/hole, $U_{c1}$ and $U_{c2}$ increases
with $J$, as $J$ reduces the effective Coulomb interaction in the
system. At all  intermediate fillings $1<n_d<N_c$, the special
non-monotonic dependence of $U_{c1}$ and $U_{c2}$ on $J$  occurs,
which has been mentioned by several previous studies (especially for
$U_{c2}$) \citep{Fanfarillo2015,Georges2013,deMedici2017,deMedici2011}.

This non-monotonic behavior can be understood to a great
extent from the local
multiplet structure of the underlying local Hamiltonian.
For $n_d=2$, the relation of the local multiplet structure 
in \Fig{fig:Hloc:mult} with the phase diagram  is discussed in \Fig{fig:4a-AW} (bright blue, white and orange regimes).
As pointed
out with \Eqs{eq:Hloc2:exc} in \Sec{sec:Hloc2:mult},  we expect a strong qualitative change in 
the physics of the \HHM once $\tilde{\omega}_h$ turns
negative.  For the local multiplet structure, this occurs at the sizeable
Hunds coupling $J \ge U/6$.  
Accordingly, in the \HHM,  
one can distinguish two regimes in the $U$-$J$ phase diagram of
\Fig{fig:4a-AW}, by relating the $U_{c1}$ phase boundary with the
reference line $J=U/6$, for which a single crossing point exists at
$(U,J) \approx (7.5,1.25)$ [cyan circle in \Fig{fig:4a-AW}].
Therefore, for the {\it sizeable} Hund's coupling
\begin{eqnarray}
 J>J^\ast_{c1}\cong 1.25
\text{ ,}\label{eq:def:Jstar:c}
\end{eqnarray}
which we define as the ``Hund regime"  in the \HHM,
the high-lying $h$- and $\eone$-multiplets have
crossed below the $\gtwo$-level.
In this regime,  a qualitatively 
different behavior occurs all the way up to $U_{c1}$
as compared to standard Mott physics.
Specifically, 
$Z$ is low in the entire ``Hund regime" 
[see color shading of symbols 
in \Fig{fig:4a-AW}].
In contrast, 
for $J<J^\ast_{c1}$, which we refer to as ``good-metal regime'',  
$Z$ reaches up to much larger values [squares are colored
in intensive red in \Fig{fig:4a-AW}]. More generally, one may
already expect the crossover to the Hund regime to set in earlier.
For example, considering the approaching $h$-level at
$\tilde{\omega}_h \sim 2 \delta\omega_g$ with $\delta\omega_g \equiv
\omega_{\gtwo} - \omega_{\gone}=2J$, this results in $J=U/14$ [also
indicated by an orange dashed line in \Fig{fig:4a-AW}]. Its crossing
point with $U_{c1}$ occurs around $(U,J)\approx(6.66,0.48)$ [orange diamond
in \Fig{fig:4a-AW}] which turns out to be in close proximity to the
point where the non-monotonic behavior of $U_{c1}$ versus $J$ reaches a
minimum, i.e. turns around at
$U_{c1}^\mathrm{min} \simeq 6.66$.
In summary, we see that as the Hund's coupling exceeds the
moderate value of $J\gtrsim J^\ast_{c1}\sim 1$, the \HHM 
is dominated by Hund physics: sizeable $J$ leads to a  qualitative change in the 
   local multiplet 
   structure and thus to a strong change in the 
   physics of the \HHM, affecting both the phase boundaries, $U_{c1}$ and  $U_{c2}$, \textit{and} 
   the regime far from the MIT at much lower $U$, where $Z$ is low.

The scaling of $U_{c1}$ and $U_{c2}$  for large $J$,
 eventually, is linked to a further stark
change in the local multiplet structure, namely when the $h$- and
$e$-levels actually become the new local ground states having
$\omega_{\eone}<0$.  Allowing for a shift by kinetic energy
this suggests $U_{c1} \simeq 2J+\mathrm{const}$. 
This scaling
is approximated by thin solid red and dashed
grey lines in Fig.~\ref{fig:P1_nd2_1fig}(a), respectively,
and will be further corroborated 
in Fig.~\ref{fig:P1_nd2_3fig}(c).
\begin{figure}
\centering 
\includegraphics[width=0.65\linewidth]{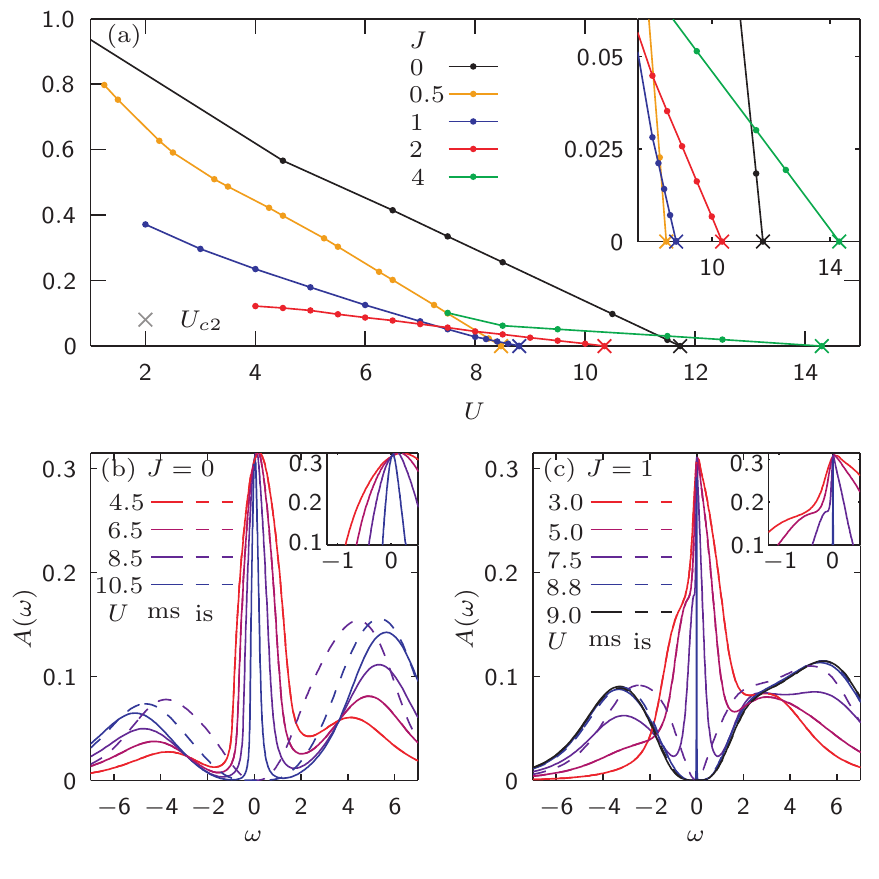}
 \caption{(a) The QP weight, $Z$, of the \HHM at $n_d=2$,
     plotted as a function of $U$,  shows Janus-faced behavior
    when $J$ is increased: on the one hand, at small to moderate
    $U$, $Z$ decreases  (metallicity worsens), on the other hand,
    $U_{c2}$ (marked by crosses) increases (metallicity improves).
    Each dot on the curves represents a DMFT+NRG data point.  The
    inset is a zoom of the $U_{c2}$-behavior. (b,c) The local
    spectral function, $A(\omega)$, shows a MIT with growing $U$ for
    (b) $J=0$  and (c) $J=1$. Solid (dashed) lines are DMFT results
    for a metallic (insulating) seed. The insets zoom into the QP.
    For $J=1$, the QP in $A(\omega)$ shows a shoulder
    characteristic of SOS.
}
\label{fig:P1_nd2_1fig}
\end{figure}

\subsection{Janus-faced behavior of $Z$}

In \Fig{fig:P1_nd2_1fig}(a) we plot $Z$ versus $U$ for various
values of $J\in[0,4]$.  In general, $Z$ is finite in the metallic
phase (with an upper limit of $Z=1$ for the non-interacting case)
and zero in the insulating phase. $U_{c2}$ is defined by the
transition point between both phases [marked by $\times$ in
\Fig{fig:P1_nd2_1fig}(a)]. We note, however, that near the MIT 
Landau's Fermi-liquid theory might break down as a valid physical
description of the excitations and $Z$ only remains as a heuristic
indicator of the MIT.  For all $J$,  we observe in
\Fig{fig:P1_nd2_1fig}(a) that  $Z$ decreases with increasing
$U$ in the metallic phase, thus strong correlation effects increase
with increasing proximity to $U_{c2}$, as known from the half-filled
one-band Hubbard MIT.  However, the strength of correlations
strongly differs for different values of $J$. For small $J$, $Z$ is
still large at small to moderate $U$, while for large $J$, $Z$ is
generally small [compare e.g. black and yellow curve to red or green
curve  in \Fig{fig:P1_nd2_1fig}(a)]. Moreover,  $J$ induces
competing effects.  While $Z$ strongly decreases with $J$ at
moderate $U$ [see e.g. black to red curve at $U=6$ in
\Fig{fig:P1_nd2_1fig}(a)], $U_c$ 
increases with $J$ (for $J\ge0.5$, after a slight decrease for very
small $J$) [see inset in \Fig{fig:P1_nd2_1fig}(a)]. We thus
observe Janus-faced behavior in our data similar to
Ref.~\citep{deMedici2011}: on the one hand $J$ promotes bad
metallicity by a loss of coherence, on the other hand it promotes
metallicity by increasing $U_{c2}$. In sum, this Janus-faced
behavior leads to a strongly reduced $Z$ for sizeable $J$ in a large
interval of $U$  (including the Hund-metal regime at $n_d=2$) [as seen e.g.
for the red or green curve in \Fig{fig:P1_nd2_1fig}(a)]. We
will clarify its physical origin and nature  in the following by
disentangling the opposing Janus-faced effects. 

\subsection{Real-frequency study of MIT at zero and finite $J$}

For each data point in our $U$-$J$ phase diagram, NRG yields  a set
of detailed frequency-dependent information  of the system, in
contrast to previous QMC or slave-boson studies. This is useful,
because $Z$ only measures the strength but not the type, Hundness or
Mottness, of strong correlations. 

Much additional information about the MIT can be gained from the
real-frequency local spectral function, $A(\omega)$, defined in \Eq{eq-Aw}. 
For example,  the dual character of strongly correlated
electrons is directly reflected in the shape of $A(\omega)$. In
\Fig{fig:P1_nd2_1fig}(b,c) we track the MIT  in $A(\omega)$, i.e.
how this dual character changes with $U$,  for $J=0$ and $J=1$,
respectively.  The metallic, delocalized behavior of electrons in
the solid is characterized by a finite spectral weight at the Fermi
level in form of  a well-defined QPP [see e.g.
solid and dashed   red curves in \Fig{fig:P1_nd2_1fig}(b,c)]. Local
Kondo-type screening processes of the ground state multiplet
dominate the low-energy physics of the self-consistent impurity
model and  lead in the \HHM to a Fermi-liquid ground state
with coherent QP excitations in the whole metallic phase,
as will be discussed in detail later.  The localized behavior of the
electrons is manifest at high energies in terms of local (atomic)
multiplet excitations which are broadened by the solid-state
environment and form the Hubbard side bands (see discussion of
\Fig{fig:P1_nd2_2fig}).  At small to moderate $U$, these incoherent
high-energy bands are close to the Fermi level and even overlap, and
the QPP is broad.  With increasing $U$, the Hubbard side bands move
to larger $|\omega|$  and the QPP narrows [compare red versus blue
curves in \Fig{fig:P1_nd2_1fig}(b,c)]. Above $U_{c1}$ or $U_{c2}$
(depending on the seed) the DMFT self-consistency opens a Mott gap 
in $A(\omega)$ around the Fermi level, the QPP vanishes and
$A(\omega)$ then consists solely of the high-energy bands [see e.g.
black curve in \Fig{fig:P1_nd2_1fig}(c)]. Heuristically, this
decrease of the QPP width  with  increasing $U$
is tracked by the QP weight, $Z$, as the peak height is
pinned to a fixed value at zero frequency (Luttinger
pinning \citep{Georges1996,Muller-Hartmann1989}) for all $U<U_c$.

As part of the MIT, we also directly observe the coexistence region
$U_{c1}<U<U_{c2}$ in \Fig{fig:P1_nd2_1fig}(b,c). While the purely
metallic and the purely insulating phase have only one solution of
the DMFT  self-consistency, independent of the seed, we find two
differing solutions in the coexistence region, an insulating for iS
and a metallic one for mS, respectively [see dashed versus solid
purple and blue  curves in \Fig{fig:P1_nd2_1fig}(b,c)].  We note
that NRG is perfectly suited for pinpointing $U_{c1}$ and $U_{c2}$
via $A(\omega)$, as its energy resolution is exponentially refined
around the Fermi level, capturing the QPP down to its smallest width.
Thus the iterative DMFT procedure does not break down  before its
solution becomes thermodynamically unstable. However, the broadening
of discrete spectral data in NRG might minimally shift  additional
spectral weight to the Fermi level, thus  artificially but only
slightly shifting the coexistence region to larger $U$ values.

At first glance, the MITs for $J=0$ and $J=1$  seem to behave
overall similarly with changing $U$. However, we find striking
differences between the spectra in \Fig{fig:P1_nd2_1fig}(b) and
\Fig{fig:P1_nd2_1fig}(c), corresponding to the black and blue lines
in  \Fig{fig:P1_nd2_1fig}(a), respectively.

As discussed above,  $Z$ is much lower for the $J=1$ MIT than for
the $J=0$ MIT. Accordingly, we observe qualitative differences in
the shape of the QPP. For finite $J$, in \Fig{fig:P1_nd2_1fig}(c),
the QPP has a shoulder at negative frequencies and a slight kink at
positive frequencies.  The shoulder (and the kink) drastically
narrow the top of the QPP while the bottom remains broad. These
features are present for all values of $U$, but they are more
pronounced for smaller $U$, for which the overall width of the QPP is
broader [see inset of \Fig{fig:P1_nd2_1fig}(c)].  At $J=0$, however,
these features are absent [see \Fig{fig:P1_nd2_1fig}(b) and its
inset]. From Ref.~\citep{Stadler2015} we know that the shoulder
emerges due to SOS, which only occurs for finite
$J>0$. \Fig{fig:P1_nd2_1fig}(c,d) thus give a first hint that
\textit{there is a direct connection between the Janus-faced low $Z$
and SOS.
}

Further, we find differences in the shape of the Hubbard side 
bands.  For $J=0$ there are two bumps in \Fig{fig:P1_nd2_1fig}(b).
The lower Hubbard band at negative frequencies is   less pronounced
than the upper Hubbard band at positive frequencies. With growing
$U$, the distance between these Hubbard bands increases, reminiscent
of the single-band Hubbard model.  For $J=1$ there are in principle
two Hubbard side bands, as well, in \Fig{fig:P1_nd2_1fig}(c),
however the band at positive frequencies consists of two bumps, so
that, at large $U$, we observe \textit{three} peaks altogether. For
small $U$, the negative frequency and the lower positive frequency
peaks are hidden in the QPP [red curve in \Fig{fig:P1_nd2_1fig}(c)]
and only one positive-frequency bump is visible. But with growing
$U\ge4$ the lower peak is shifted to lower frequencies and the
two-peak structure at positive frequencies clearly develops [see
purple, blue and black curves in \Fig{fig:P1_nd2_1fig}(c)].

\subsection{Peak structure of Hubbard bands: Hubbard-I analysis}

The peak structure of the Hubbard bands (at zero temperature) can be
fully understood in terms of  a Hubbard-I approximation  of  the
lattice Green's function, i.e. from its local multiplet
excitations,  as demonstrated in \Fig{fig:P1_nd2_2fig}. (This was
stated in previous studies but never demonstrated explicitly,  due
to the lack of reliable real-frequency
data \citep{,Georges2013, deMedici2017,deMedici2011}. So far, a similar
real-frequency analysis was only carried out for a three-band Hund model at $n_d=1$
using Fork Tensor Product States as real-time DMFT
solver \citep{Bauernfeind2017}).

To obtain the local multiplet excitations spectrum of the underlying
atomic problem, i.e in the ``atomic limit", $t=0$, we diagonalize
the local Hamiltonian as discussed in \Sec{sec:Hloc2:mult} with \Eq{eq:Hloc2} and
schematically depicted in \Fig{fig:Hloc:mult}.
\begin{figure}
\centering
\includegraphics[width=0.65\linewidth, trim=0mm 45mm 0mm 0mm, clip=true]{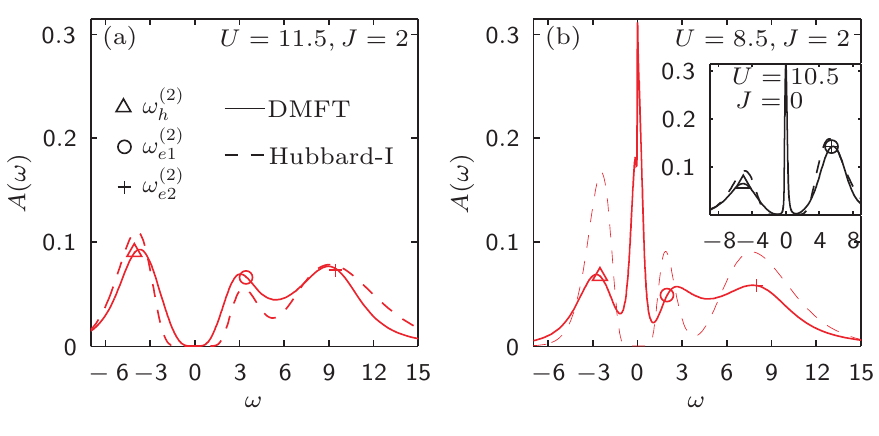}
\caption{
    (a) Insulating  and (b) metallic   local spectral function,
    $A(\omega)$, for $J=2$, obtained from DMFT+NRG (solid) and via
    Hubbard-I approximation (dashed).
           The symbols, as specified in the legend, 
           correspond to the local multiplet excitations
    listed in \Eqs{eq:Hloc2:exc:g1}.
   The inset in (b) shows results for $J=0$. Here,
   the symbols correspond additionally to the transition frequencies in
   \Eqs{eq:Hloc2:exc:g2},
   i.e triangles and pluses also
   correspond to  $\tilde\omega_h$ and $\tilde\omega_{\etwo}$.
In order to directly compare the Hubbard-I approximation with the 
log-Gaussian broadened  DMFT+NRG results, 
we convoluted the Hubbard-I spectral function with a 
log-Gaussian broadening Kernel of width $\alpha=0.4$, as defined in 
Ref.~\citep{Weichselbaum2007}.
}
\label{fig:P1_nd2_2fig}
\end{figure}

The positions of the peaks in the Hubbard bands 
shown in \Fig{fig:P1_nd2_2fig} are well captured by
the discrete multiplet excitations
indicated by the symbols provided with \Eqs{eq:Hloc2:exc}.
Thus 
\textit{the structure of
the incoherent side-bands can 
be understood from atomic physics.}
In order to explicitly demonstrate this, i.e. to reproduce
the form of the Hubbard bands, we use the Hubbard-I approximation
around the atomic limit  to disperse  the atomic eigenstates by
embedding them in a lattice environment. In this approximation the
lattice self-energy is replaced in Eq.~(\ref{eq:Gimpr}) by the
purely atomic self-energy corresponding to the limit $t=0$ in
Eq.~(\ref{eq:HU-Hloc}): $\Sigma(\omega)=\Sigma_{\rm atom}(\omega)$.
The atomic self-energy is given by $\Sigma_{\rm
atom}(\omega)=\omega+\mu-G_{\rm atom}^{-1}(\omega)$ in terms of the
atomic Green's function, $G_{\rm atom}(\omega)=\sum_{M}
p_M/(\omega-\omega_M+i0^+)$, summing over the atomic multiplet
excitation  poles
with $p_M$ the
probability for a one-particle excitation from the ground state into
the excited state $M$.

The resulting Hubbard-I spectral functions are plotted with dashed
lines in \Fig{fig:P1_nd2_2fig}. The insulating DMFT spectral
function for $U=11.5$ and $J=2$  is reproduced very well
[\Fig{fig:P1_nd2_2fig}(a)]. The structures of the Hubbard bands in
the metallic states for $U=8.5$ and $J=2$  [\Fig{fig:P1_nd2_2fig}(b)]
and for $U=10.5$ and $J=0$ [inset of \Fig{fig:P1_nd2_2fig} (b)] are
still matched reasonably well, but the QPP is not captured at all
within the Hubbard-I approximation because finite-lifetime effects
are not contained in the purely real  atomic self-energy. 
For smaller $U$ in the metallic regime, thus for a broader QPP
in the spectral functions, the deviations between the DMFT and the
Hubbard-I results therefore naturally increase.
 
The atomic excitation energies listed in \Eqs{eq:Hloc2:exc}
fully explain the qualitatively different structure of the
corresponding Hubbard bands: while two bumps are well-separated and
pronounced at $J=0$  (with a larger peak at positive frequency due
to the higher degeneracy of the corresponding atomic excitation),
the three-peaked Hubbard bands  form a broad incoherent background
for sizeable $J$, because $J$ shifts the inner side-peaks at
$\omega_{\eone}=-\omega_{h}=\frac{U}{2}-J$  towards the Fermi level, while the
peak at $\omega_{\etwo}=\frac{U}{2}+2J$ is shifted to higher frequencies.
This difference was also recently revealed for two  archetypal
correlated materials, the  Mott material V$_2$O$_3$ and the Hund
material Sr$_2$RuO$_4$ \citep{Deng2018}.  We note that additional
structures at the low-energy edges of the  Hubbard bands with
doublon-holon origin \citep{Lee2017}  are principally expected, but
presumably a higher resolution using adaptive
broadening \citep{Lee2016} and/or extensive
z-averaging \citep{Zitko2009} would be needed to resolve them.

\subsection{The ``bare gap" as a measure of Mottness}

In a next step we use the atomic excitation spectra for
sizeable $J\gtrsim J^*_{c1}$ 
to derive a measure of Mottness.  Following
Refs.~\citep{Georges2013,deMedici2017,deMedici2011,deMedici2011a}, we define
the ``bare gap", $\Delta_b \equiv 
\omega_{\eone}-\omega_{h}=U-2J$, as the
distance between the lowest atomic excitations at positive and
negative frequencies.
[Incidentally, $\Delta_b$ is equal to the atomic
interaction of the energetically most favored atomic configuration
in line three of \Eq{eq:Hloc1}].
Up to an offset, $\Delta_b$ measures the distance to the MIT.
In this sense it is similar to the true Mott
insulating gap $\Delta$ which closes at the MIT. Here
$\Delta=\omega^+ -\omega^-$ is defined
from  the criterion that $A(\omega)<10^{-3}$ holds for $\omega^-<
\omega<\omega^+$.

In the inset of Fig.~\ref{fig:P1_nd2_3fig}(a) we plot $\Delta$
versus $U$ (for iS) for various values of $J$  and derive $U_{c1}$
from the closure of the Mott insulating gap,
$\Delta(U_{c1})=0$ (marked by crosses) using a
well-suited linear extrapolation 
to the data points.
Obviously,  $U_{c1}$ strongly depends on $J$ (as seen already in
Fig.~\ref{fig:P1_nd2_1fig}). However, when $\Delta$ is plotted
versus $\Delta_b$  [see Fig.~\ref{fig:P1_nd2_3fig}(a)]  the
different lines lie ever closer to each other at large $J$ and the
critical value of the bare gap,
$\Delta_b^{c1}\equiv U_{c1}-2J$, approaches a constant value, $W_1=4.8$.
This is also demonstrated  in Fig.~\ref{fig:P1_nd2_3fig}(b). For
large 
$J \gg J^\ast_{c1}$,  
the critical interaction $\Delta_b^{c1}$  (solid red
line) is $J$-independent. Consequently, 
\textit{$\Delta_b$ serves as measure for Mottness}, 
in the sense that $W_1-\Delta_b$ quantifies the distance
to the MIT at $U_{c1}$ ($\Delta_b^{c1}$). 
Thus, the larger $\Delta_b$, the closer  the system is to  the MIT 
and the stronger the influence of Mottness.
 We demonstrate that this idea also works for an mS: for
$J>J^\ast_{c1}$, $\Delta_b^{c2}=U_{c2}-2J$ approaches a constant value
$W_1=6.3$ [see dashed black line in Fig.~\ref{fig:P1_nd2_3fig}(b)
and $\times$-signs in Fig.~\ref{fig:P1_nd2_3fig}(c)].  
We thus switch from $U$ to $\Delta_b(J,U)$ as independent
parameter in the following to quantify Mottness.
However, we note that for
$J<J^\ast_{c1}$, $\Delta_b^{c1}$ and $\Delta_b^{c2}$ do still depend on  $J$,
thus $\Delta_b$ breaks down as a simple measure for Mottness for
small $J$ in the above sense.

The reason for the $\Delta_b^{c,i}$ with $i=1,2$ becoming
a constant for large $J$ can again be roughly understood by
simply looking at the local multiplet structure, where for $J>U/2$
the excited levels $h$ and $\eone$ actually pass across
$g1$ (see discussion at end of \Sec{sec:Hloc2:mult}).
Therefore one may expect a qualitative change of
behavior at $\Delta_b = U-2J \sim \mathrm{const}$, 
as already mentioned in \Sec{sec:UJphasediagram}. 

The finite offset for $\Delta_{b}^{c1}$
can be explained with the Hubbard
criterion \citep{Hubbard1963} 
for the breakdown of the Mott insulating
state, which uses  $\Delta=\Delta_b-\tilde{W}(J)\equiv 0$ to conclude
that $\Delta^{c1}_b\equiv \tilde{W}(J)$: the system becomes metallic
when the effective kinetic energy in the system, $\tilde{W}(J)$, is
large enough to overcome the energy cost of hopping, given by the
energy scale of the bare gap $\Delta_b$. $\tilde{W}(J)$ sets the
scale for the dispersion of  the Hubbard bands and can be regarded
as the effective bandwidth of the system.  As shown in
Fig.~\ref{fig:P1_nd2_3fig}(b), $\tilde{W}(J)$ has a large value
$\tilde{W}(0)=7.3$ at $J=0$ and decreases  with increasing but small
$J$, approaching a constant  $W_1=4.8$ for sizeable $J>J^\ast_{c1}$.
From \Fig{fig:Hloc:mult} we know that at
the SU(6) symmetric point $J=0$ the atomic excitation
spectrum becomes more degenerate:
$\gtwo$ becomes degenerate with $\gone$,
and thus also a true ground state; furthermore,
all three excited levels $h$, $\eone$, and $\etwo$
become degenerate.
Accordingly, the widths of the Hubbard
bands, i.e. $\tilde{W}(J)$, are larger at small $J$,  because more
hopping processes are allowed than for $J>J^\ast_{c1}$. 
In contrast,
sizeable $J$ favors  high-spin states,  reducing  the atomic ground
state degeneracy by quenching its orbital fluctuations  and blocking
many excitations. 
We note that a similar analysis was performed in
Refs.~\citep{Georges2013,deMedici2017}.
\begin{figure}
\centering 
\includegraphics[width=0.65\linewidth]{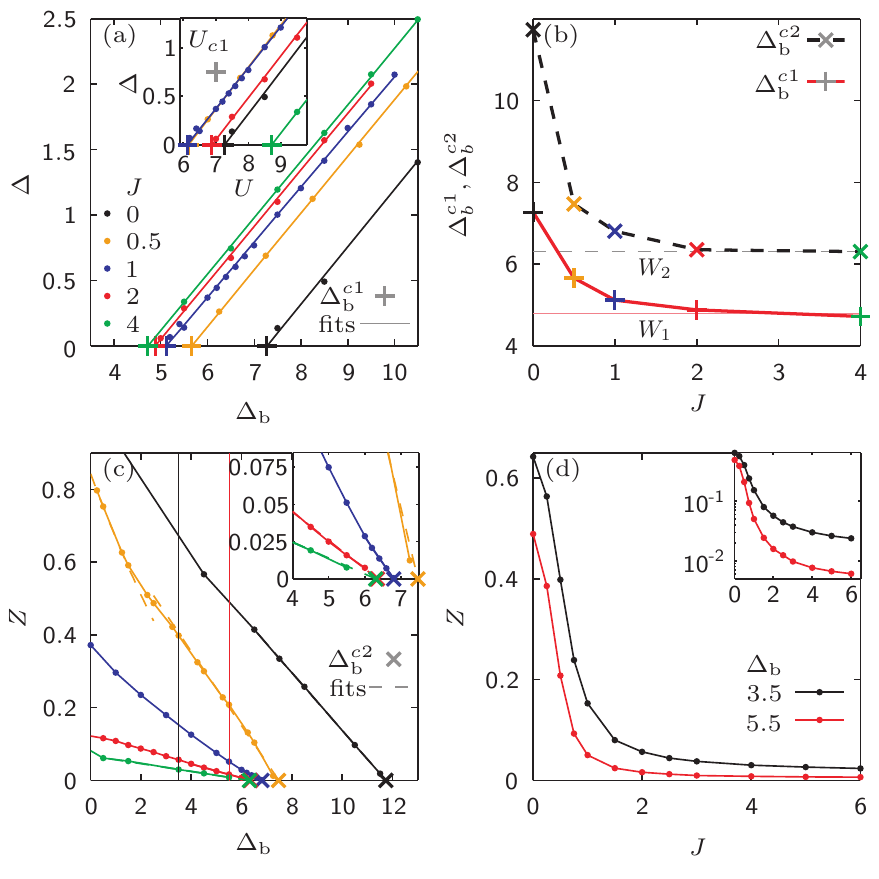}
\caption{
   (a) Mott insulating gap, $\Delta$, as
    a function of the bare gap, $\Delta_b=U-2J$, for several values
    of $J$.  Each dot on the curves represents a DMFT+NRG data
    point using iS. 
    The lines are linear
    fits from which the critical $\Delta_b^{c1}$ values (pluses) are
    defined as $\Delta(\Delta_b^{c1})=0$.  The inset shows the same
    data as a function of $U$.
(b) $\Delta_b^{c1}$ and
    $\Delta_b^{c2}$ as  functions of $J$: both  first decrease
    roughly exponentially 
    at small $J<J^\ast_{c1}$
    [see also \Sec{sec:sosandUJphasediagram}]
    and then approach fixed
    values, $W_1=4.8$ (thin solid red line) and $W_2=6.3$ (thin
    dashed grey line), respectively, at large $J>J^\ast_{c1}$.
(c) $Z$ is
    plotted as a function of $\Delta_b$ to disentangle the
    Janus-faced behavior of \Fig{fig:P1_nd2_1fig}(a): the slope
    of $Z $ decreases with increasing $J$, while $\Delta_b^{c2}=W_2$
    is $J$-independent for sizeable $J>J^\ast_{c1}$ (and grows with decreasing
    $J$ for $J<J^\ast_{c1}$). Thus $Z$ is small far away from the MIT due to
    Hundness rather than Mottness. The dashed yellow lines are
    quadratic 
    and linear fits to the $J=0.5$ behavior of $Z$ at
    small $U$ and larger $U$, respectively. The inset is a zoom of
    the $\Delta_b^{c2}$-behavior.
(d) $Z$ is plotted as a function
    of $J$, for two fixed values of $\Delta_b$, indicated by the
    thin black and red lines in (c). Inset: same data in a
    semilog-plot of $Z$, revealing its roughly exponential decrease with
    increasing $J$ for $J<J^\ast_{c1}$, whereas $Z$ is very small but rather
    constant for $J>J^\ast_{c1}$.
}
\label{fig:P1_nd2_3fig}
\end{figure}

As in Ref.~\citep{Georges2013}, we conclude that the
non-monotonic behavior of $U_{c1}$ can be summarized as follows:
with growing $J$, $U_{c1}$ decreases at small $J$ due the reduction
of the kinetic energy by orbital blocking, whereas it increases
again at large $J$, due to the reduction of $\Delta_b$ by reducing
the energy cost for the double occupancy of \textit{different}
orbitals. The turnaround occurs around $J\sim 1$,
i.e. when $J$ is on the order of the lattice hopping, $t=1$.
At the same time, as we point out at the end of
\Sec{sec:Hloc2:mult}, the non-monotonic behavior in $U_{c1}$
can also be directly linked to a qualitative change in the underlying
multiplet structure: the turn-around of $U_{c1}$ coincides 
with the point in the parameter regime where the `excited' levels
$h$ and $\eone$ pass across the `low-energy' level $\gtwo$
in the metallic regime $J>U_{c1}/6$. This occurs when $J\gtrsim 1$.
The behavior of $U_{c2}$, which  is similar to  $U_{c1}$,
will be  revisited and explained in Sec.~\ref{sec:sosandZ} in the
context of SOS.

\subsection{Hundness as origin of strong correlations}

In contrast to previous studies, we now use $\Delta_b$ as a measure
for Mottness in Fig.~\ref{fig:P1_nd2_3fig}(c,d) to disentangle the
Janus-faced effects of $J$ in $Z$ and to analyze the ``pure" effect
of Hundness for strong correlations.

Fig.~\ref{fig:P1_nd2_3fig}(c)  shows $Z$ versus $\Delta_b$ for
various values of $J$. We observe that, as visible for $J=0.5$, 
the reduction in $Z$ with $\Delta_b$ first follows a quadratic behavior 
for small $\Delta_b <4J$ (which coincides with $U<6J$)
followed, as visible for all values of $J$, by
a linear behavior for
moderate $\Delta_b$ up to $\Delta_b^{c2}$ (for $J=0.5$ this behavior is
illustrated by fits, shown as the upper and lower dashed yellow
lines, respectively).
For $J>J^\ast_{c1}$, $\Delta_b^{c2}$ is $J$-independent  [see inset of
Fig.~\ref{fig:P1_nd2_3fig}(c)], and  $W_2-\Delta_b$ again
measures the distance to the MIT.

For fixed $\Delta_b$, we observe in \Fig{fig:P1_nd2_3fig}(c)
that increasing $J$ reduces $Z$,
with the decay in $Z$ significantly slowed down
for $J>J^\ast_{c1}$ (see inset).

The data along the thin
red and black vertical lines is further summarized
in \Fig{fig:P1_nd2_3fig}(d).
Note that the curve for $\Delta_b=5.5$
already proceeds midway in between $U_{c1}$ and $U_{c2}$
in the coexistence region in \Fig{fig:4a-AW}
for large $J$ (e.g., see intercept at $J=0$ for their
linear extrapolation), whereas $\Delta_b=3.5$ is still
in the metallic phase.

Interestingly, for fixed $\Delta_b$,  the overall suppression of $Z$ 
with increasing $J$ is more pronounced 
for smaller $\Delta_b$, where the values of $Z$ are still very large for small $J$, 
but strongly reduced for large $J$ [compare e.g. the $Z$ values following the 
thin vertical lines for $\Delta_b=3.5$ and $\Delta_b=5.5$ in Fig.~\ref{fig:P1_nd2_3fig}(c) or 
compare black and red curve in \Fig{fig:P1_nd2_3fig}(d)].
This behavior can be
inferred from the important insight that increasing $J$ reduces the \textit{slope} of $Z$ 
when plotted as a function of $\Delta_b$ (or $U$)  
in Fig.~\ref{fig:P1_nd2_3fig}(c) for \textit{all} $J>0$, while $\Delta_b^{c2}$ is 
first reduced and then approaches a fixed value. As another
major result of this work we thus summarize: \textit{for sizeable
$J$, $Z$ is strongly lowered also far from the MIT, at small
$\Delta_b$, because Hundness promotes the reduction of the slope of $Z$.}
The latter effect holds for  \textit{any} nonzero $J$
[yellow, blue, red and green curve in
Fig.~\ref{fig:P1_nd2_3fig}(b)], even independently of the fact
whether $\Delta_b$ is a valid measure of Mottness (green and red
curve) or not (yellow curve).  Therefore, \textit{
Hundness, i.e. scenario (ii),
is the origin  of strong correlations in the Hund-metal
regime far from the MIT at $n_d=2$.}

In the next section, we 
focus also on small $J<J^\ast_{c1}$. As seen in Fig.~\ref{fig:P1_nd2_3fig}(d),
in this regime, $Z$ is reduced roughly exponentially
with increasing $J$ (see also inset). However, here, we cannot
fully disentangle the Janus-faced behavior of $Z$ using $\Delta_b$. 

\subsection{Spin-orbital separation in the $U$-$J$ phase diagram}
\label{sec:sosandUJphasediagram}

In order to better understand the strong 
reduction of $Z$ at
small $J$ and to reveal  the physical nature causing the low $Z$ for
$J>J^\ast_{c1}$, we now systematically analyze the underlying DMFT+NRG
real-frequency spectral data in the metallic (and coexistence)
region of the $U$-$J$ phase diagram.  In particular, we consider
$\chi''_{\orb}(\omega)$ and $\chi''_{\spin}(\omega)$, the imaginary
parts of the dynamical impurity orbital and spin susceptibilities,
defined in \Eqs{eq:chi}, the local spectral function
$A(\omega)$, and the imaginary part of the self-energy,
$\imag \Sigma(\omega)$, 
defined in \Eqs{eq:sigma}.  Similar to Ref.~\citep{Stadler2015}, we plot
$\chi''_{\orb}(\omega)$ and $\chi''_{\spin}(\omega)$ in
\Fig{fig:PRB-SOS:J}(a) and
\Fig{fig:PRB-SOS:U}(a) to deduce $\Tkorb$ and $\Tkspin$
from their respective maxima.
$A(\omega)$ is plotted in
\Fig{fig:PRB-SOS:J}(b-d) and
\Fig{fig:PRB-SOS:U}(b-d), and $\imag \Sigma(\omega)$
in \Fig{fig:PRB-SOS:J}(e,f) and \Fig{fig:PRB-SOS:U}(e,f).
 In
\Fig{fig:PRB-SOS:J}
$\Delta_b=3.5$ is fixed and $J$ is varied, while in
\Fig{fig:PRB-SOS:U} $J=2$ is fixed and $U$ ($\Delta_b$) is varied
[the latter is similar to \Fig{fig:P1_nd2_1fig}(c),
there for $J=1$].
\begin{figure}
\centering 
\includegraphics[width=0.65\linewidth, trim=0mm 15mm 0mm 0mm, clip=true]{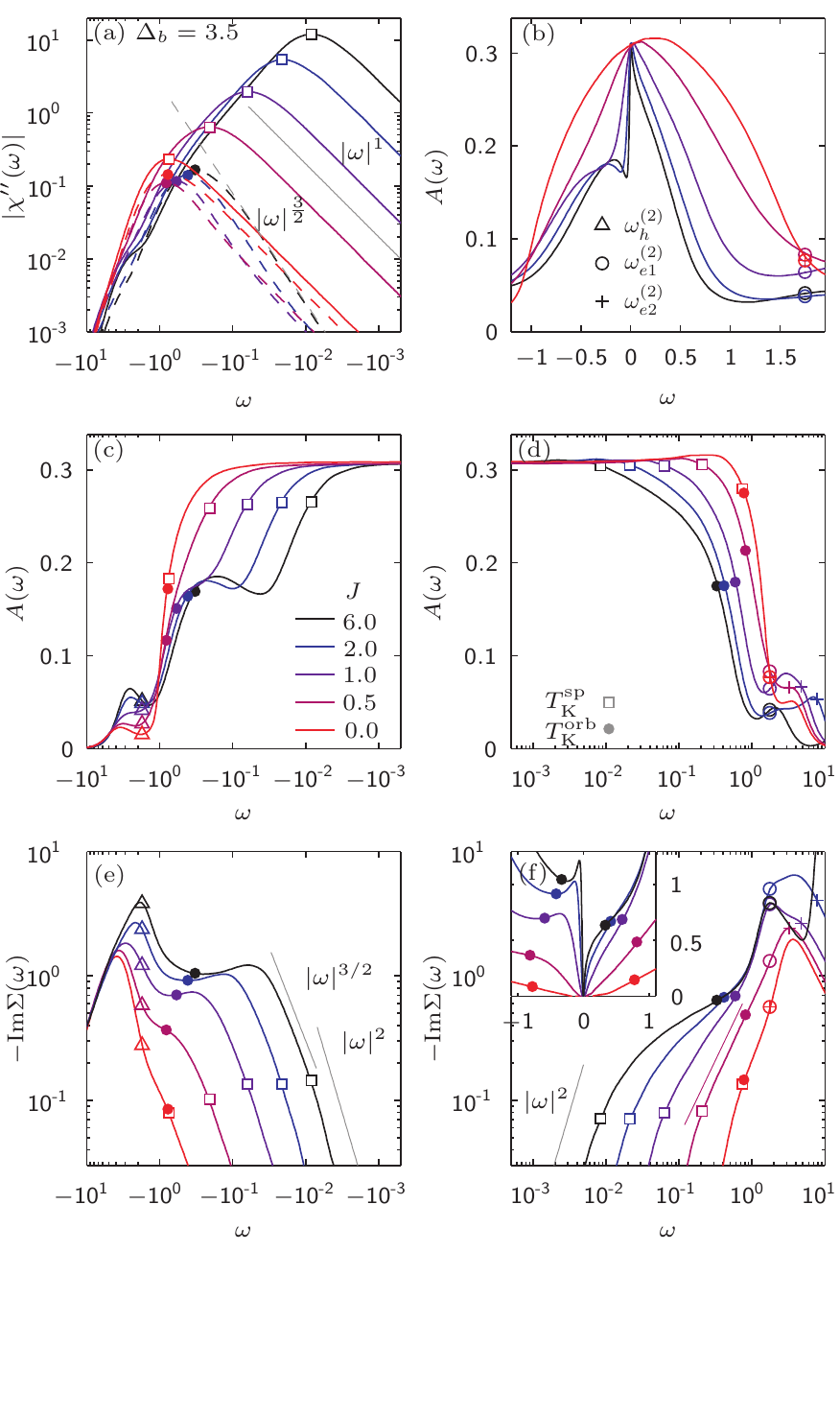}
\caption{ (a) The imaginary parts of the dynamical impurity
    orbital and spin susceptibilities,  $|\chi''_{\orb}(\omega)|$
    (dashed) and $|\chi''_{\spin}(\omega)|$ (solid), (b-d)  the
    local spectral function $A(\omega)$, and (e,f) the imaginary 
    part of the self-energy, $\imag \Sigma(\omega)$, for fixed $\Delta_b=3.5$ and
    various choices of $J$. (a) $\Tkorb$ (filled circles) and
    $\Tkspin$ (open squares) are defined from the maxima of
    $\chi''_{\orb}(\omega)$ and $\chi''_{\spin}(\omega)$,
    respectively. With increasing $J>0$, an SOS
    regime clearly develops, $\Tkorb>|\omega|>\Tkspin$, with complex
    NFL behavior. 
    $\chi''_{\orb}(\omega)$ follows an apparent
    $|\omega|^{3/2}$ power law in the SOS regime
    (dashed grey guide-to-the-eye line),
    which we believe is just a cross-over behavior (see discussion
    in \Sec{sec:flow}). Below $\Tkspin$, the expected
    $|\omega|^1$ FL power-law behavior sets in, indicated by a solid
    grey guide-to-the-eye line. (b,c,d) With increasing $J$  a
    $SU(6)$ Kondo resonance in $A(\omega)$ splits into a
    $SU(3)$ Kondo peak (shoulder for $\omega<0$ and kink for
    $\omega>0$) and a sharp $SU(2)$ Kondo QPP, reflecting
    two-stage screening of orbital and spin degrees of freedom due
    to SOS.  These features are shown on (b)
    linear  and (c,d) logarithmic frequency scales for (c) negative
    and (d) positive frequencies.
   (e,f) $\imag \Sigma(\omega)$ is
plotted versus (e) negative and (f) positive frequencies. Solid grey guide-to-the-eye lines indicate  
    $|\omega|^{2}$ FL power-law behavior and apparent $|\omega|^{3/2}$ behavior at $\omega<0$, 
    the magenta guide-to-the-eye line in (f) shows an apparent fractional-power law at $\omega>0$ for $J=0.5$. 
    The latter fractional power laws presumably originate just from a cross-over behavior. 
     The symbols, as specified in the legend in (b), correspond to the local multiplet excitations
    listed in \Eqs{eq:Hloc2:exc:g1}.
   For $J=0$,  triangles and pluses also correspond 
   to the transition frequencies in
   \Eqs{eq:Hloc2:exc:g2}, i.e to  $\tilde\omega_h$ and $\tilde\omega_{\etwo}$.
}
\label{fig:PRB-SOS:J}
\end{figure}
\begin{figure}
\centering 
\includegraphics[width=0.65\linewidth, trim=0mm 15mm 0mm 0mm, clip=true]{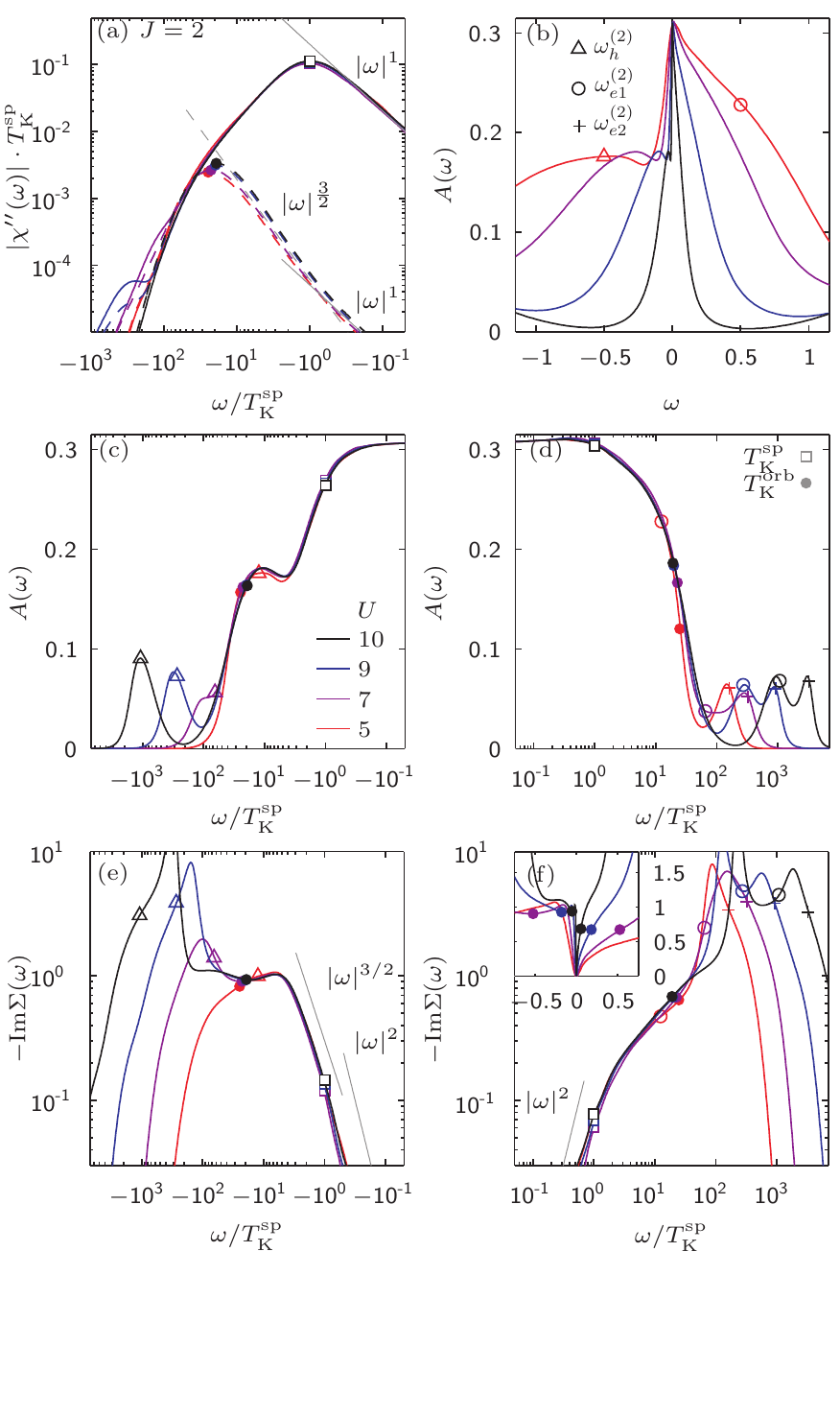}
\caption{
   Similar data as in
   \Fig{fig:PRB-SOS:J}, but for fixed
   $J=2$ and various choices of $U$ ($\Delta_b$), plotted as
   a function of  $\omega/\Tkspin$  on a logarithmic frequency
   scale in (a,c,d,e,f),  and in (b) as a function of  $\omega$
   on a linear frequency
   scale. All curves are identical for
   $|\omega/\Tkspin|<\Tkorb/\Tkspin\approx20$
   while, nevertheless, the low-energy physics
   moves to smaller energies with increasing $U$
   on a linear scale (panel b). (c-f) Thus, ``QP Hund
   features" in $A(\omega/\Tkspin)$ and $\imag \Sigma(\omega/\Tkspin)$ are independent of $U$ in both
   the rescaled SOS regime, and the rescaled FL regime for $|\omega|< \Tkspin$
   (narrow, sharp peak in $A(\omega\Tkspin)$). 
The symbols, as specified in the legend in (b), correspond to the local multiplet excitations
   listed in \Eqs{eq:Hloc2:exc:g1}. 
}
\label{fig:PRB-SOS:U}
\end{figure}

SOS, i.e $\Tkorb \gg \Tkspin$,
occurs in the whole metallic regime for nonzero $J$,
as seen in \Fig{fig:PRB-SOS:J}(a) and
\Fig{fig:PRB-SOS:U}(a).
It is a generic consequence
of finite Hund's coupling in particle-hole asymmetric
\textit{multi}-band systems, as anticipated early
on \citep{Okada1973}. 
Since $\Tkspin$ is finite,
the  ground state is a FL  [see thin grey
$|\omega|^1$-guide-to-the-eye lines in
Fig.~\ref{fig:PRB-SOS:J}(a) and
Fig.~\ref{fig:PRB-SOS:U}(a)] for all
values of $U$ and $J$ at $n_d=2$, independently of the proximity to
the MIT. This strongly contradicts
the spin-freezing phase scenario proposed in
Ref.~\citep{Werner2008}, but confirms 
the expectations of
Refs.~\citep{Haule2009,Yin2012, Werner2012,deMedici2011}. 

For fixed $\Delta_b$, the SOS
regime opens up with increasing $J$ [the
maxima of $\chi''_{\spin}(\omega)$ are shifted to smaller
$|\omega|$ in \Fig{fig:PRB-SOS:J}(a)].
This effect is accompanied  by the formation of a
shoulder at $\omega<0$, and a weak 
kink at $\omega>0$ in
$A(\omega)$, which narrow the top of the QPP [see
\Fig{fig:PRB-SOS:J}(b-d)], and
reveal a strong particle-hole asymmetry in the system.
Accordingly, the imaginary part of the self-energy, $\imag \Sigma(\omega)$, develops 
a pronounced shoulder (bump) in the SOS regime at $\omega<0$ [\Fig{fig:PRB-SOS:J}(e)], and
a kink at $\omega>0$ [\Fig{fig:PRB-SOS:J}(e)], as well. Note that the kink is only visible for $J>1$, while at smaller $J$,
$\imag \Sigma(\omega)$ seems to follow  apparent power-laws (as indicated by the magenta
guide-to-the-eye line for $J=0.5$ in \Fig{fig:PRB-SOS:J}(f) 
and observed in Fig.~3(b,e) of Ref.~\citep{Stadler2015}).
For $J=0$, the QPP is formed by one broad $SU(6)$ Kondo resonance.
With increasing $J$, this Kondo resonance is split into a
narrow $SU(2)$ 
spin Kondo resonance on top of a
wider 
$SU(3)$ 
orbital Kondo resonance (e.g., the shoulder),
corresponding to spin
and orbital screening, respectively [see \Fig{fig:sos_sketch}(a)
for a schematic sketch]. The orbital features become
strongly particle-hole asymmetric with increasing $J$,
with lesser effects on the spin resonance.
Thus, SOS is manifest in a two-tier QPP
with a wide base 
and a narrow ``needle'' 
of \mbox{(half-)} width $\Tkorb$ and $\Tkspin$, respectively.
We see from the behavior of $\Tkorb$ in
\Fig{fig:PRB-SOS:J}(a) that the ``full"  width  of the QPP
is rather stable with increasing $J$ (at least for negative frequencies).
In contrast, the width of the needle
strongly reduces with $J$ [compare  e.g. red and black curves in
Fig.~\ref{fig:PRB-SOS:J}(b-d)].  

We
note that the orbital and spin screening in the \HHM are
non-trivial screening processes that differ from  standard SU(N)
Kondo-type screening processes. The  Kondo model corresponding to
the \HHM with specific representations of the impurity
spin and orbital operators has been worked out 
in Refs.~\citep{Aron2015,Walter2018}, e.g. resulting
in a ferromagnetic bare spin coupling.
In particular, a  complex, protracted RG flow has been revealed where
orbital and spin degrees of freedom are \textit{not} decoupled,
leading to a subtle spin-orbital Kondo effect
(see also \Fig{fig:sos_sketch}):
first, at higher
energies, the  intermediate-coupling NFL fixed point of an
underlying effective 2 (spin)-channel $SU(3)$
Coqblin-Schrieffer model is reached, where the ferromagnetic spin
coupling is quenched. Then, at much lower energies, the spin
coupling renormalizes to an anti-ferromagnetic value and the RG flow
results in a strong-coupling FL fixed point. For $J=0$, the Kondo
model reduces to the single-channel antiferromagnetic $SU(3 \times
2)$ Coqblin-Schrieffer model. Therefore, when for $J>0$, we refer to a
$SU(3)$ orbital  and  a $SU(2)$ spin Kondo resonance,
or, for $J=0$, to a $SU(6)$ Kondo resonance, we have this
non-trivial spin-orbital Kondo effect in mind.
 
Fig.~\ref{fig:PRB-SOS:U} shows similar
data as in Fig.~\ref{fig:PRB-SOS:J},
but now for a fixed 
$J$ and different values 
of $U$ ($\Delta_b$),  
plotted as a function of $\omega/\Tkspin$ in (a,c,d,e,f)
and $\omega$ in (b).
Here, $U$ affects $\Tkorb$ and $\Tkspin$ in the same way:
their ratio, $\Tkorb/\Tkspin\approx20$, is
essentially independent of $U$, such that
the curves in \Fig{fig:PRB-SOS:U}(a)
lie on top of each other
for $|\omega|<\Tkorb$ (see  also the
discussion of Fig.~\ref{fig:AoP_ZTK_T0}, and the expressions for the
orbital and spin Kondo scales derived in
Ref.~\citep{Aron2015}). As a consequence, the shapes of
the QPPs in $A(\omega)$ and the self-energies $\imag\Sigma(\omega)$
are scale invariant for $|\omega|\leq\Tkorb$, too, when plotting both quantities as a function of
$\omega/\Tkspin$ [see
Fig.~\ref{fig:PRB-SOS:U}(c,d) and (e,f), respectively],
reminiscent of the universal behavior in the
single-band Hubbard model. The reason for this 
is that the ratio $\Tkorb/\Tkspin$ is constant
in the underlying Kondo model  \citep{Aron2015} of the \HHM  (for a
fixed $n_d=2$ corresponding to a certain spin and orbital operator
representation). This universal behavior of the Kondo scales 
is not changed by the DMFT self-consistency: 
the 
SOS is characteristic of impurity physics, i.e. it also emerges in  the
impurity AHM in the absence of an MIT \citep{Stadler2015}.  
The DMFT self-consistency just
adjusts the overall width of the QPP, by affecting the value of
$\Tkorb$, but not its internal structure, governed by
$\Tkorb/\Tkspin$.  In Fig.~\ref{fig:PRB-SOS:U}(b), on a
linear frequency scale, the SOS features are
more pronounced for larger $\Tkorb$, i.e. smaller $U$,
when compared to bare energy scales in the system.
\begin{figure}
\centering 
\includegraphics[width=0.65\linewidth, trim=0mm 0mm 0mm 0mm, clip=true]{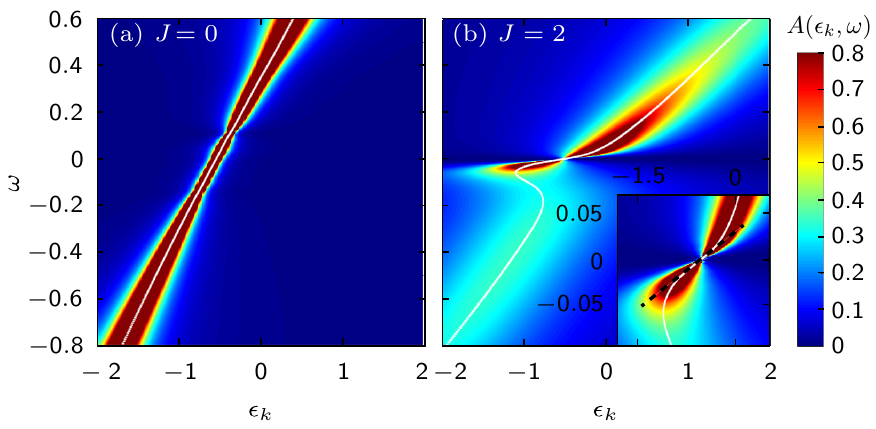}
\caption{
The structure factor, $A(\epsilon_k, \omega)$,
at $\Delta_b=3.5$ and $T=0$ for (a) $J=0$ and (b) $J=2$. 
The white curves show 
the QP dispersion, $E$ (see text for a definition).
The inset in (b) 
zooms into the FL regime at $J=2$. FL behavior is indicated by the black dashed guide-to-the-eye line.
}
\label{fig:PRB-SOS:Akw}
\end{figure}

We summarize the effect of spin-orbital separation at
$n_d=2$ in \Fig{fig:PRB-SOS:Akw}.  There we show the
structure factor $A(\epsilon_k, \omega)$, as experimentally
accessible by angle-resolved photoemission spectroscopy
(ARPES), for $J=0$ [panel (a)] and $J=2$ [panel (b)].
Within DMFT, $A(\epsilon_k, \omega)$ is directly obtained
from the self-energy $\Sigma(\omega)$: $A(\epsilon_k,
\omega) = -\tfrac{1}{\pi}\imag\left[\omega + \mu - 
\epsilon_k-\Sigma(\omega)\right]^{-1}$.
The QP dispersion 
(white curve) is defined as the solution to the equation 
$\omega+\mu-\epsilon_k-\real \Sigma(\omega)=0$ \cite{Deng2015}.
For fixed $\omega$, this
trivially yields a single value for $\epsilon_k$,
but not necessarily a unique value for $\omega$
for fixed $\epsilon_k$. 
Considering the latter solution(s), $E(\epsilon_k)$,  for given $\epsilon_k$,
then for $J=0$,
$E$ shifts linearly with $\epsilon_k$, i.e.\ the
band corresponding to the QPP is fully  characterized by a
linear FL dispersion relation with constant slope
$\tfrac{\partial E}{\partial\epsilon_k}\sim\tfrac{1}{m^*_{J=0}}\sim
Z\sim\Tkspin$, in the whole frequency regime plotted in
\Fig{fig:PRB-SOS:Akw}(a).  In contrast, for $J=2$,
$\Tkspin$ is reduced by more than one order of magnitude
compared to $J=0$. Thus
$\tfrac{\partial E}{\partial\epsilon_k}\sim\tfrac{1}{m^*_{J=2}}$
is constant only in a very small energy regime [as
indicated by the black dashed line in the inset of
\Fig{fig:PRB-SOS:Akw}(b)]. Further, this slope
 is much smaller than for $J=0$,
indicating a strong reduction of the effective mass, $m^*$,
for finite $J$ (due to Hund's-coupling-induced strong
correlations). Interestingly, when entering the SOS regime
for frequencies $|\omega|$ above the FL regime, the slope
becomes steeper: the spin degrees of freedom become
unscreened, the QPs thus ``undressed" and the effective
mass smaller.  For $\omega>0$, this change in the slope is
manifest in a slight kink, followed by
a rather constant behavior of
$\tfrac{\partial E}{\partial\epsilon_k}$.  For
$\omega<0$, the shoulder (bump), observed in $A(\omega)$
and $\imag\Sigma(\omega)$, leads to a somewhat artificial
s-shaped dispersion, $E$,
including a divergence
in the slope and negative
effective masses (due to the Bethe lattice). In this
regime, three maxima are observed
in $A(\epsilon_k, \omega)$ at fixed $\epsilon_k$.
All these SOS features of
$A(\epsilon_k, \omega)$ are completely absent for $J=0$.

\subsection{Spin-orbital separation as origin of low $Z$}
\label{sec:sosandZ}

We are now ready to reveal the connection of SOS
and $Z$. We corroborate and summarize our findings of
the previous
\Sec{sec:sosandUJphasediagram} by directly analyzing the
behavior of $\Tkorb$ and $\Tkspin$ as functions of $\Delta_b$
and $J$.
Importantly, we expect, as pointed out earlier
[see Luttinger pinning \citep{Georges1996,Muller-Hartmann1989},
here with $A(\omega=0)=1/\pi$],
that the width of the Kondo resonance
scales linearly with the QP weight $Z$. As we will demonstrate below,
in the Hund regime of $J$, this holds for the spin Kondo
scale, i.e. $Z \propto \Tkspin$ for $J>J^*_{c1}$.  

We replot the data of \Fig{fig:P1_nd2_3fig}(c,d)
in \Fig{fig:AoP_ZTK_T0},
but now with focus on $\Tkorb$ and
$\Tkspin$ instead of 
$Z$ on a linear
[Fig.~\ref{fig:AoP_ZTK_T0}(a,c)]  and a semi-logarithmic
[Fig.~\ref{fig:AoP_ZTK_T0}(b,d)] scale. For reference,
we also replot our $Z$ 
data, but rescale 
it by a factor $a(J)\equiv \Tkspin/Z $ [indicated by the
dotted grey curve in Fig.~\ref{fig:AoP_ZTK_T0}(c)], which is 
essentially the same for all values of $\Delta_b$.
Fig.~\ref{fig:AoP_ZTK_T0}(a,b) show that for fixed $J$, $\Tkspin$
and $Z$ have the same dependence on $\Delta_b$, i.e.
$\Tkspin=a(J)Z$, with a proportionality factor, $a(J)\simeq 0.36$,
for
$J>J^\ast_{c1}$ and increasing values of $a(J)>0.36$ for  decreasing $J<J^\ast_{c1}$
[see $a(J)$ in Fig.~\ref{fig:AoP_ZTK_T0}(c)].
Analogously, for fixed $\Delta_b$ and  varying but sizeable $J>J^\ast_{c1}$ in
Fig.~\ref{fig:AoP_ZTK_T0}(c,d), we find that $\Tkspin \approx 0.36\,Z$.

We thus conclude, as a major result of this work, that
\textit{the reduction of $Z$ in the Hund-metal regime} of
Fig.\ref{fig:sketch_phasediagram} at $n_d=2$  \textit{is directly
linked to the reduction of $\Tkspin$ due to SOS},
and that  all insights gained  for $Z$ hold for
$\Tkspin$, and vice versa, specifically so 
for sizeable $J$.  Based on
the knowledge that the \HHM at $n_d=2$ has a FL ground
state, it is of course expected that $Z$ is a measure of the
coherence scale using Landau's FL theory
 (see Luttinger theorem above), as e.g. also
pointed out in Refs.~\citep{Georges2013, deMedici2017,deMedici2011}.
In this work, we have now demonstrated \textit{quantitatively} that
and how  $Z$  and $\Tkspin$ are connected. Additionally, we have
conclusively identified the origin of low $Z$ and the physical
mechanism causing the bad-metallic transport -- spin-orbital
separation.

Fig.~\ref{fig:AoP_ZTK_T0}(a,b) demonstrate again the important
insight that SOS is absent for $J=0$ for all
values of $\Delta_b$ ($U$): $\Tkorb=\Tkspin$ (black filled big
circles and black open squares lie approximately on top of each
other; the small difference is due to the fact that
$|\chi''_{\orb}(\omega)|$ was obtained form a calculation
with different NRG parameters, i.e. stronger truncation
due to numerical cost; we checked that using the same (stronger) 
truncation leads to exactly $\Tkorb=\Tkspin$.
But also here, $\Tkspin=a(J)Z$ with $a(J)>1$ [see dotted grey line in 
Fig.~\ref{fig:AoP_ZTK_T0}(c)] 
due to the FL ground state. In contrast, for nonzero $J$, SOS
with $\Tkorb\gg\Tkspin$ occurs: with increasing $J$, $\Tkorb$ is only
moderately reduced, while $\Tkspin$ and thus $Z$ are strongly
reduced (at fixed $\Delta_b$). More importantly, \textit{the slope
of the linear function $\Tkspin(\Delta_b)$ and thus $Z(\Delta_b)$
is strongly reduced with increasing $J$} [solid lines in
Fig.~\ref{fig:AoP_ZTK_T0}(a)],  \textit{while the slope of the
linear function $\Tkorb(\Delta_b)$ is approximately $J$-independent}
[dashed lines in Fig.~\ref{fig:AoP_ZTK_T0}(a)].  \textit{Far away
from the MIT, at small to moderate $\Delta_b$, this leads to a broad
SOS regime which  is extended from very low up
to very large energy scales} (comparable to the  bare atomic
excitations). When approaching the MIT with increasing $\Delta_b$,
both $\Tkorb$ and $\Tkspin$ decrease linearly, but with different
slopes: the SOS regime shrinks and is shifted to
lower energies [compare values of $\Tkorb$ and $\Tkspin$ at
$\Delta_b=3.5$ (black vertical line) and $\Delta_b=5.5$ (red
vertical line) for a fixed $J>0$ in Fig.~\ref{fig:AoP_ZTK_T0}(a);
see also the black ($\Delta_b=3.5$) and red ($\Delta_b=5.5$) curves
in Fig.~\ref{fig:AoP_ZTK_T0}(c): for  $J>J^\ast_{c1}$, the  distance between
dashed and solid line  is smaller for larger $\Delta_b=5.5$].
During this process the ratio $\Tkorb/\Tkspin$ first remains
constant, as can be observed on a semi-logarithmic scale in
Fig.~\ref{fig:AoP_ZTK_T0}(b) (blue curves). Very close to the MIT
both $\Tkorb$ and $\Tkspin$ (and thus also $Z$) vanish together.
Clearly,  \textit{the DMFT self-consistency
affects the QPP as a whole} and finally destroys the QPP -- including
its internal structure -- at the MIT.
\begin{figure}
\centering 
\includegraphics[width=0.65\linewidth]{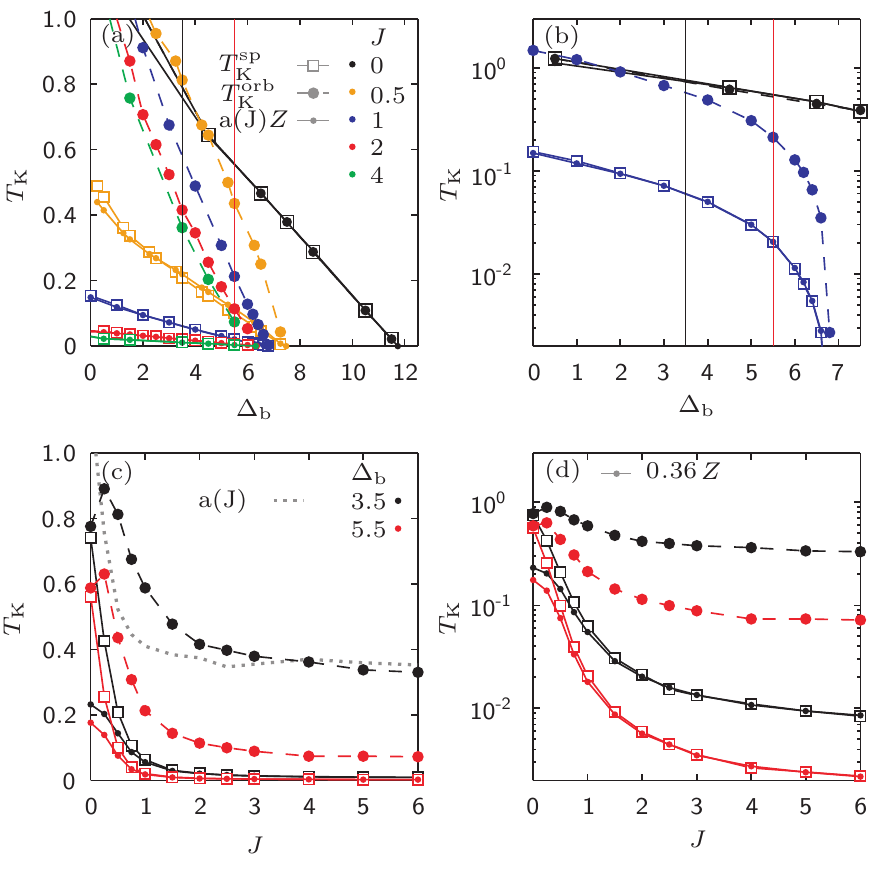}
\caption{
   The orbital Kondo scale, $\Tkorb$ (dashed curves with big filled
   circles), the spin Kondo scale, $\Tkspin$ (solid curves with open
   squares), and the rescaled QP weight, $a(J)Z$ (dotted
   grey curve),  plotted as a function of $\Delta_b$ for various
   values of $J$ using  (a) linear scale and (b) logarithmic scales
   for the y-axis.  Both $\Tkorb$ and $\Tkspin$ decrease linearly
    with $\Delta_b$, with a larger slope for $\Tkorb$ if $J>0$. The
    slope of  $\Tkspin$ strongly decreases with $J$, whereas the
    slope of  $\Tkorb$ is rather $J$-independent.  
    SOS, $\Tkorb\gg\Tkspin$, occurs for all $\Delta_b\le \Delta_b^{c2}$ at $J>0$, 
     but is more
    prominent at smaller  $\Delta_b$.  (c,d) Same quantities as in
    (a,b) now plotted as a function of $J$ for two values of
    $\Delta_b$ [indicated by vertical lines in (a)]. When $J$ is
    turned on, both $\Tkorb$ and $\Tkspin$ decrease strongly, but
    differently, opening up the SOS regime at
    small $J<J^\ast_{c1}$, and saturating at $J>J^\ast_{c1}$.
}
\label{fig:AoP_ZTK_T0}
\end{figure}

We now also discuss in more detail the behavior of the Kondo scales
and $Z$ for fixed $\Delta_b$ and varying $J$ [see
Fig.~\ref{fig:AoP_ZTK_T0}(c,d)].  At small $J$, spin-orbital
separation is turned on. The broad $SU(6)$ Kondo QPP with large
$\Tkorb=\Tkspin$ splits very abruptly with increasing $J$ into a
$SU(3)$ and a $SU(2)$ Kondo resonance,  reducing, after a slight decrease of
$\Tkorb$, 
both 
$\Tkspin$ and $\Tkorb$. As $\Tkspin$ is affected much stronger,
the ratio $\Tkorb/\Tkspin$ grows with increasing $J$, eventually
saturating for sizeable $J>J^\ast_{c1}$. In the latter large-$J$ regime, we
observe that both $\Tkorb$ and $\Tkspin$ ($Z$) are only slightly
reduced with increasing $J$ [as already observed in
Fig.~\ref{fig:P1_nd2_3fig}(d) for $Z$], and $a(J)\approx0.36$
is $J$-independent, i.e. SOS  is fully
developed and quite stable for sizeable $J$, and thus $Z$ is low.
Therefore, the main reason for lowering $\Tkspin$ and $Z$ upon turning on
$J$ can be heuristically ascribed to the following effect:  the
ground state multiplet degeneracy is lifted  by blocking orbital
fluctuations through the selection of high-spin multiplets,
as discussed in \Sec{sec:Hloc2:mult}. The
resulting orbital degeneracy is still much larger than the spin
degeneracy. Consequently, local Kondo-type screening of orbital
degrees of freedom occurs at much higher scales than spin screening.
$\Tkorb$ is only moderately whereas $\Tkspin$ and thus $Z$ are
strongly lowered. As mentioned before, a quantitative analysis for
a corresponding Kondo model is given in
Refs.~\citep{Aron2015,Walter2018}. 
As the degeneracy of the FL ground state changes when $J$ is turned
on, the factor $a(J)$ is strongly reduced, as well, in the small-$J$
regime  [see grey dotted curve in Fig.~\ref{fig:AoP_ZTK_T0}(c)]. The
reduction of $Z$ with increasing $J$ is thus less severe than the
reduction of $\Tkspin$ (compare solid  lines with small dots to
solid lines with open squares). 

Since $Z\propto \Tkspin$, also the behavior of $\Delta_b^{c2}$ ($U_{c2}$) is 
determined by SOS. For $J\ll J^\ast_{c1}$, $\Tkspin$ 
and thus $\Delta_b^{c2}$  ($U_{c2}$)  first decrease with increasing $J$
[see Fig.~\ref{fig:AoP_ZTK_T0}(c) and Fig.~\ref{fig:P1_nd2_3fig}(b), respectively]. 
For $J>J^\ast_{c1}$, $\Tkspin$ plotted as a function of $\Delta_b$ essentially saturates,
accordingly also $\Delta_b^{c2}$ saturates
[see black dashed curves in Fig.~\ref{fig:P1_nd2_1fig}(a) and
\Fig{fig:P1_nd2_3fig}(b)].  This explains why  $U_{c2}$ behaves
non-monotonously, similar to $U_{c1}$, and shows that the bare gap,
$\Delta_b$, can be used as a measure of Mottness at sizeable $J$
both for a mS and an iS.

Let us summarize the main conclusion of Sec.~\ref{sec:Janus}.
\textit{The  main effect to induce strong correlations in the Hund
metal regime of the \HHM at $n_d=2$ is Hundness rather
than Mottness, 
i.e. the very abrupt turning-on of spin-orbital
separation in the presence of nonzero (sizeable) $J$, independently
of the value of $\Delta_b$, thus also far from the MIT. } The MIT
itself, which is purely induced by the DMFT self-consistency, is an
additional but  subleading effect in the system, that only  further
lowers the spin and orbital Kondo scales with increasing $U$.  The
formation of $J$-induced  large spins  is, in principle, a local
process occurring on individual lattice sites. In contrast, the
formation of a charge gap is a highly non-local process that needs
to self-consistently incorporate the whole lattice dynamics (via a
gapped hybridization function). As a consequence of Hundness, the
nature of the incoherent transport is governed by ``Hund metal
physics" in the SOS regime at $n_d=2$:  large slowly
fluctuating spins are non-trivially coupled to screened orbitals
(see definition in Sec.~\ref{sec:sosrevisited}).  

But when
SOS is a generic effect in the metallic regime
of the \HHM (and presumably of all particle-hole
asymmetric degenerate multi-band Hund models), in which sense do
Hund- and Mott-correlated systems then differ in nature?

\subsection{Hund- versus Mott-correlated bad metals} 

Indeed, for the \HHM at fixed and sizeable $J$, the
features occurring for instance in $A(\omega)$, differ, in principle,
only  quantitatively when $U$ is varied: the Kondo scales shift as a
function of $U$, but the qualitative structure of the QPP does not
change.  However, we argue that the ratio of the bare atomic scales
and the Kondo scales (in particular $\Tkorb$), or phrased
differently, the ratio of the characteristic energy scale of the
Hubbard bands and the overall width of the QPP, sets the framework
for a meaningful characterization of Mott- and Hund-correlated
systems: this ratio is much larger in Mott than in Hund systems
(see Fig.~\ref{fig:sos_sketch}), leading to qualitative different
signatures, as demonstrated for temperature-dependent quantities in
Ref.~\citep{Deng2018}. 

Hund metals (characterized by moderate $U$, but sizeable $J$) are by
definition  far from the MIT. Their lowest bare atomic excitation
scales, $\omega_h$ and $\omega_{\eone}$ are small
[see discussion following \Eqs{eq:Hloc2:exc:g2}]. 
The Hubbard bands still overlap for moderate values of $U$ and form a
broad incoherent background in a range estimated by
$\omega_{\etwo}-\omega_{h}$,
having $\omega_h<0<\omega_{\eone}<\omega_{\etwo}$.
While $\Tkspin$ and thus $Z$ are
considerably reduced, $\Tkorb$ is  comparable to the bare atomic
excitation scales. This implies a ratio of order one between
$\Tkorb$ and the bare atomic excitation scales
[see Fig.~\ref{fig:sos_sketch}(b)]. As a consequence, the incoherent
SOS window, $\Tkspin< |\omega|< \Tkorb$, is broad
and ``Hund metal physics''  is relevant in a large energy window in
Hund metals. For instance, the  temperature-dependent local spin
susceptibility of a Hund metal shows Curie-like behavior in the
incoherent regime revealing large localized spins \citep{Deng2018}.
The low $Z$ of Hund metals thus implies spin localization but
\text{no} charge localization. Impurity physics dominates.

Multi-band Mott systems (characterized by $U$ being large compared
to $J$) are by definition close to the MIT. Their lowest bare atomic
excitation scales, $\omega_h$ and $\omega_{\eone}$ are large,  thus the
Hubbard bands are pronounced and well separated. Both Kondo scales
are small and thus the QPP narrow. Together this
implies that the bare atomic scales are much larger than $\Tkorb$
[see Fig.~\ref{fig:sos_sketch}(c)]. Further, the incoherent
SOS window, $\Tkspin<|\omega|<\Tkorb$, is very
small and ``Hund metal physics" is almost not observable. 
Similar to one-band Mott systems, $Z$ is low
because charge fluctuations are suppressed. In sum, typical Mott
physics, i.e. the DMFT self-consistency, dominates.

Finally, we note that the physics of Hund metals also strongly
differs from that of generic one-band (or multi-band) Hubbard models
(with $J=0$) which are far from the MIT. First, the latter are
weakly correlated, whereas a Hund system is strongly correlated,
despite being far away from the MIT. Second, SOS
and thus incoherent ``Hund metal physics"  only occurs for
particle-hole asymmetric \textit{multi}-orbital systems with at
least three-bands, fillings of $1<n_d<2N_c-1$ with $n_d\neq N_c$, 
and, most importantly, \textit{nonzero} $J$.

\section{Proximity to the half-filled MIT: Hundness  versus Mottness at $2<n_d<3$}
\label{sec:half-filled-MIT}

\begin{figure}
\centering
\includegraphics[width=0.65\linewidth,
trim=73mm 50mm 35mm 9mm, clip=true]{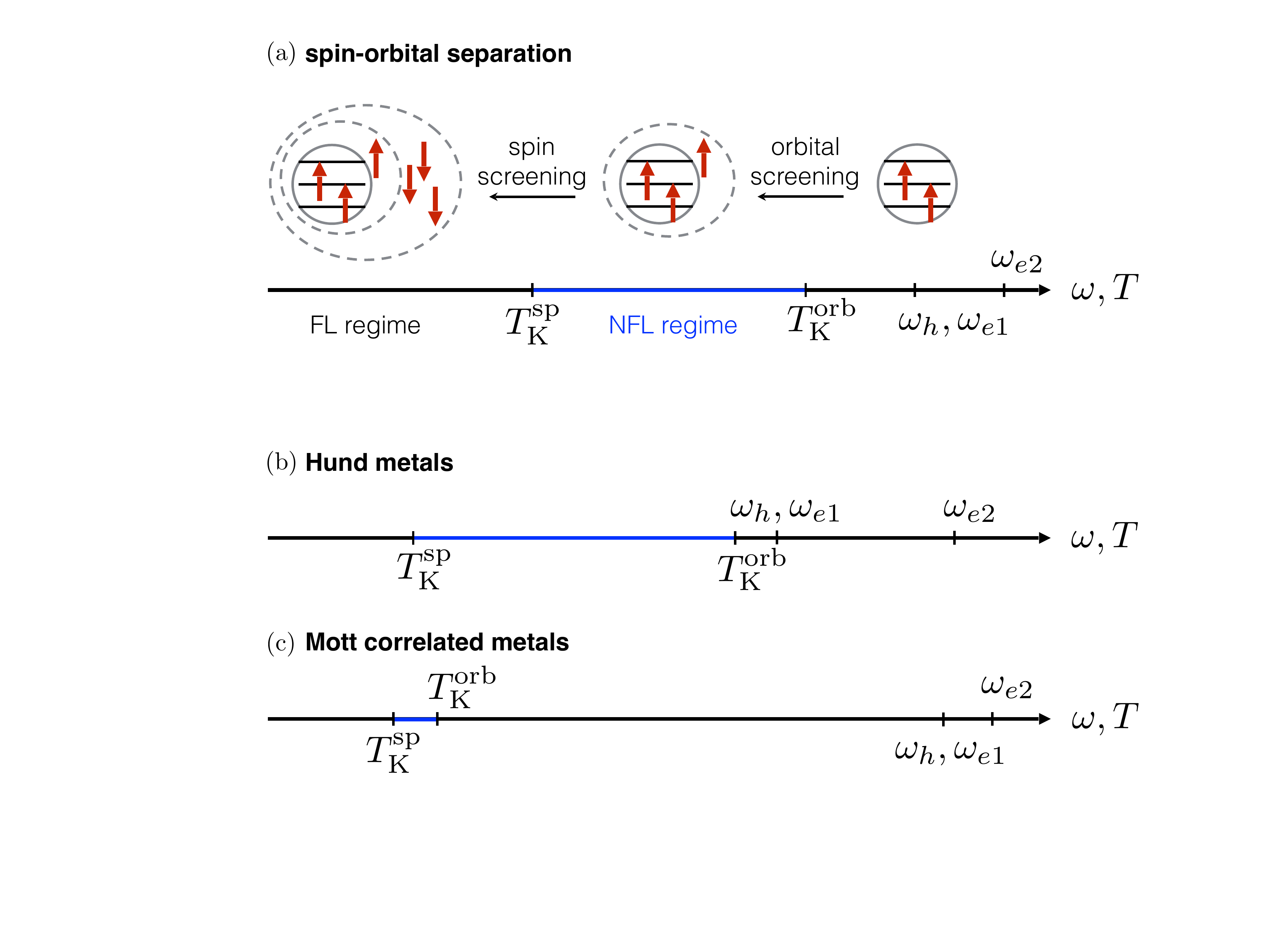}
\caption{
(a) Schematic depiction of the two-stage screening process
    of SOS at filling $n_d=2$.  First the orbital degrees of
    freedom are screened below  the orbital Kondo scale, $\Tkorb$,
    by the formation of a large, effective, Hund's-coupling induced
    $\rfrac{3}{2}$-spin including a bath spin degree of freedom.
    Then, at a lower spin Kondo scale, $\Tkspin$, this effective
    $\rfrac{3}{2}$-spin is fully screened by the three bath channels
    of the \HHM (see also the discussion
    for Fig.~\ref{fig:P1_nd3_3fig} in Sec.~\ref{sec:flow}).
    Incoherent NFL behavior is found for $\Tkspin<|\omega|,T<\Tkorb$,
    and  FL  behavior at energies below $\Tkspin$.
(b) In Hund metals,  bare atomic excitation scales, $\omega_h$ and
    $\omega_{\eone}$, and the overall width of the QPP, $\Tkorb$,
    are comparable in magnitude, while $\Tkspin$ and thus $Z$  are
    much smaller, opening a large relevant NFL regime in the system.
(c) In Mott-correlated metals, we find
    $\Tkspin \sim \Tkorb\ll \omega_h,\omega_{\eone}$, 
    such that $Z$ is reduced while SOS is not important.
}
\label{fig:sos_sketch}
\end{figure}

We now study the doping-dependence of the QP weight, $Z$, 
and of the electronic compressibility, $\kappa_{\rm el}
\equiv \frac{\partial {n}_d}{\partial \mu}$.
In particular, we 
demonstrate that SOS also occurs
for $2<n_d<3$, and that it determines the low $Z$-behavior there,
as well.
In particular, we focus on the question how Mottness of type (iii), i.e. the 
MIT at $n_d=3$, affects SOS and whether (i) 
Hundness or  (iii) Mottness is the key player to induce strong 
correlations in the Hund-metal regime for $n_d\gtrsim2$. 
Further, we will show that, for all parameters studied, no
Hund's-coupling-induced 
Fermi-liquid
instabilities (negative compressibilities) 
occur near the half-filled MIT of the \HHM, in contrast to suggestions in 
Ref.~\citep{deMedici2016}.

\subsection{MIT at $n_d=3$}

As mentioned before, at half-filling $n_d=3$,
$U_{c2}^{(3)}$ is much
smaller than at other fillings. This is now explicitly demonstrated in  
Fig.~\ref{fig:P1_nd3_1fig}(a), where we plotted $A(\omega)$ at  
$n_d=3$, and $J=1$ for various values of $U$, revealing the MIT at 
$n_d=3$. Starting from an mS and using $J=1$, we 
deduce from our real-frequency data the
extrapolated value $U_{c2}^{(3)} \sim 2.1 \pm 0.1$ at $n_d=3$, which is
strongly lowered compared to $U_{c2}^{(2)}=8.8$ at $n_d=2$.
While the region of low $Z$ around $n_d=2$ reaches down to
moderate values of $U$ far below $U_{c2}^{(2)}$, i.e. far
away from the MIT at $n_d=2$ in
Fig.\ref{fig:sketch_phasediagram}, these $U$ values are
still larger than $U_{c2}^{(3)}$. Therefore,
Refs.~\citep{Fanfarillo2015,deMedici2017} have argued
that the MIT at $n_d=3$ might be the reason for the low $Z$ at
moderate $U \ll U_{c2}^{(2)}$ (even at $n_d=2$) -- a
statement that will be investigated in this section. 

Further, we observe that also the structure of the Hubbard bands at 
$n_d=3$  differs completely from those at $n_d=2$ [compare red and 
black curves in Fig.~\ref{fig:P1_nd3_1fig}(b)]. Specifically, 
in contrast to the $n_d=2$ results of Sec.~\ref{sec:Janus},
the spectral functions of Fig.~\ref{fig:P1_nd3_1fig}(a) are particle-hole 
symmetric and the QPP has no shoulder, only slight kinks (see inset).
In a pictorial language, in the case of $n_d=3$ for larger $J$,
the only local multiplet is the $\rfrac{3}{2}$ spin,
with a singlet orbital character. Hence  orbital Kondo physics
is absent (or quenched up to energies on the order of
the local multiplet excitations, i.e. the Hubbard bands).
Therefore SOS features, as revealed for $n_d=2$, are 
absent at half-filling.

\subsection{Peak structure of Hubbard bands at $2\le n_d\le3$}

At integer filling 
$n_d=2$ [red curve in \Fig{fig:P1_nd3_1fig}(b)] $A(\omega)$ consists of three peaks
away from $\omega=0$, while at $n_d=3$ it has only two
pronounced peaks [black curve in \Fig{fig:P1_nd3_1fig}(b)] that are particle-hole
symmetric with respect to $\omega=0$.
The peak positions at finite frequency can be understood
simply from the underlying atomic multiplet transition
energies listed in \Eqs{eq:Hloc2:exc} for $n_d=2$ and
\Eqs{eq:Hloc3:exc:g1} for $n_d=3$, assuming sizeable $J$.
\begin{figure}
\centering
\includegraphics[width=0.65\linewidth, trim=0mm 0mm 0mm 0mm, clip=true]{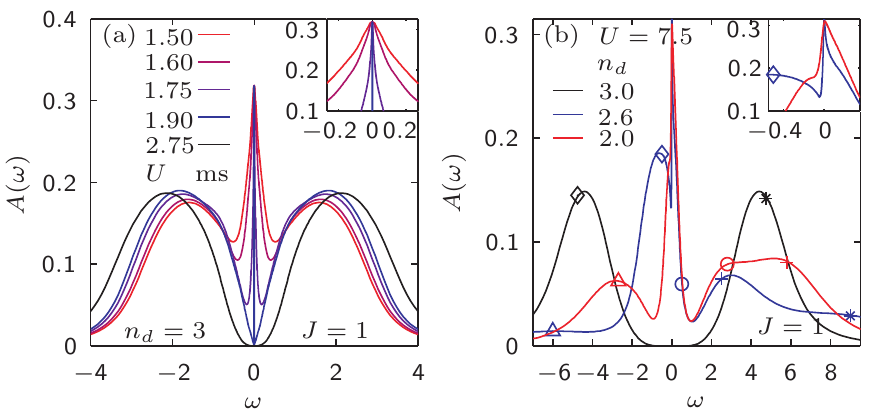}
\caption{
    The zero-temperature local spectral function, $A(\omega)$,
(a) for $n_d=3$, $J=1$ and various values of $U$, revealing an MIT
    with very small $2<U_{c2}^{(3)}<2.25$, and
(b) for $U=7.5$, $J=1$ and varying $n_d$, revealing how the structure
    of the Hubbard side bands changes with filling. The five different
    markers represent the energy of the atomic multiplet excitations
    at given $n_d$ [for $n_d=2$, see \Eqs{eq:Hloc2:exc};
    for $n_d=3$, see \eqref{eq:Hloc3:exc:g1}
    for details and an assignment of the markers;
    for $n_d=2.6$, the excitation energies are adapted to
    $\mu(n_d)$].
    The insets in (a,b) zoom into the QPP. 
}
\label{fig:P1_nd3_1fig}
\end{figure}

In order to study scenarios (i) and (iii) 
at intermediate fillings, $2<n_d<3$, we start by investigating
the structure of the Hubbard side bands for a filling,
$n_d=2.6$ [blue curve in Fig.~\ref{fig:P1_nd3_1fig}(b)].
We find that they are composed of all five types of atomic multiplet
excitations from both the $n_d=2$ and $n_d=3$ ground states
(5 peaks altogether) with their excitation energies adapted to $\mu(nd=2.6)$.
Overall,
at intermediate fillings, $n_d=2\rightarrow3$, we find a
smooth crossover in the  structure of the Hubbard bands
between their shape at $n_d=2$ and $n_d=3$, respectively,
caused by the smooth level transformation of eigenstates in
the spectrum of the local Hamiltonian with changing
$\mu(n_d)$, interchanging the ground state and varying the
probability of one-particle multiplet excitations. In
contrast, the shape of the Kondo resonances at $\omega=0$
change drastically when moving from $n_d=2$ to $n_d=3$.

\subsection{Spin-orbital separation at $2< n_d<3$ 
as the origin of low $Z$}
Next we gain insights from the structure of the QPP with varying $n_d$.
Similar to Fig.~\ref{fig:AoP_spin_freezing_ndvar},  we study the
filling dependence of $\Tkorb$ (dashed curves) and $\Tkspin$
(solid curves) in  Fig.~\ref{fig:P1_nd3_2fig}(a) and its inset, now for three
different values of $U$. With increasing $n_d$ (decreasing distance to half-filling, $3-n_d$), we observe an
increasing separation of both Kondo scales,  i.e. an increasing
ratio of $\Tkorb/\Tkspin$, for all values of $U$.
Thus 
SOS emerges for all fillings $1< n_d<3$
in the metallic phase (as already indicated in
Sec.~\ref{sec:spinfreezing} and the inset of Fig.~3(f) in
Ref.~\citep{Stadler2015}). We will show, however,
that the ``nature" of SOS changes with $n_d$. We remark that the  
behavior of $\Tkspin$  plotted versus $n_d$ in the inset 
of Fig.~\ref{fig:P1_nd3_2fig}(a) 
corroborates earlier results of Ref.~\citep{Yin2012}. 

\begin{figure}
\centering 
\includegraphics[width=0.65\linewidth]{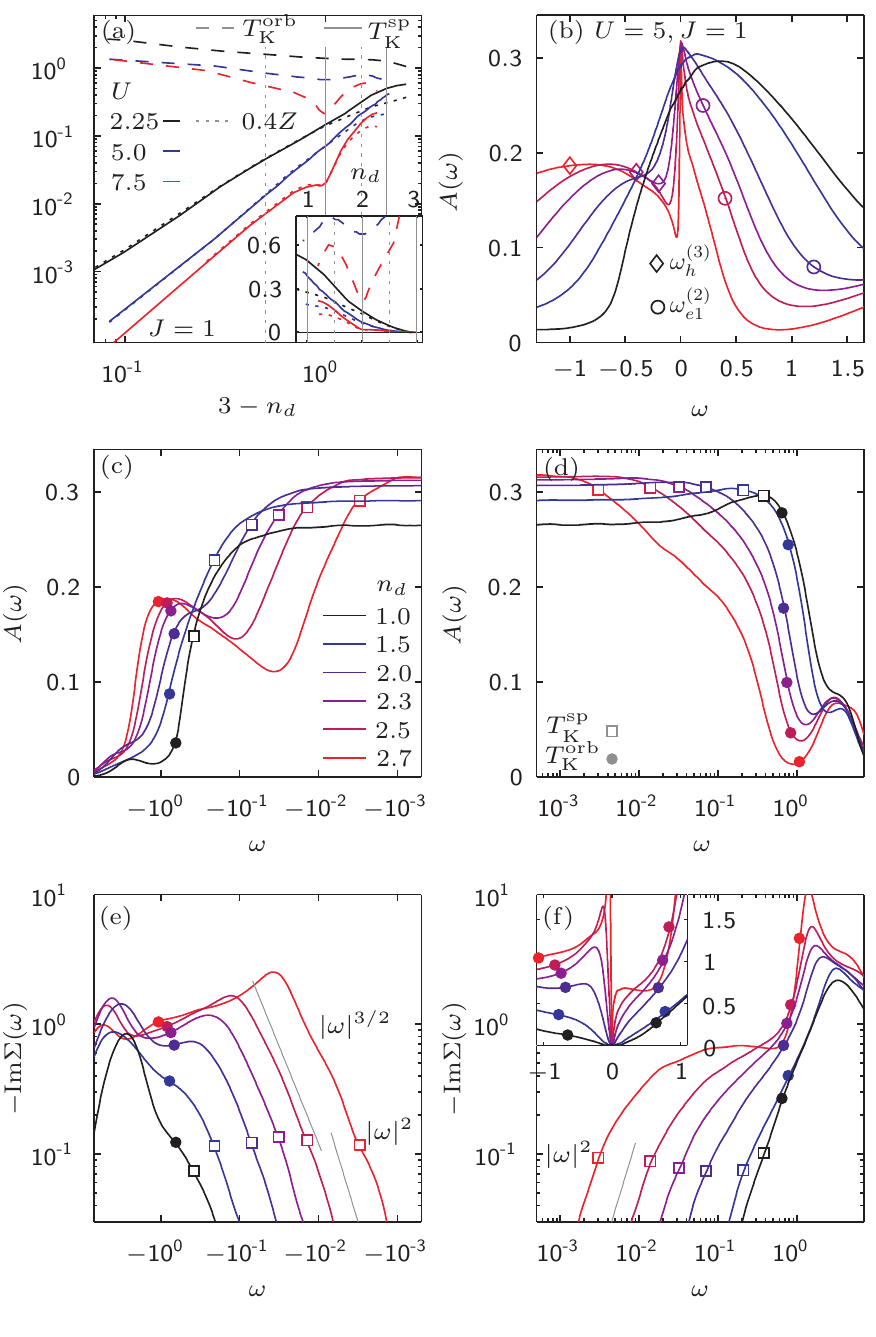}
\caption{
(a) The orbital and spin Kondo scales, $\Tkorb$ (dashed)
    and $\Tkspin$ (solid), on a log-log plot
    versus the distance to half-filling, $3-n_d$, 
    reveal the  filling-dependence of SOS.  The low QP weight $Z$
    (dotted curves) essentially follows the behavior of
    $\Tkspin$ for $2\le n_d<3$, and is thus determined by SOS.
   The inset shows the same data plotted
    versus $n_d$ on a linear scale.
(b-d) The local spectral function $A(\omega)$ for $U=5$,
    $J=1$ and various choices of $n_d$, shown on (b)  linear and
(c,d) logarithmic frequency scales for (c) negative and (d)
    positive frequencies. The symbols 
    in (b) indicate atomic multiplet excitations
    at given $n_d$ [for $n_d=2$, see \Eqs{eq:Hloc2:exc};
    for $n_d=3$, see \eqref{eq:Hloc3:exc:g1}
    for details and an assignment of the markers; 
    for $2<n_d<3$, the excitation energies are adapted to $\mu(n_d)$].
    For $n_d\rightarrow 3$,  the $\omega_h^{(3)}$
    excitations (diamonds) gain weight and replace the SOS
    shoulder in $A(\omega)$, which is clearly present as a
    pure QP-like feature at $n_d=2$.       
    (e,f) The imaginary part of the self-energy, $\imag\Sigma(\omega)$,
plotted versus (e) negative and (f) positive frequencies. Solid grey guide-to-the-eye lines indicate  
    $|\omega|^{2}$ FL power-law behavior and apparent $|\omega|^{3/2}$ behavior at $\omega<0$. 
    The latter fractional power-law presumably originates just from a cross-over behavior.  
}
\label{fig:P1_nd3_2fig}
\end{figure}

We begin by considering $n_d=1$.  We note that, in the
absence of charge fluctuations, i.e. for the pure Kondo limit
of the AHM, and if the energy scale of charge fluctuations
is much larger than the Kondo scales in the \HHM (or AHM),
the Hund's coupling $J$ just becomes an energy offset
and hence irrelevant, such that the SU(6) symmetry remains
intact. Therefore it holds at $n_d=1$ that
$\Tkorb=\Tkspin$ independent of $J$ (as demonstrated for the
impurity AHM in the inset of Fig.~3(f) in
Ref.~\citep{Stadler2015} and for a Kanamori model in
Fig.~6 of Ref.~\citep{Horvat2016}). 
In the presence of charge fluctuations at higher
energies, it still holds $\Tkorb \approx \Tkspin$.
For example, 
in \Fig{fig:P1_nd3_2fig}(a) for the self-consistent \HHM,
$\Tkorb$ is shifted by about a factor of $2$ towards larger
values compared to $\Tkspin$, especially for lower values of
$U$ which encourages larger charge fluctuations (see e.g black curves).
For $n_d$ near 1, the Kondo scales are large in
energy and comparable to the bare atomic multiplet
excitations scales. Thus,  signatures of the QP and
of bare atomic physics merge in $\chi''_{\orb}$ and
$\chi''_{\spin}$ [see \Fig{fig:P1_nd3_3fig}(d)].
As both quantities are affected
differently by the charge fluctuations due to Hund's
coupling, their maxima, $\Tkorb$ and $\Tkspin$, become
shifted in energy with respect to each other.

As the local occupation
increases towards $n_d=2$, SOS is turned on, i.e
the impurity's ground state $SU(6)$ symmetry is split, and $\Tkspin$
decreases by more than a factor of $2$ for $U=2.25$ (solid black
curve), of $5$ for $U=5$ (solid blue curve) and of $10$ for $U=7.5$
(solid red curve). At the same time, $\Tkorb$ first slightly
increases, reaching a maximum at around $n_d=1.5$, and then
(slightly) decreases again. For the largest $U=7.5$,
this leads to a reduction of $\Tkorb$ by a factor of about $4$
(dashed red curve; see also inset). 
There at $n_d=2$,
a strong minimum develops in $\Tkorb$ and a shoulder in
$\Tkspin$, respectively,  with increasing $U$ (red curves) due to
the growing influence of the MIT at $n_d=2$, lowering both Kondo
scales (as explained in Sec.~\ref{sec:Janus}).
For $n_d\rightarrow3$, similar to the behavior in the inset of
Fig.~3(f) in Ref.~\citep{Stadler2015} for the impurity AHM,
$\Tkspin$ drops
below the lowest relevant energy scale.
On the contrary, $\Tkorb$ grows up to energy scales 
comparable to the bare atomic scales in the system.
This shows that 
orbital fluctuations are suppressed right 
away together with charge fluctuations. Hence 
no orbital Kondo physics can develop.
What is left at half-filling, is a large spin $S{=}\rfrac{3}{2}$
on the impurity that needs to be screened dynamically.

\FIG{fig:P1_nd3_2fig}(a) also shows $Z$ (dotted curves) as a 
function of $n_d$. We find that, similar to the case of $n_d=2$ in 
Sec.~\ref{sec:Janus}, $Z$ essentially follows the behavior of $\Tkspin$ 
for $2\le n_d<3$ with $\Tkspin/Z \approx 0.4$,  
reflecting the fact that  the 
ground state is a FL. Throughout this regime, 
\textit{the small values of $Z$ can be  
understood, via their proportionality to $\Tkspin$, to be a direct 
consequence of SOS, which ensures that $\Tkspin \ll \Tkorb$.}
For $n_d\rightarrow 1$ the ratio $\Tkspin/Z$ changes,
due to strong changes in the ground state degeneracy  [see deviations 
between dotted  and solid curves for $n_d<2$ in the inset of
Fig.~\ref{fig:P1_nd3_2fig}(a)], reminiscent of the behavior of $Z$ for 
small $J$ in Fig.~\ref{fig:AoP_ZTK_T0}(c).

We remark that from the behavior of $\Tkspin(n_d)$ we cannot deduce any indication for a 
 relation between the physics at $n_d=2$ and the physics at 
$n_d=3$.
On the contrary, we see markedly different physical behavior
for $n_d=3$ as compared to $n_d=2$, e.g. with the absence of
Kondo physics in the orbital sector, and in this sense the absence
of SOS for $n_d=3$.
Further, the Hund-metal regime, (hatched area in Fig.~\ref{fig:sketch_phasediagram})
is special in that there we have not only SOS with $\Tkspin \ll \Tkorb$,
but in addition also a dynamically generated,  
fairly small value of $\Tkorb$. Thus, 
conditions there are optimal for the Hund's coupling to align
spins in different orbitals without forming an orbital singlet from the 
outset, allowing for a non-trivial interplay between both spin and orbital 
degrees of freedom, which induces SOS. 
We thus argue that 
\textit{the MIT at $n_d=3$ does not trigger the low $Z$ around 
$n_d=2$.}

\subsection{Spin-orbital separation at $2\leq n_d<3$: QPP structure}

Next we study the qualitative change in the structure of the low-energy
quasi-particle peak due to SOS with filling in more detail. 
In Fig.~\ref{fig:P1_nd3_2fig} (b,c,d) we plotted $A(\omega)$ with focus on
the QPP, and in Fig.~\ref{fig:P1_nd3_2fig} (e,f) $\imag\Sigma(\omega)$
for $U=5$, $J=1$  and various fillings, $1\le n_d < 3$.

In Fig.~\ref{fig:P1_nd3_2fig} (b), for $n_d>2$, $A(\omega)$
is shown on a linear frequency scale and we marked the multiplet
excitations of \Sec{sec:Hloc2:mult} and \Sec{sec:Hloc3:mult} [with 
the excitation energies adapted to $\mu(n_d)$],
as some of these (diamonds and circles) are rather low
in energy  and therefore might influence the shape of the QPP.
Complementary to this, in Fig.~\ref{fig:P1_nd3_2fig} (c,d),
$A(\omega)$ [and in Fig.~\ref{fig:P1_nd3_2fig} (e,f) $\imag\Sigma(\omega)$]
is shown on a logarithmic frequency scale and $\Tkorb$
and $\Tkspin$ are marked by open squares and filled circles,
respectively [see legend in (d)].
Clearly, with increasing $n_d$, the SOS regime opens
up: while there is no substructure in the QPP in $A(\omega)$ for $n_d\lesssim1.5$ [black and
blue curve in Fig.~\ref{fig:P1_nd3_2fig} (b,c,d)],  a pronounced shoulder develops with increasing $n_d\gtrsim2$
for $\omega<0$ and a  kink for $\omega>0$. 
Accordingly, a shoulder (kink) emerges in $\imag\Sigma(\omega)$ for $n_d>1.5$
at $\omega<0$ ($\omega>0$) which develops to a pronounced bump (plateau)
 for $n_d>2.5$ [see Fig.~\ref{fig:P1_nd3_2fig} (e,f)]. In a sense, the behavior of the 
 SOS features with increasing $1<n_d<3$ seems  reminiscent of their behavior with increasing $J$.
We note however
that the character of the shoulder in $A(\omega)$ changes for $n_d$ well beyond $2$:
the shoulder gradually transforms into a Hubbard side band
at the atomic hole excitation $\omega_h^{(3)}$ for $\omega<0$
[diamonds in magenta and red  curve in
Fig.~\ref{fig:P1_nd3_2fig} (b); see also inset of
Fig.~\ref{fig:P1_nd3_1fig}]. In contrast,
the QPP substructure narrows significantly, e.g. for $\omega>0$,
giving rise to a single albeit still strongly asymmetric Kondo peak
at $n_d=2.7$.
A true QP-like shoulder  only occurs for fillings
$n_d\lesssim2.5$, which we have checked in pure impurity AHM
calculations, where the Kondo scales can be tuned to lower values
and QP-like and atomic-like features are well separated.

\subsection{Spin-orbital separation at $2\leq n_d<3$: 
NRG flow diagrams} \label{sec:flow}

The nature of SOS is best revealed by the 
RG flows accessible to NRG via finite-size level spectra,
aka. energy flow diagrams
[see \Fig{fig:P1_nd3_3fig}(a-c)].
Technically, they show how the
lowest-lying rescaled eigenlevels of a length-$l$ Wilson 
chain \citep{Wilson1975,Bulla2008}
evolve with $l$, where ``rescaled'' means given 
in units of $\omega_{l} \propto \Lambda^{-l/2}$
(in the convention of Ref.~\citep{Weichselbaum2012a},
where $\Lambda >1$ is the NRG discretization parameter;
see supplement of Ref.~\citep{Stadler2015}).  
Conceptually, these levels represent the finite-size spectrum of the 
impurity+bath put in a spherical box of radius 
$R_{l} \propto \Lambda^{l/2}$, centered on the 
impurity \citep{Wilson1975,Delft1998}: as $l$ increases, the finite-size level 
spacing $\omega_{l} \propto 1/R_{l}$ decreases exponentially. The 
corresponding flow of the finite-size spectrum is stationary 
($l$-independent) while $\omega_{l}$ lies within an energy regime 
governed by one of the fixed points, but changes when $\omega_{l}$ 
traverses a crossover between two fixed points.
As the rescaled ground state energy of a  Wilson chain differs for even 
and odd numbers $l$ of sites due to fermionic parity,
the RG flow of the system is separated into an ``even'' and
``odd'' NRG 
flow diagram,
both reflecting the same physics of the system.
In Fig.~\ref{fig:P1_nd3_3fig}(a-c), we purely concentrate
on the even flow, since this permits the energetically favored
global (Kondo) singlet ground state as $l\to\infty$.
We fully exploited the symmetries
$\text{U(1)}_\charge\times \text{SU(2)}_\spin \times\text{SU(3)}_\orb$ 
of the \HHM in our NRG. Hence each line represents a multiplet 
and the color of each line specifies
a well-defined symmetry sector $(Q,S,q_1q_2)$,
where the total charge $Q$ is measured relative to half-filling,
$S$ is the total SU(2) spin multiplet sector, and $q\equiv(q_1 q_2)$
is the SU(3) orbital label.

The multiplets with significant spin or orbital character
behave qualitatively differently 
in the flow diagrams in \Figs{fig:P1_nd3_3fig}(a-c)
at finite $J$ at the crossover scales
$\Tkspin$ and $\Tkorb$ (vertical dashed lines).
The energy range in between defines the SOS regime.
We emphasize
that \textit{the SOS regime is an entirely new 
intermediate phase, which is absent for {$J=0$}}
[see inset in Fig.~\ref{fig:P1_nd3_3fig}(a)],
\textit{and opens up right at the Kondo scale in the NRG flow 
diagram when turning on $J$, while the energy flow at large energies 
and the FL fixed point towards $\omega_{l}\rightarrow0$ remains exactly 
the same.} At $n_d=2$, the spacing between $\Tkorb$ and $\Tkspin$, 
though, only spans about an order of magnitude which
is too small for the level flow to display a
stationary intermediate fixed point.

Above $\Tkorb$ the spectra 
correspond to the high energy physics of the Hubbard bands. 
Below $\Tkspin$ the excitation spectra reach a FL-fixed point  
with qualitatively identical multiplet
eigenlevel structures 
 for all values of $n_d$, $U$, and $J$: they can be interpreted in terms of
non-interacting single-particle excitations
[see also the $|\omega|^1$-scaling of $\chi''_{\orb}$ and $\chi''_{\spin}$ 
in Fig.~\ref{fig:P1_nd3_3fig}(d)].

\begin{figure}
\centering
\includegraphics[width=0.65\linewidth, trim=0mm 28mm 0mm 0mm, clip=true]{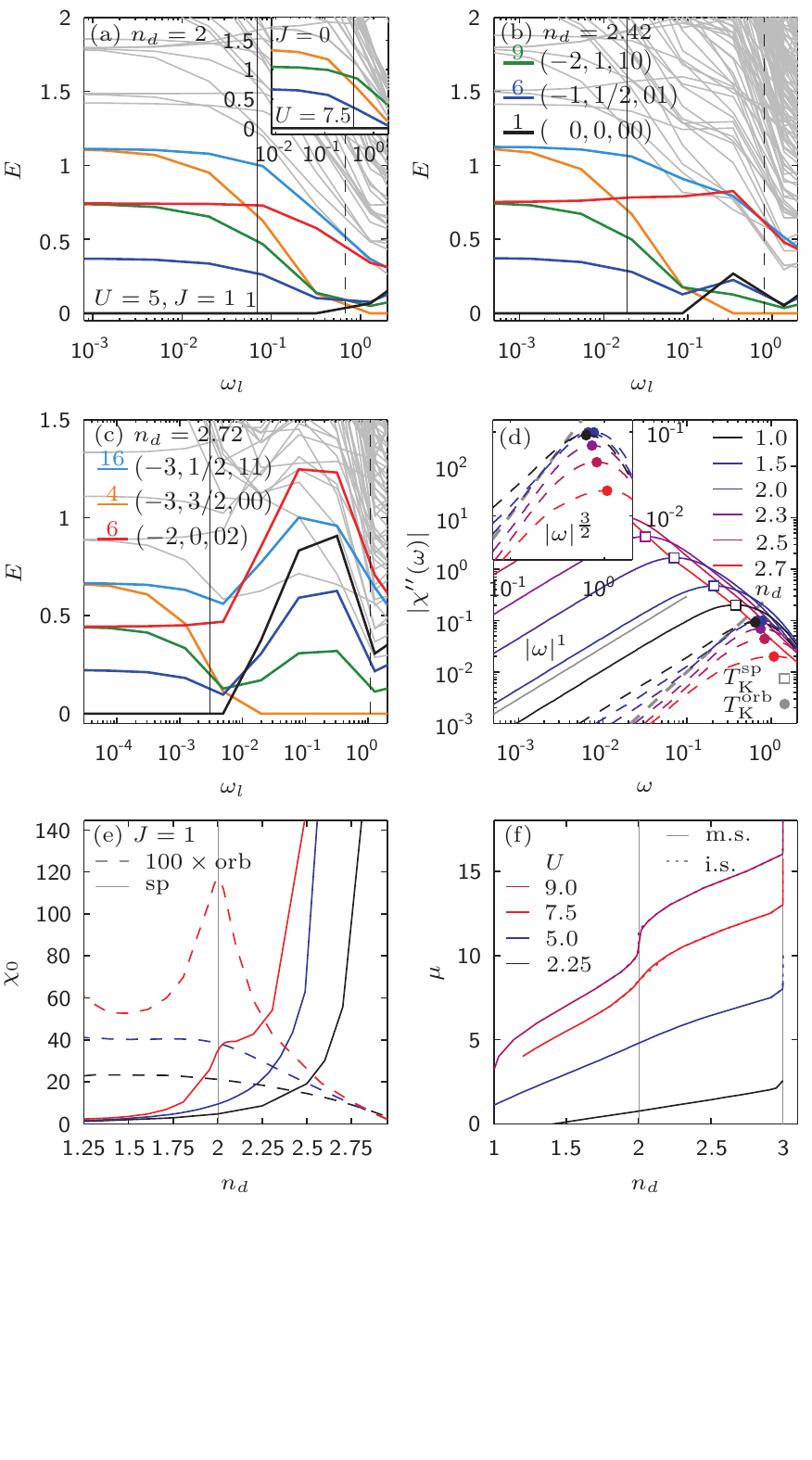}
\caption{ (a-c) Even NRG flow diagrams for different fillings, (a)
  $n_d=2$, (b) $n_d=2.42$, and (c) $n_d=2.7$.  The data represents
  rescaled energies of the lowest-lying eigenmultiplets of a Wilson
  chain of length $l$ plotted versus the characteristic level spacing
  $\omega_{l} \propto \Lambda^{-l/2}$ (see text).  NRG
    parameters: $\Lambda=4$, $E_{\mathrm{trunc}}=9$,
    thus keeping up to
    $D^\ast\lesssim 5,000$ $\mathrm{U(1)}_\mathrm{charge}\times
    \mathrm{SU(2)}_\mathrm{\spin}\times
    \mathrm{SU(3)}_\mathrm{\orb}$
    multiplets (corresponding to about $D=155,000$ 
    states)    \citep{Weichselbaum2012a,Weichselbaum2012b}. 
   The color specifies the symmetry sectors $(Q,S,q_1q_2)$
  (see text) as given in the legend.  Numbers above lines in the
  legend give multiplet degeneracies.  Solid (dashed) vertical lines
  mark the spin (orbital) Kondo scale, $\Tkspin$ ($\Tkorb$),
  respectively, where the range $\Tkspin< |\omega|< \Tkorb$ represents
  the SOS regime.  The inset of (a) shows, for comparison, the NRG
  flow for $J=0$ at $U=7.5$.  (d) The imaginary parts of the dynamical
  impurity orbital and spin susceptibilities,
  $|\chi''_{\orb} (\omega)|$ (dashed) and $|\chi''_{\spin}(\omega)|$
  (solid) for $U=5$, $J=1$ and various choices of $n_d$.  $\Tkorb$
  (filled circles) and $\Tkspin$ (open squares) are defined from the
  maxima of $\chi''_{\orb} (\omega)$ and $\chi''_{\spin}(\omega)$,
  respectively.  $\chi''_{\orb}(\omega)$ follows an apparent
  $|\omega|^{3/2}$ power law in the SOS regime (dashed grey
  guide-to-the-eye line) for fillings $2\lesssim n_d\lesssim 2.5$, which is
  likely just a cross-over behavior as seen from the flows in
  (a,b). Below $\Tkspin$, the $|\omega|^1$ FL power-law behavior sets
  in, indicated by a solid grey guide-to-the-eye line.  The inset
    is a zoom of $\chi''_{\orb} (\omega)$, revealing different
    ``slopes" of $\chi''_{\orb} (\omega)$ in the SOS regime for
    different $n_d$.  (e) The static local orbital and spin
  susceptibilities, $\chi_{0}^{\orb}$ (dashed) and $\chi_{0}^{\spin}$
  (solid) 
 are plotted as a function of
  $n_d$ for three different values of $U$ and $J=1$.
  (f) The chemical potentials, $\mu$, are plotted as functions of the
  filling $n_d$, for $J=1$ and various values of $U$ to study the
  behavior of the electronic compressibility $\kappa_{\rm el}$. 
 }
\label{fig:P1_nd3_3fig}
\end{figure}

We now focus on Fig.~\ref{fig:P1_nd3_3fig}(a) for $n_d=2$, $U=5$ and
sizeable $J=1$ (similar to Fig.~3(g)  in Ref.~\citep{Stadler2015}).
As $\omega_{l}$ drops below $\Tkorb$, orbital
screening sets in, favoring orbital singlets $q=(00)$
[black and orange curves], hence other multiplets rise in energy. 
For the same charge $Q$, 
large-spin multiplets lie lower in energy (green curve lies
below red one  for $Q=-2$,  and orange below bright blue for
$Q=-3$). As $\omega_{l}$ drops below $\Tkspin$, spin
screening sets in, favoring spin singlets and pushing up
multiplets with $S \neq 0$.  Now, multiplets with same
particle number but different spins become degenerate
(compare again green and red curves for $Q=-2$,  and orange
and bright blue curves for $Q=-3$).

Interestingly, with increasing $n_d$, where the spin-orbital
regime becomes wider, a new flow behavior slowly emerges at
energies entering from (just above) $\Tkorb$:
the multiplet with large spin $S=\rfrac{3}{2}$
and singlet orbital character $q=(00)$ 
[orange curve \Figs{fig:P1_nd3_3fig}(a-c)], which is
still outside the SOS regime at $n_d=2$ [\Fig{fig:P1_nd3_3fig}(a)],
moves into the SOS regime at $n_d=2.42$ [\Fig{fig:P1_nd3_3fig}(b)], and
takes over the `SOS regime' at $n_d=2.72$ [\Fig{fig:P1_nd3_3fig}(c)].
At the same time, the $\Tkorb$ moved upward and merges
with the bare atomic energy scales.
At $\Tkspin$, finally, a FL develops: the large spin
$S=\rfrac{3}{2}$ is screened and moves upward,
crossing multiple lines. The new ground state at energies
below $\Tkspin$ is the Kondo spin singlet (black line).

Note that the crossing of the large spin state (orange line)
starts just above $\Tkorb$ at $n_d=2$, and has moved all the way down
to $\Tkspin$ at $n_d=2.72$. In particular, we also emphasize
that the shoulder in $A(\omega)$ for $n_d=2$ in \Fig{fig:P1_nd3_2fig}(c)
emerges precisely around 
this crossing region. 
Therefore this qualitative change in the energy flow diagram is 
responsible that \textit{the intermediate SOS regime strongly changes
its character as the filling is increased
towards $n_d=3$}. At $n_d=3$ the SOS becomes trivial
in the sense that the orbital blocking is immediately
present due to the given filling. 

Importantly, the structure of the flow  below the crossing
region, i.e the transition behavior with decreasing
$\omega_{l}$ from the NFL into the FL fixed point is the same for all
fillings $2\le n_d<3$. It is therefore natural to assume
that also in the SOS regime at $n_d=2$, the
physics is governed by an underlying NFL fixed point (i.e. a fixed point that 
would show up for a larger SOS region as observed 
in a new analysis \citep{Walter2018} of the Kondo limit of the \HHM),
which also enforces the reversion of the lowest few multiplets
compared to the FL fixed point and has a $S=\rfrac{3}{2}$
and $(q_1q_2) = (00)$ multiplet as ground state. 

From the NRG flow analysis we deduce  the following generic
screening mechanism of SOS, which is visualized in
Fig.~\ref{fig:sos_sketch}(a) for $n_d=2$.

\textit{SOS is a two-stage screening process.} \textit{First
the orbital degrees of freedom are quenched below  the
orbital Kondo scale, $\Tkorb$.} In a Kondo-screening
language, described in the following  for $n_d=2$, we have
$S=1$  in the spin sector, while in the
orbital sector, we have the fundamental representation
$q=(10)$ with dimension $3$ [green lines in \Figs{fig:P1_nd3_3fig}(a-c)],
coupled to the $3$ channels, leading to full
orbital screening.  As a result of  this screening process
\textit{the impurity binds one electron from the bath to
form an orbital singlet.} This electron has a spin
$\rfrac{1}{2}$, which combines with the local spin $1$ --
due to ferromagnetic  Hund's coupling -- to a  spin
$\rfrac{3}{2}$.  \textit{Then, at a lower spin Kondo scale,
$\Tkspin$, this effective $\rfrac{3}{2}$ spin is fully
screened by the three bath channels of the \HHM.} The
formation of the orbital singlet causes the orbital
susceptibility to reach a maximum. The resulting free spin enhances the
spin susceptibility as the frequency decreases [see 
Fig.~\ref{fig:PRB-SOS:J}(a), 
Fig.~\ref{fig:PRB-SOS:U}(a) and also 
Fig.~\ref{fig:P1_nd3_3fig}(d)].
Since a bath electron with a specific orbital degree of
freedom is included in the orbital screening process,
\textit{ spin and orbital degrees of freedom are still
coupled, leading to a highly intertwined NFL in the SOS
regime at $n_d=2$.} The same screening process occurs, in
principle, for $2\le n_d<3$ as well, but the details vary
with filling. For $n_d$ approaching $3$, the  $\rfrac{3}{2}$
spin  is increasingly composed purely from the impurity
spin, which facilitates the formation of the orbital singlet
[$\Tkorb$ grows in  Fig.~\ref{fig:P1_nd3_2fig}(a)], but is
harder to be screened [$\Tkspin$ decreases in
Fig.~\ref{fig:P1_nd3_2fig}(a)]. Thus the contribution of the
bath electron in the screening process becomes less
important, and  the dynamics of the spin and orbital degrees
of freedom get more and more decoupled. For $n_d=3$, the
orbital singlet is directly and locally formed from the
impurity  $\rfrac{3}{2}$ spin without any involvement from
bath degrees of freedom.  Accordingly, in a  weak coupling
analysis \citep{Aron2015} of the \HHM, it is
emphasized that the spin Kondo scale depends
\textit{explicitly} on the representations of the spin and
the orbital isospin, which is unusual and only occurs for
complex Kondo models in which spins and orbitals are
coupled.

\subsection{Spin-orbital separation at $2\leq n_d<3$: susceptibilities}

In \Fig{fig:P1_nd3_3fig}(d), we analyze the behavior
of the imaginary parts of the dynamical impurity orbital
and spin susceptibilities
$\chi''_{\orb}$ and $\chi''_{\spin}$, 
for various fillings $n_d$ at $U=5$, $J=1$, and
in \Fig{fig:P1_nd3_3fig}(e) the behavior of the static local orbital
and spin susceptibilities $\chi_0\equiv \chi(0)$ 
for various $U$ at fixed $J=1$.
As already seen in \Fig{fig:AoP_spin_freezing_ndvar}(c),
with increasing filling between $1\le n_d <3$
in \Fig{fig:P1_nd3_3fig}(d),
the maxima of $\chi''_{\spin}$ ($\Tkspin$, marked by open
squares) increase in height and decrease in $|\omega|$,
and accordingly $\chi_{0}^{\spin}$ [solid curves in
Fig.~\ref{fig:P1_nd3_3fig}(e)] grows with $n_d$ for all
values of $U$. For $n_d\le2$, the enhancement of
$\chi_{0}^{\spin}$ is small and just part of 
an upward trend if $U\ll U_{c2}^{(2)}$ (black
und blue curves), but develops into a shoulder
if $U$ is close to the MIT at $n_d=2$ (red curve).
For $n_d>2$,  $\chi_{0}^{\spin}$
increases very strongly with growing $n_d$, almost diverging. 
In contrast, with increasing filling, $n_d\le2$, the maxima
of $\chi''_{\orb}$  almost coincide [see filled circles in
Fig.~\ref{fig:P1_nd3_3fig}(d)], and $\chi_{0}^{\orb}$ is
approximately constant for $U\le5$ [see  dashed black and
blue curves in Fig.~\ref{fig:P1_nd3_3fig}(e)]. Only for
$U=7.5$ much closer to $U_{c2}^{(2)}$,  $\chi_{0}^{\orb}$
first decreases and then strongly increases near the MIT at
$n_d=2$, indicating the presence of strong orbital
fluctuations. With increasing filling, $n_d>2$, the height
of the maxima of $\chi''_{\orb}$ declines  [see filled
circles in Fig.~\ref{fig:P1_nd3_3fig}(d)] and
$\chi_{0}^{\orb}$ drops to zero when approaching $n_d=3$,
for all values of $U$ [see dashed curves in
Fig.~\ref{fig:P1_nd3_3fig}(e)], reflecting the absence of
orbital fluctuations at this point. We remark that the
occurrence of a maximum in $\chi_{0}^{\orb}$  has also been
shown in DMFT+QMC calculations \citep{Yin2012}. 

In Fig.~\ref{fig:P1_nd3_3fig}(d), $|\omega|^1$-FL-scaling is
clearly observed in $\chi''_{\orb}$ and $\chi''_{\spin}$
below $\Tkspin$ for all values of $n_d$, as indicated by the
solid grey guide-to-the-eye line. Within the SOS regime
$\Tkspin<\omega<\Tkorb$, $\chi''_{\orb}$ shows NFL behavior
(no $|\omega|^1$-scaling) [see also inset of Fig.~\ref{fig:P1_nd3_3fig}(d)].  
With increasing $n_d>1$ and
widening SOS regime,  the ``slope" of $\chi''_{\orb}$ (on a
log-log plot) becomes steeper than in the FL regime, i.e. an
approximate  power-law would have a power larger than $1$.
For $2\lesssim n_d \lesssim 2.5$, $\chi''_{\orb}$ reaches an
approximate power of ${\frac{3}{2}}$.
This, however, is presumably not a pure power law, since
the SOS regime is not wide enough, i.e.
the RG flows of Fig.~\ref{fig:P1_nd3_3fig}(a-c) are yet far
from reaching a stationary fixed point in the SOS regime.
For $n_d>2.5$, however, the slope 
is again lowered to almost $1$. 
 
\textit{ Based on these observation and the RG flows we
argue that intriguing NFL behavior with relevance for Hund
metals occurs mainly in the filling regime of approximately
$1.5\lesssim n_d\lesssim2.5$.} Only there, a complex  two-stage
screening process \textit{couples} the dynamics of spin and
orbital degrees of freedom by the formation of a large,
effective Hund's-coupling induced $\rfrac{3}{2}$ spin
including a bath spin degree of freedom. Although fully
screened, the orbital degrees of freedom  still ``feel" the
slowly fluctuating, large local moments, which is reflected
in the fact that,  in the SOS regime in
Fig.~\ref{fig:P1_nd3_3fig}(d), the ``slope" of
$\chi''_{\orb}$ is increased compared to FL scaling. 

To summarize, \textit{we argue that  the suppression of $Z$
in the Hund metal regime  around $n_d\gtrsim2$ at moderate
$U\ll U_{c2}^{(2)}$ is mainly caused by SOS,} and thus by
the presence of a sizeable Hund's coupling in the system. It
is not triggered by Mottness (iii),  the proximity to the
MIT at half-filling, $n_d=3$.  Of course, as also known from
the MIT in the one-band Hubbard model, $Z$ is
\textit{further} lowered  by the proximity to the MIT at
$n_d=3$, but this effect is strong only  close to $n_d=3$
and is subleading in the Hund-metal regime. Further, the
physics close to $n_d=3$ is dominated by fully blocked
orbital degrees of freedom while for Hund metals the orbital degrees
of freedom play a subtle role in the nature of the NFL
physics.

We remark that our insights might be relevant to better
understand the physics of iron pnictides with hole and
electron doping \citep{Werner2012, Werner2016b}. For instance,
for BaFe2As2 (with a nominal $d6$ occupation in the parent
compound) correlations are enhanced upon approaching half
filling with hole-doping, achieved  by replacing Ba with
K, and reduced upon electron doping, achieved by
replacing Fe with Co \citep{deMedici2014}.

\subsection{Filling dependence of the compressibility, $\kappa_{\rm el}$}
We finish this section with a discussion of the compressibility  in 
Fig.~\ref{fig:P1_nd3_3fig}(f). We plot $\mu$ versus $n_d$ to access the zero-temperature
behavior of the electronic compressibility,  $\kappa_{\rm el}=\frac{\partial n_d}{\partial \mu}$, 
for finite $J=1$ and for several values of $U$, varying from slightly 
above $U_{c2}^{(3)}$ to slightly above $U_{c2}^{(2)}$.
Solid (dashed) lines are the results for a mS (iS), 
respectively.  Normally, $\kappa_{\rm el}$ has finite, positive values for 
metals and vanishes for insulators. 
We would like to investigate whether $\kappa_{\rm el}$ remains positive 
throughout, or  becomes negative for $n_d$ close to the MIT at $n_d=2$ 
or close to the MIT at $n_d=3$. The latter scenario, a zone of 
Hund's-coupling-induced negative compressibility in the $n_d$-$U$ 
phase diagram, has been observed in a slave-boson 
study \citep{deMedici2016} of degenerate and non-degenerate multi-band 
Hund models, for nonzero $J$ and $U\ge U_c$ at $T=0$. The divergence of 
$\kappa_{\rm el}$, when $\kappa_{\rm el}$ changes sign,  has been  
assumed to be connected to the enhanced critical $T_c$ of HTCS.
However, for the \HHM, for all parameters studied in Fig.~\ref{fig:P1_nd3_3fig}(f),  
$\mu$ clearly  increases monotonically  with $n_d$.
Hence the slope, 
$\kappa_{\rm el}$, is positive for all non-integer fillings,
also close to the insulating 
phase at $n_d=2$ and $n_d=3$, where $n_d$ is fixed and thus 
incompressible for varying $\mu$, i.e.
$\kappa_{\rm el}=0$.
We summarize that, for our study, \textit{no negative (or divergent) compressibility has been observed for the \HHM}. We note, though, that 
 in principle a compressibility divergence can occur very close to a MIT in certain situations \cite{Kotliar2002}.


\section{Conclusion} 
\label{sec:Conclusion}

In this work, we studied the full phase diagram of the \HHM at zero temperature with real-frequency 
DMFT+NRG data. Our main goal was to reveal the origin of the bad-metallic 
behavior (characterized by a low quasiparticle weight $Z$) 
in the Hund-metal regime (hatched area in 
\Fig{fig:sketch_phasediagram}) and to establish a 
global picture of SOS.

As a main result we demonstrated that, for nonzero 
$J$ and for fillings $1<n_d<3$, SOS 
is a generic feature  in the \textit{whole} metallic (and coexistence) 
phase of the \HHM, independently of $U$: turning on $J$ opens up 
a new incoherent energy regime, 
$\Tkorb>|\omega|,T>\Tkspin$,  in the system. 
Interestingly, for fillings around $n_d=2$ (i.e approximately 
in the regime $1.5\lesssim n_d\lesssim2.5$)  the SOS is special,
 as has been pointed out in Ref.~\citep{Yin2012}. 
There, orbital and spin degrees of freedom are \textit{coupled} 
and thus behave very distinctly: orbital degrees of freedom 
are (mostly) quenched below $\Tkorb$
and fluctuate rapidly, whereas
spin degrees of freedom are unquenched, form large local 
moments, and  fluctuate extremely slowly. Below, the strongly 
reduced spin Kondo scale, $\Tkspin$, both orbital 
\textit{and} spin degrees of freedoms are fully screened 
and FL behavior sets in. 

We confirm in detail that the suppression of $\Tkspin$
 with increasing $J$ can be explained from a qualitative
  change in the underlying local multiplet spectrum, 
  involving a reduction in the atomic ground state degeneracy. 
 $Z$ is explicitly shown to be proportional to $\Tkspin$, 
 and thus small due to SOS.
 
 In agreement with the analysis in the Kondo regime of the 
 \HHM \citep{Aron2015}, we argue that SOS is a non-trivial
  two-stage screening process, in which orbital and spin
   degrees of freedom are explicitly coupled: below 
   $\Tkorb$, the orbital degrees of freedom form an orbital singlet through
 the formation of a large, effective, Hund's-coupling induced
$\rfrac{3}{2}$ spin -- \textit{including} a bath spin degree of freedom; and 
below $\Tkspin$, the latter is fully screened by the three bath channels
of the \HHM. 

In the real-frequency spectral function, SOS results in a "two-tier" 
QPP peak with a narrow needle (width $\propto\Tkspin$) 
on top of a wide base (width $\propto\Tkorb$). 

Based on the SOS analysis we conclude, as major result of this work, that
in the Hund-metal regime, at sizeable $J$, moderate $U$ well below
$U_{c}^{(2)}$
 and fillings close to $n_d=2$, i.e far from any MIT, 
 Hundness, i.e scenario (i), is the origin of bad-metallic behavior and  governs the 
 physics of Hund metals. This constitutes a new route towards 
 strong correlations very distinct from Mottness: while in the latter case
 charges are localized in close proximity to an MIT, Hundness implies the
 localization of spins but not the localization of charges.
For Hund-correlated metals, $\Tkorb$ is comparable in magnitude to bare 
atomic energy scales of the system, while $\Tkspin$ (and thus $Z$) is 
strongly reduced, leading to low FL coherence scales and to a broad 
incoherent SOS regime. Hundness is thus physics governed 
by the QP needle being narrow, while the QP base remains wide.
 Importantly, this regime is characterized by 
the non-trivial interplay  of orbital and spin degrees of freedom, 
induced by the special two-stage SOS screening process, which essentially dominates 
 the normal-state incoherence of Hund metals.
We remark that Mottness of type (ii) does affect the SOS 
when the distance to the MIT  is decreased at fixed $n_d=2$, by 
further lowering $\Tkorb$ and $\Tkspin$, while their ratio remains constant.
Whereas $\Tkorb$ governs the Mott transition  (which requires the full QPP to
disappear), $\Tkspin$, being proportional to $Z$, governs the strength of
correlations. 

Mott-correlated metals, close to the MIT at $n_d\approx2$, are dominated by 
Mottness, while the SOS regime is strongly downscaled  and 
becomes negligible.

Close to the MIT at $n_d=3$, 
the SOS regime widens up because the orbital 
degrees of freedom get blocked by the formation of a $\rfrac{3}{2}$ impurity spin, 
but its nature changes: the orbital and spin dynamics get decoupled.
Thus, Mottness of type (iii) does not mediate the low $Z$ in the Hund-metal regime.

In sum, our DMFT+NRG results corroborate the physical picture of Hund metals 
established  in Refs.~\citep{Haule2009, Yin2012, Aron2015, Georges2013} and enabled the 
quantitative analysis of the real-frequency properties of their unusual incoherent 
SOS regime. 
We showed that the spin-freezing phenomenon \citep{Werner2008} and the 
Janus-faced influence of Hund's rule coupling can be consistently explained 
in the framework of SOS. We also explicitly demonstrated that 
no Hund's-coupling-induced FL instabilities 
(negative compressibilities) \citep{deMedici2016} occurs in our study of the
\HHM phase diagram.



\appendix

\section{Methods}
\label{appendix}

We treat the \HHM of Eq.~(\ref{eq:HU-Hloc}) with 
single-site DMFT and use  full-density-matrix (fdm)NRG  \citep{Weichselbaum2007} 
as real-frequency impurity solver. 

\subsection{Single-site Dynamical Mean-Field Theory}

Single-site DMFT  is a widely-used  non perturbative
many-body approach to strongly correlated systems \citep{Georges1996}.
Its basic idea is to approximate the full non local 
self-energy of the correlated lattice model by the purely local, 
but still frequency-dependent self-energy,  $\Sigma (\omega)$,
of the corresponding  self-consistently determined  quantum
impurity model. In our case, we iteratively map the
lattice \HHM of Eq.~(\ref{eq:HU-Hloc}) onto a
three-band Anderson-Hund model (AHM) of the form 

\begin{subequations}
\label{eq:AHM}
\begin{eqnarray}
\hat H_\AHM &=& \hat{H}_{\rm imp} +\hat{H}_{\bath} , 
\\
\hat{H}_{\rm imp}&=& \varepsilon_d
\,\hat{N}+\hat{H}_\interact [\hat d^\dag_\nu]
\label{eq:Himp}
\end{eqnarray} 
\end{subequations}
with the same local interaction term, $\hat H_\interact$, as
in Eq.~(\ref{eq:Hloc1}). Within this mapping process,  the
hybridization function $\Gamma(\varepsilon) = \pi \sum_k
|V_k|^2 \delta(\varepsilon - \varepsilon_k)$ is determined
self-consistently and eventually fully characterizes the
interplay of the impurity and the non-interacting three-band
spinful bath, 
\begin{eqnarray}
\label{eq:hyb}
H_\bath = 
\sum_{k\nu}\left(\varepsilon_{k} 
c^{\dagger}_{k\nu}\hat{c}^\pdag_{k\nu} + 
V_{k}\bigl[\hat{d}^{\dagger}_{\nu}\hat{c}^\pdag_{k\nu} 
+ \hat{c}^{\dagger}_{k\nu} \hat{d}^\pdag_{\nu} \bigr] \right ) . \qquad 
\end{eqnarray}
Here $d^\dag_\nu$ creates a local (``impurity'') electron of
flavor $\nu$ with energy $\varepsilon_d=-\mu$. 
The total spin operator $\hat {\mathbf{S}}$ (and $\hat {\mathbf{S}}_i$, respectively)
are lattice sums over
$(\hat n_i-N_c)$, i.e. charge relative to half-filling.
The average local site occupation number
$n_d \equiv \langle \hat{n}_i \rangle$
is a measure of the lattice
filling per site. 

The lattice dynamics is fully captured by the local retarded
lattice Green's function, $G_{\rm {latt}}(\omega)$ , which
is -- after the self-consistent mapping -- equal to the
retarded impurity Green's function,  $G_{\rm imp}(\omega)=
\langle \hat{d}^\pdag_\nu \mbox{$\parallel$}\, \hat{d}_\nu^\dag
\rangle_\omega$,  imposing the self-concistency condition:
$G_{\rm {latt}}(\omega)= G_{\rm imp}(\omega)\equiv
G(\omega)$. Note that we consistently drop the flavor index
$\nu$ for all  correlation functions as they are identical
by symmetry for all spins and orbitals.

In this work, we  study Hund metals only on the Bethe
lattice, i.e. we use the semi-elliptic density of states
that occurs in this limit of infinite lattice coordination
and neglect realistic band-structure effects, to investigate
the pure correlation effects of multi-orbital Mott and Hund
physics. The self-concistency condition can then be
simplified to,  
\begin{equation}
\Gamma(\omega)=-t^2 \imag G(\omega).
\label{eq-selfconcond}
\end{equation}

The approximation of a purely local self-energy in
single-site DMFT is strictly  valid only in the artificial
limit of infinite lattice coordination number. However, if
interactions act only locally in a lattice system with
finite coordination number, as in the case of Hund's rule
coupling which is adopted from local  atomic physics,
single-site DMFT  is assumed to be an appropriate method to
reproduce the correct physics. This assumption is supported
by recent cluster-DMFT calculations 
for Hund metals \citep{Semon2017}.
Further, single-site DMFT is in general able to capture
basic strong correlations effects  of finite dimensional
systems (like the MIT)  due to its non-perturbative
character: through the energy-dependence of the local
self-energy both the itinerant and localized nature of
electrons, and thus both  weak and strong correlations,
can be handled on equal footing. 

This is considered to be of utmost importance for the
description of iron-based HTSCs and other Hund metals, as
very likely, neither pure atomic physics nor pure band
theory does apply. In these bad-metallic multi-orbital
systems, the existence and interplay of itinerant, but
strongly renormalized electrons \textit{and} strongly, but
not fully localized large spin moments have to be analyzed
without any method-induced bias -- even far from any Mott
insulating state  \citep{Werner2016b,Yin2011,
Bascones2015a,Manella2014}.


\subsection{Numerical Renormalization Group}

In each step of the DMFT self-consistency loop, we solve the
quantum-impurity problem Eq.~(\ref{eq-selfconcond}) with
fdmNRG, a  powerful  impurity solver that offers numerically
exact real-frequency spectral resolution at arbitrarily low
energies and temperatures for multi-band impurity
models \citep{Weichselbaum2007,Weichselbaum2012b,Stadler2016}
\textit{and} lattice models in the DMFT
context \citep{Stadler2015,Lee2017,Lee2018}.

NRG \citep{Weichselbaum2012b,Wilson1975,Bulla2008} has a
longstanding and successful history as the standard tool to deal with
impurity models. Its basic idea goes back to Wilson's
fundamental insight \citep{Wilson1975} to introduce a
logarithmic discretization of the noninteracting bath
Eq.~(\ref{eq:hyb}) of an impurity model Hamiltonian and map
the discretized bath   onto a 1D semi-infinite,
tight-binding chain, a ``Wilson chain", with the interacting
impurity site coupled to one end.  The hopping matrix
elements then decay exponentially down the Wilson chain and
introduce an energy-scale separation that allows for an
iterative RG solution scheme based on successive
diagonalization and truncation of  high-energy states. The
size of the Fock state space can thus be kept fixed with
increasing chain length while still obtaining an
exponentially increased resolution of the low-energy part of
the spectrum.  The resolution at high energies is, however,
more coarse-grained.  Nevertheless, our approach captures
all essential high-energy features \citep{Weichselbaum2007}.

In recent years, significant progress has been made in
developing NRG into an efficient  high-quality multi-band
DMFT impurity solver \citep{Stadler2015,Lee2017, Stadler2016,
Lee2018}.  Our fdmNRG solver is implemented based on the 
QSpace tensor library \citep{Weichselbaum2012b} applied to
matrix product states (MPS)  \citep{Weichselbaum2012a,Schollwoeck2011} 
as generated in NRG. In the QSpace
tensor library, Abelian and non- Abelian symmetries are
implemented on a generic level: the state space is organized
into symmetry multiplets, and tensors ``factorize'' into two
parts, acting in the reduced multiplet space and the Clebsch
Gordon coefficient space, respectively. Diagonalization of
the NRG Hamiltonian at each iteration step can then be done
in multiplet space rather than state space, significantly
reducing the matrix sizes and hence computational cost.
NRG calculations with three and even more \textit{degenerate}
bands  \citep{Weichselbaum2012b,Walter2018,Stadler2016}
became feasible, also in the DMFT context  \citep{Stadler2015,Lee2018}.
For solving
our \HHM in Eq.~(\ref{eq:HU-Hloc}), we explicitly exploit its
$\text{U(1)}_\charge\times \text{SU(2)}_\spin
\times\text{SU(3)}_\orb$ symmetries.  We note that also
models with three (or even more) \textit{non-degenerate}
bands are within the reach of NRG, using iNRG, the
``interleaved" version of NRG \citep{Stadler2016}. It is thus
also possible to study orbital differentiation with
DMFT+iNRG, as will be demonstrated
elsewhere
 \citep{Kugler2018}.  
 
The  fdmNRG solver is established on a complete basis
set \citep{Anders2005,Anders2006}, constructed from the
discarded states of all NRG iterations.  Spectral functions
for the discretized model are given from the Lehmann
representation as a sum of poles, and can be calculated
accurately directly on the real-frequency axis in sum-rule
conserving fashion \citep{peters2006} at zero or arbitrary
finite temperature.  Continuous spectra are obtained by
broadening the discrete data with a standard log-gaussian
Kernel of frequency-dependent
width \citep{Weichselbaum2007,Bulla2008}.

To improve the resolution of spectral data, we ``$z$-average"
over the results obtained from several, differing NRG runs,
for which the logarithmic discretization  of the bath has
been uniformly shifted with respect to each other
 \citep{Zitko2009,Oliveira1994}. 
We note that, within DMFT, the NRG discretization scheme
(originally developed for the flat hybridization function
$\Gamma(\varepsilon)=\Gamma\Theta(D-|\varepsilon|$) of
quantum impurity models with half-bandwidth D=1) has to be
adapted to optimally discretize the frequency-dependent
hybridization functions that emerge in every step of the
self-consistency loop. Here, we use a numerically stable
implementation \citep{Lee2018a} of the scheme in
Refs.~\citep{Zitko2009,Zitko2009a} to accurately
represent the nontrivial continuous baths in terms of
discrete bath states.

Within the DMFT+NRG approach, the resolution of spectral
data can be further improved  by applying the so-called
self-energy trick \citep{Bulla1998}. In every step of the
iterative mapping,  the self-energy is calculated as the
ratio of two NRG correlation functions  \citep{Bulla1998}
\begin{equation}
\label{eq:sigma}
\Sigma(\omega)=\frac{F(\omega)}{G(\omega)},
\end{equation}
 where
$F(\omega)=\langle [\hat{d}^\pdag_\nu, \hat{H}_\interact
[\hat{d}^\dag_{\nu} ]] \mbox{$\parallel$}\, \hat{d}_\nu^\dag
\rangle_\omega$. The imaginary parts of both correlators,
${F(\omega)}$ and ${G(\omega)}$,  are fdmNRG spectral
functions while the real-parts are obtained from their
Kramers-Kronig transformations, respectively.  Instead of
using the raw NRG result ${G(\omega)}$ for the
self-consistency condition Eq.~(\ref{eq-selfconcond}), an
improved version of the (lattice) Green's function is
calculated via  the simple analytic form 
\begin{eqnarray} 
\label{eq:Gimpr}
G_{\rm {impr}}(\omega) = \frac{1}{2t^2}\left(\xi-\sqrt{\xi^2-4t^2}\right)
\end{eqnarray} 
with $\xi=\omega+\mu-\Sigma(\omega)$, valid only for the
Bethe lattice. In this work we only refer to the improved
Green's function and therefore drop the index from now on:
$G(\omega)\equiv G_{\rm {impr}}(\omega)$.

From the improved Green's function, we have direct
access to the real-frequency spectral function, also called
local density of states:
\begin{equation}
A(\omega) = -
\tfrac{1}{\pi} \imag G^R(\omega).
\label{eq-Aw}
\end{equation}

All computational parameters and further details of our
DMFT+NRG calculations are listed in the Supplementary
material of Ref.~\citep{Stadler2015}.

In Ref.~\citep{Stadler2015} we have already
demonstrated that DMFT+NRG is perfectly suited for the
investigation of the \HHM. The exponentially enhanced
resolution around the Fermi level resolves spectral features
down to the lowest relevant energy scale of the system. In
contrast to QMC solvers, the NRG solver thus reaches the
strongly reduced FL ground state in a $T=0^{+}$ simulation of
the model.  At the same time atomic-like features which
constitute the Hubbbard side bands are well 
reproduced, 
e.g. as 
shown in Sec.~\ref{sec:Janus} and
Sec.~\ref{sec:half-filled-MIT}. The access to real-frequency
quantitities helps us to understand the nature of the
incoherent regime together with NRG eigenlevel
renormalization group (RG) flow diagrams that reveal the
relevant physics at all energy scales.



\bibliographystyle{elsarticle-num-names} 
\bibliography{biblio-HundvsMott1}

\end{document}